\documentclass[10pt]{iopart}

\usepackage{graphicx}
\usepackage{amssymb,bm}
\usepackage{mathrsfs}
\usepackage{geometry}
\usepackage[hidelinks]{hyperref}
\usepackage[outdir=]{epstopdf}
\usepackage{color}
\usepackage{cite}
\usepackage[title]{appendix}
\usepackage[hidelinks]{hyperref}
\usepackage{url}

\usepackage{xcolor} 

\definecolor{darkgreen}{RGB}{34, 139, 34}

\begin{document}




\title[Tokamak edge-SOL turbulence simulations in H-mode conditions]{Tokamak edge-SOL turbulence in H-mode conditions simulated with a global, electromagnetic, transcollisional drift-fluid model}

\author{Wladimir Zholobenko, Kaiyu Zhang, Andreas Stegmeir, Jan Pfennig, Konrad Eder, Christoph Pitzal, Philipp Ulbl, Michael Griener, Lidija Radovanovic$^{1)}$, Ulrike Plank, and the ASDEX Upgrade Team$^{2)}$}

\address{Max-Planck-Institut f\"ur Plasmaphysik, Boltzmannstra{\ss}e 2, 85748 Garching, Germany

$^{1)}$ Institute of Applied Physics, TU Wien, Fusion@ÖAW, Wiedner Hauptstr. 8-10, 1040 Vienna Austria

$^{2)}$ see author list of H. Zohm \textit{et al}, 2024 \textit{Nucl. Fusion}, https://doi.org/10.1088/1741-4326/ad249d

\medskip
Email: Wladimir.Zholobenko@ipp.mpg.de}

\begin{abstract}

The design of commercially feasible magnetic confinement fusion reactors strongly relies on the reduced turbulent transport in the plasma edge during operation in the high confinement mode (H-mode). 
We present first global turbulence simulations of the ASDEX Upgrade tokamak edge and scrape-off layer (SOL) in ITER baseline H-mode conditions. 
Reasonable agreement with the experiment is obtained for outboard mid-plane measurements of plasma density, electron and ion temperature, as well as the radial electric field. 
The radial heat transport is underpredicted by roughly 1/3. 
These results were obtained with the GRILLIX code implementing a transcollisional, electromagnetic, global drift-fluid plasma model, coupled to diffusive neutrals. 
The transcollisional extensions include neoclassical corrections for the ion viscosity, as well as either a Landau-fluid or free-streaming limited model for the parallel heat conduction. 
Electromagnetic fluctuations are found to play a critical role in H-mode conditions. 
We investigate the structure of the significant $E \times B$ flow shear, finding both neoclassical components as well as zonal flows. But unlike in L-mode, geodesic acoustic modes are not observed. 
The turbulence mode structure is mostly that of drift-Alfvén waves. However, in the upper part of the pedestal, it is very weak and overshadowed by neoclassical transport. At the pedestal foot, on the other hand, we find instead the (electromagnetic) kinetic ballooning mode (KBM), most clearly just inside the separatrix. 
Our results pave the way towards predictive simulations of fusion reactors.

\end{abstract}

\noindent Published in \textit{Nuclear Fusion} under \url{https://doi.org/10.1088/1741-4326/ad7611}\\
\noindent Last edited: \today

\maketitle

\section{Introduction}

High confinement of heat is critical for a fusion reactor to keep the fusion fire burning. Under most circumstances, magnetic confinement tends to degrade with the applied heating power, but it improves with machine size \cite{Zohm2019, ConfinTransport1999}, setting a minimum required machine size to achieve a burning plasma. Technologically, machine size can be compensated by a stronger magnetic field, e.g.~by using high temperature superconductors \cite{Sorbom2015}. However, both the achievable magnetic field strength and machine size are limited, and both strongly increase the costs of the fusion reactor. Therefore, it is of paramount interest for fusion research to find regimes of operation that have the maximum achievable confinement at given engineering constrains.

One of the most successful regimes discovered to date is the so-called H-mode \cite{Wagner1982,Wagner2007}, a high-confinement regime with a layer of suppressed turbulence in the very edge of the confined plasma region, the pedestal. It occurs naturally beyond a power threshold, especially in diverted tokamaks, and boosts the plasma confinement by about a factor 2 compared to the standard, low-confinement (L-mode) conditions. Of practical importance in current research are especially the scaling of the heating power required for H-mode access \cite{Martin2008,Ryter2013,Bourdelle2020}, avoidance of large edge-localised mode (ELM) instabilities when the edge pressure and current gradients become too steep \cite{Snyder2002,Viezzer2018}, as well as the integration with a dissipative divertor solution \cite{Bernert2020,Wang2021}. Beyond empirical scaling laws, first-principle models are desired to project the strategies to address these requirements from current machines to reactors.

Many theories have been proposed to explain the L-H transition \cite{Connor1999,Bourdelle2020}. Some clearly identified ingredients include the edge ion heat flux \cite{Ryter2014}, but not the total or electron heat flux. The formation of a strong radial electric field shear $\partial_r E_r$ and flow shear in general are important \cite{Biglari1990,Groebner1990,Burrell1997}. Thereby, different mechanisms for the $E\times B$ flow generation can compete. In the confined region, the flow is determined by neoclassical friction and viscous forces \cite{Hinton1976,Stroth2011,Cavedon2016} as well as zonal turbulence-driven contributions \cite{Diamond2005,DifPradalier2009,Zholobenko2021,Grover2024}. In the SOL, the dominant effects are parallel currents and sheath physics \cite{Stangeby2000,Brida2020}. Of course, the reduction of turbulence intensity depends on the mechanisms of turbulence drive and saturation in L- and H-mode. Turbulence has been proposed to be driven by a combination of drift-Alfvén-waves (DAW), kinetic and resistive ballooning modes (K/RBM), ion temperature gradient (ITG), trapped electron (TEM), and microtearing modes (MTM) on ion scales $k_y \rho_\mathrm{i} \gtrsim 1$ \cite{Scott1998,Rogers1998,Snyder2011,Dickinson2012,Hatch2016,Ku2018,Zocco2018,Bonanomi2019,Kotschenreuther2019,Scott2021,Scott2021a,Eich2021,Leppin2023,Bonanomi2024}. On electron scales $k_y \rho_\mathrm{e} \gtrsim 1$, electron temperature gradient (ETG) driven turbulence is thought to contribute to electron heat transport \cite{Hatch2016,Kotschenreuther2019,Scott2021a,Leppin2023}. Most authors stress the importance of electromagnetic fluctuations and $E\times B$ shear \cite{Eich2021,Bonanomi2024}. Edge turbulence can also interact with larger scale MHD activity such as ELMs \cite{Kendl2010,Snyder2011,Li2022}. The main challenge for H-mode predictions thus remains the general understanding of edge plasma turbulence, which is also difficult to simulate due to the strong variation of magnetic geometry \cite{Kendl2006,Fedorczak2012,Joffrin2017}, the interaction with neoclassical and SOL flows \cite{DifPradalier2009,Zholobenko2021}, and neutral gas \cite{Stotler2017,Zholobenko2021a}.


In the present work, we approach the challenge of H-mode turbulence simulations with the GRILLIX code \cite{Stegmeir2019}. 
GRILLIX is built on the flux-coordinate independent (FCI), locally field-aligned approach \cite{Stegmeir2016,Stegmeir2017,Stegmeir2018,Michels2021,Stegmeir2023}: this allows to perform efficient turbulence simulations in (also advanced) diverted geometry \cite{Body2019,Zholobenko2021FEC}. GRILLIX has been previously validated on the linear LAPD device \cite{Ross2019}, in attached L-mode TCV \cite{Oliveira2022} and ASDEX Upgrade \cite{Zholobenko2021a} tokamaks. Our model is based on global drift-reduced Braginskii equations \cite{braginskii65,Zeiler1997,Zholobenko2021}: together with the FCI discretisation, this allows to perform simulations in a domain spanning from the pedestal top to the divertor, with proper variation of geometry, plasma background profiles and parameters. 
To maintain realistic background profiles, the plasma is coupled to a diffusive neutral gas model \cite{Zholobenko2021a}. 
Insulating sheath boundary conditions are implemented at the divertor with the penalization method \cite{Stegmeir2019}. 
Fluctuations of arbitrary amplitude are permitted and evolved together with the plasma background. This is particularly important in the SOL \cite{Zholobenko2023}, but it also provides a self-consistent evolution of $E \times B$ flows \cite{DifPradalier2009,Zholobenko2021}. 
Since the drift-fluid plasma model is still missing important kinetic effects, in particular trapped electron modes, the gyrokinetic code GENE-X \cite{Michels2021,Michels2022,Ulbl2023} is being developed in tandem with GRILLIX. Also electron scale ETG and current gradient driven peeling modes remain to be included for future work. 

\begin{figure}[htb]
\centering
    \centering
    \includegraphics[trim=0.0cm 0.0cm 0.0cm 0.0cm, clip, width=0.995\linewidth]{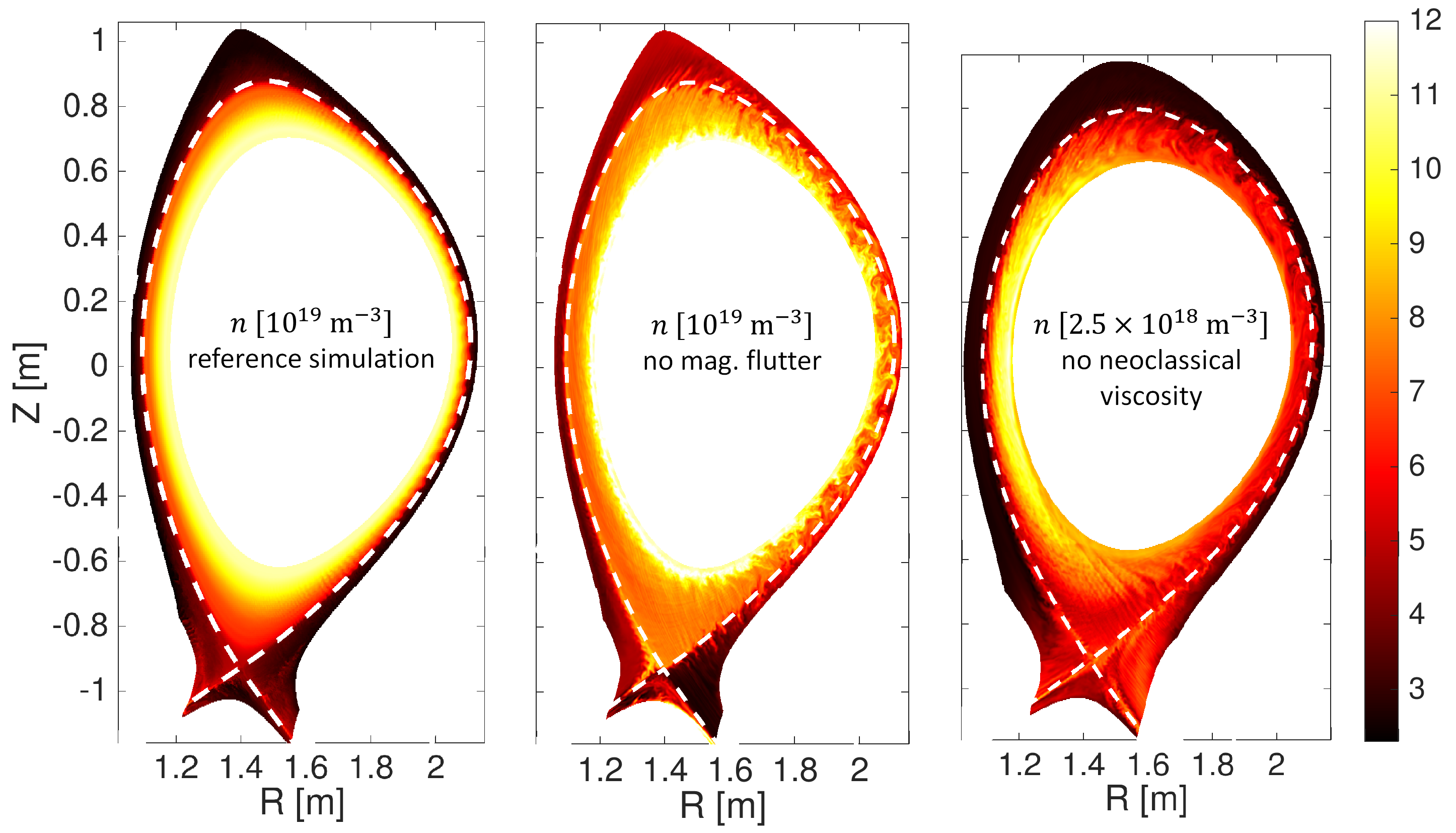}
\caption{2D snapshots of the plasma density in the poloidal cross section for 3 simulations, from left to right: the (high resolution) reference simulation for this paper (see section \ref{sec:setup}), a simulation without magnetic flutter (discussed in section \ref{sec:flutter}), and a simulation without neoclassical viscosity corrections (at lower collisionality, discussed in section \ref{sec:neocl_visc}). 
The white dashed line highlights the separatrix. The plasma core at $\rho_\mathrm{pol}<0.9$ is not simulated (there is an artificial boundary).}
\label{fig:snapshot}
\end{figure}

Here we present, to our knowledge, the first global turbulence simulations in H-mode conditions with a drift-fluid model. We compare our results with experimental data, analyze the electric field shear, and characterize the turbulence. But additionally, we emphasise what made these simulations possible in the first place: figure \ref{fig:snapshot} highlights the importance of electromagnetic fluctuations and low-collisionality extensions of the fluid closure. 
Magnetic flutter \cite{Zhang2024} in particular has been found crucial for the stabilization of drift-wave turbulence in H-mode. We also stress the need in global ``full-$f$'' turbulence models \cite{Scott2006,Hager2020,Giacomin2022,Zhang2024} for a consistent treatment of the Shafranov shift, which is known to be important for the pedestal stability \cite{Snyder2007} and transport \cite{Lackner2000}. 
While a gyrokinetic description is generally more adequate for the plasma edge, it is interesting to investigate what exactly goes wrong with the collisional fluid closure, and how it can be amended. Drift-fluid models are useful because they are very well suited for the description of the highly collisional divertor, and they are also generally computationally more affordable. But a collisional closure, e.g. Braginskii \cite{braginskii65}, overestimates parallel heat and viscosity fluxes by orders of magnitude at lower collisionality. Our solutions are firstly the approximation of the conductive heat flux with either a Landau-fluid model \cite{hammett_perkins,Umansky2015,Chen2019,Zhu2021,Pitzal2023}, or free-streaming limited expressions \cite{Thyagaraja1980,Scott1997,Stangeby2000,Fundamenski2005,Xia2015,Zholobenko2021a}. In particular, we highlight the connection between the two. And secondly, neoclassical corrections are implemented for the ion viscosity, approximating flow damping by trapped ions \cite{Hirshman1981,PerHelander2002,Rozhansky2009}. 

The paper is organized as follows. In section \ref{sec:validation}, we discuss the simulation setup of the ASDEX Upgrade (AUG) discharge $\#40411$, and compare the simulated and measured outboard mid-plane (OMP) profiles. 
Then, in section \ref{sec:transcollisional}, we demonstrate the role of the recent electromagnetic and transcollisional model extensions (the full model is detailed in appendix \ref{chap:Braginskii_equations}). 
In section \ref{sec:Er}, we investigate the composition of the radial electric field in terms of neoclassical and zonal flow components. Section \ref{sec:characterisation} is devoted to a closer characterisation of the kind of turbulence observed in our simulations. 
Finally, conclusions are drawn in section \ref{sec:conclusions}, and an outlook is given.

\section{AUG H-mode simulation setup and profiles}
\label{sec:validation}

Towards predictions of H-mode turbulence, our first step is to compare our simulations against experiments of the ASDEX Upgrade (AUG) tokamak. In this section, first, the simulation setup and the saturation behaviour are described. Then, the numerical resolution and computational costs are discussed. Finally, outboard mid-plane profiles of density, electron and ion temperature, as well as the radial electric field are compared against experimental measurements. 

\subsection{Simulation setup, sourcing and saturation}
\label{sec:setup}

In the present work, our computations are based on the AUG discharge $\#40411$, an ITER baseline attached H-mode \cite{Zohm2015,Ryter2021}. The magnetic equilibrium is reconstructed at $t = 2.64$ s. An impression of the geometry is given in figure \ref{fig:snapshot}: 
we are simulating turbulence globally in a domain extending from the pedestal top to the divertor. Compared to most AUG discharges, ITER baseline has a relatively high triangularity of $\delta = 0.36$. In fact, it is close to a double-null configuration, with a second separatrix at $\rho_\mathrm{pol} \approx 1.037$, where $\rho_\mathrm{pol}$ is the poloidal flux radius \cite{Zholobenko2021}. For simplicity, we limit our domain to $\rho_\mathrm{pol} \in (0.9, 1.036)$. No flux (Neumann) boundary conditions are applied on the density and temperatures at the limiting flux surfaces, forcing the plasma to flow out only through the lower divertor. The elongation is $\kappa = 1.76$. A second important characteristic of the ITER baseline setup is the relatively low $q_{95} = 3.3$. For ITER, it is foreseen to operate at maximum achievable plasma current, which due to the Greenwald density limit allows to maximize the plasma density \cite{Greenwald2002,Eich2021}. This leads to low $q_{95}$. For the ITER baseline scenario at ASDEX Upgrade, this low $q_{95}$ value is achieved by a plasma current of $I_p = 1.1$ MA, in combination with a reduced toroidal magnetic field of $B_\mathrm{tor} = -1.9$ T on axis, with the minus sign indicating a favourable configuration with $\mathbf{B} \times \nabla B$ pointing towards the primary (lower) X-point. This reduced magnetic field is advantageous for the comparison with numerical simulations, since it allows to perform them at higher spatial resolution in comparison to the (larger) Larmor radius, and time resolution in comparison to Alfvénic frequencies, at lower computational costs. The heating in the discharge is composed of roughly 1.2 MW Ohmic, 4.5 MW neutral beam injection (NBI) and 2.4 MW ion cyclotron resonance heating (ICRH). Subtracting 2.5 MW of radiation, the total power crossing the separatrix is roughly 5.6 MW. Thereby, the stored energy variation $\mathrm{d}W/\mathrm{d}t$ due to ELMs is estimated to be 2.5 MW, leaving 3.1 MW of heat transport for the inter-ELM phase. We note that for technical reasons, NBI heating was pulse-width modulated by $\pm1$ MW. The impurity content was very low, with $Z_\mathrm{eff} < 1.5$, justifying simulations with just two charged species: deuterons and electrons. Naturally, quantitative differences can be expected nonetheless when impurities are included in future. 

\begin{figure}[htb]
\centering
\begin{minipage}{0.46\textwidth}
	\centering
    \includegraphics[trim=1.0cm 0.2cm 1.0cm 1.0cm, clip, width=1.0\linewidth]{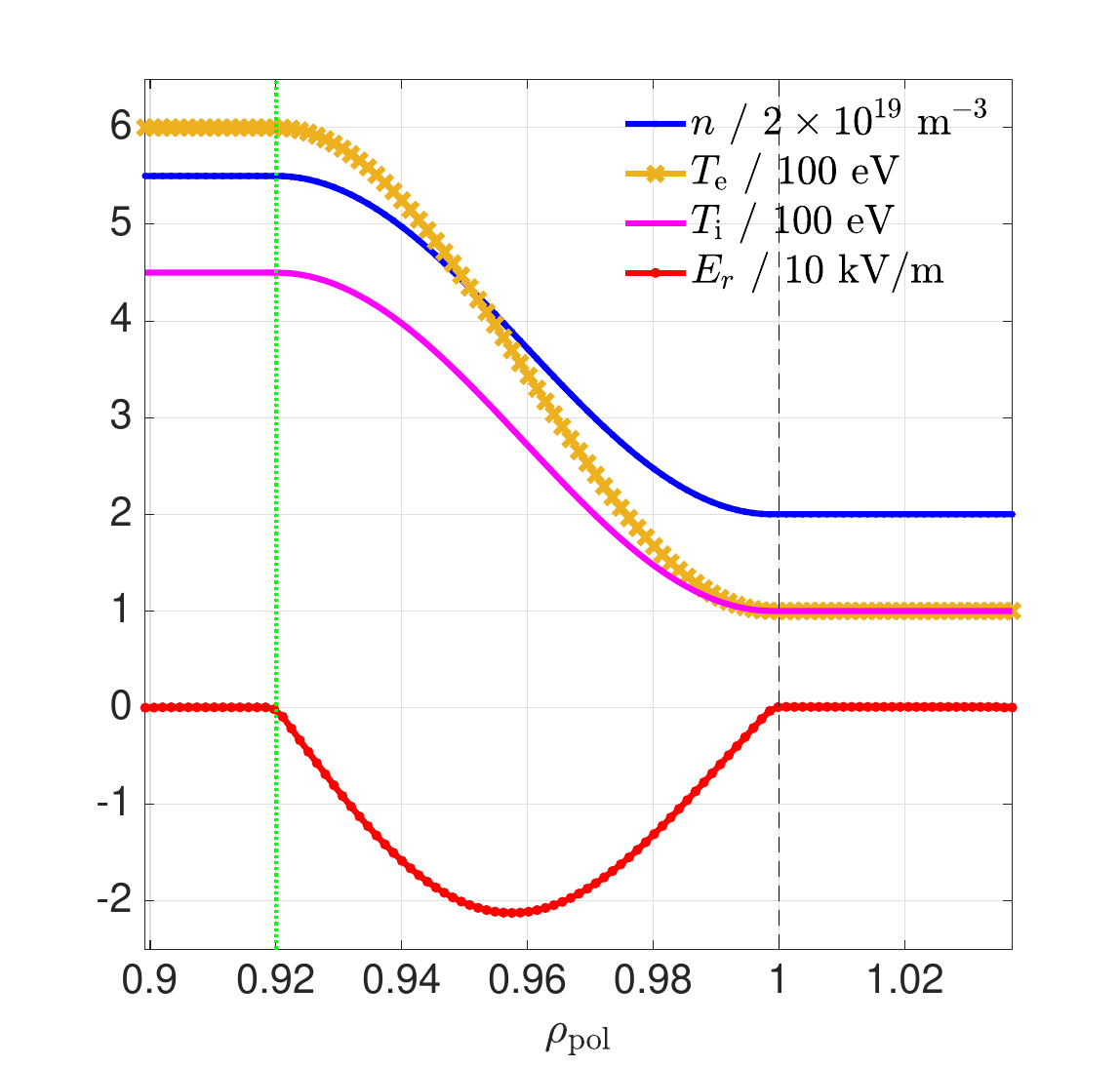}
    \caption{Initial density, temperature and electric field profiles for our simulations. The green line indicates the sourcing region.}
    \label{fig:initial}
\end{minipage}
\begin{minipage}{0.51\textwidth}
	\centering
    \includegraphics[trim=0.0cm 0.0cm 1.0cm 0.0cm, clip, width=1.0\linewidth]{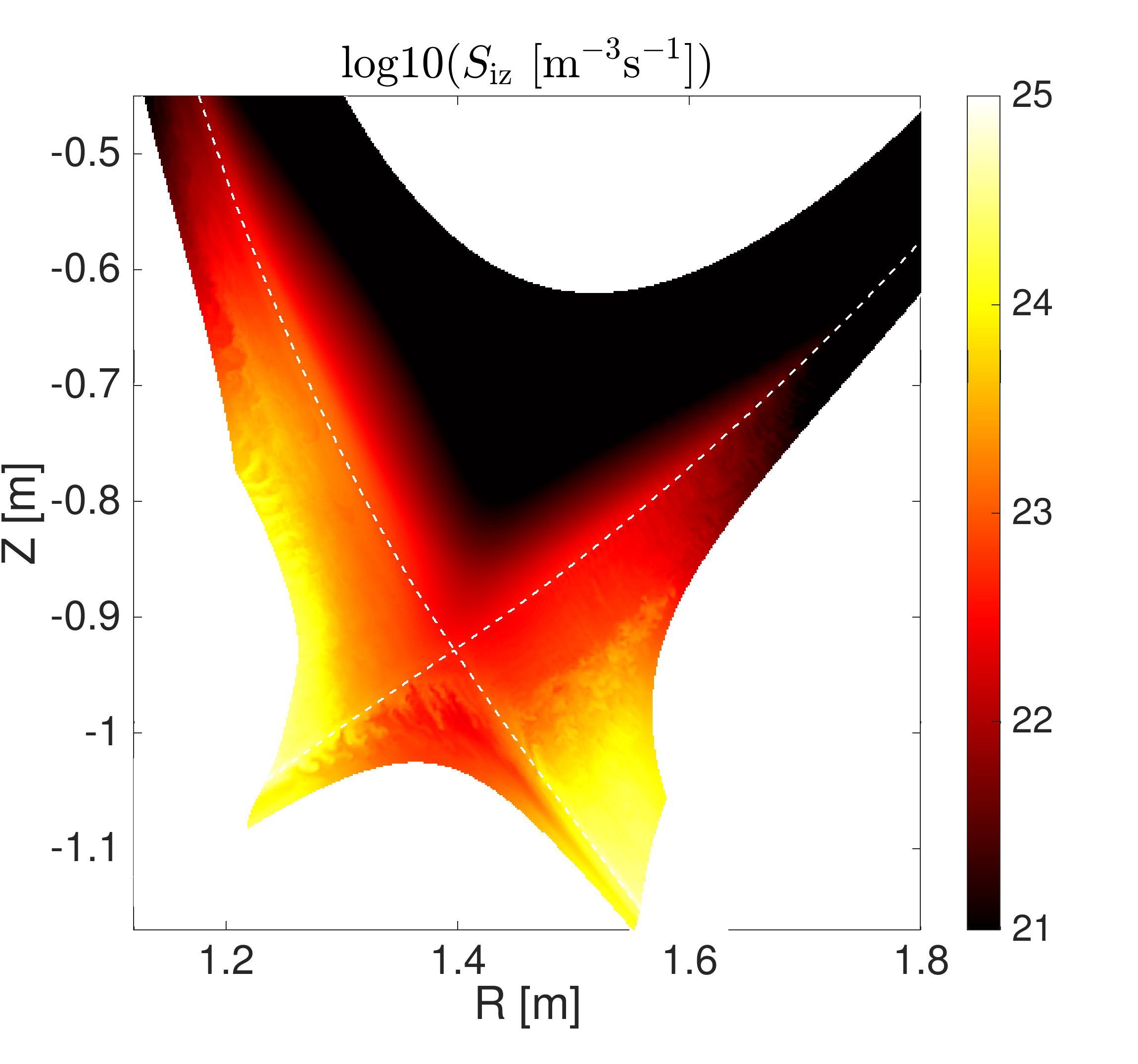}
    \caption{Distribution of the ionization source density (a snapshot).}
    \label{fig:sizrz}
\end{minipage}
\end{figure}

In global ``full-$f$'' simulations, due to the concomitant evolution of the background profiles, care needs to be taken in the setup to allow simulations to reach a steady state on time scales of the turbulence, and not global transport time scales. 
The lower bound for the saturation time is set by the SOL transit time $\tau_\mathrm{SOL}\approx L_c/u_\parallel \lesssim 1$\,ms, with the connection length $L_c \approx 20$\,m and parallel velocity at the separatrix $u_\parallel \approx 30$\,km/s. The upper bound is set by the confinement time $\tau_E = W/P_\mathrm{tot} \approx 30$\,ms, estimated from the total heat transport $P_\mathrm{tot}$ and the stored energy $W$. 
A reduced domain ($\rho_\mathrm{pol}>0.9$) lowers $W$ and thus $\tau_E$ in the simulations compared to the experiment. Additionally, reasonable initial profiles and core boundary treatment help the saturation to occur on time scales closer to $\tau_\mathrm{SOL}$ and not $\tau_E$. 
The initial profiles are shown in figure \ref{fig:initial}: they are parameterized by a simple sine function \cite{Zholobenko2021} to avoid a bias towards experimental profiles, the final results do not strongly depend on the initial state. But, since the saturation time does, core boundary and SOL values (and $W$) are chosen close to the experiment. 
At $0.9 < \rho_\mathrm{pol} < 0.92$, our simulations are ``adaptively'' flux-driven: a source keeps the plasma density fixed to $n_\mathrm{core} = 1.1 \times 10^{20}$\,m$^{-3}$, the electron temperature fixed to $T_\mathrm{core}^\mathrm{e} = 600$\,eV and the ion temperature to $T_\mathrm{core}^\mathrm{i} = 450$\,eV, based on experimental measurements. 
This avoids large profile variations in the initial transient phase, especially towards numerically problematic (too steep or hollow) profiles, and also accelerates saturation. 
Finally, the particle source is mostly determined by the ionization of neutral gas $S_\mathrm{iz}$, similarly as in \cite{Zholobenko2021a}. The neutral gas density at the divertor has been fixed to $N_\mathrm{div} = 5\times10^{18}$ m$^{-3}$. A snapshot of $S_\mathrm{iz}$ in the reference simulation is shown in figure \ref{fig:sizrz}. In the reference simulation, the adaptive core particle source $S_n^\mathrm{core}$ (which can be considered to be due to NBI) saturates at $S_n^\mathrm{core}/S_\mathrm{iz}\approx 0.3\%$. 

\begin{figure}[htb]
\centering
\begin{minipage}{0.49\textwidth}
	\centering
    \includegraphics[trim=0.5cm 0.0cm 1.0cm 0.0cm, clip, width=1.0\linewidth]{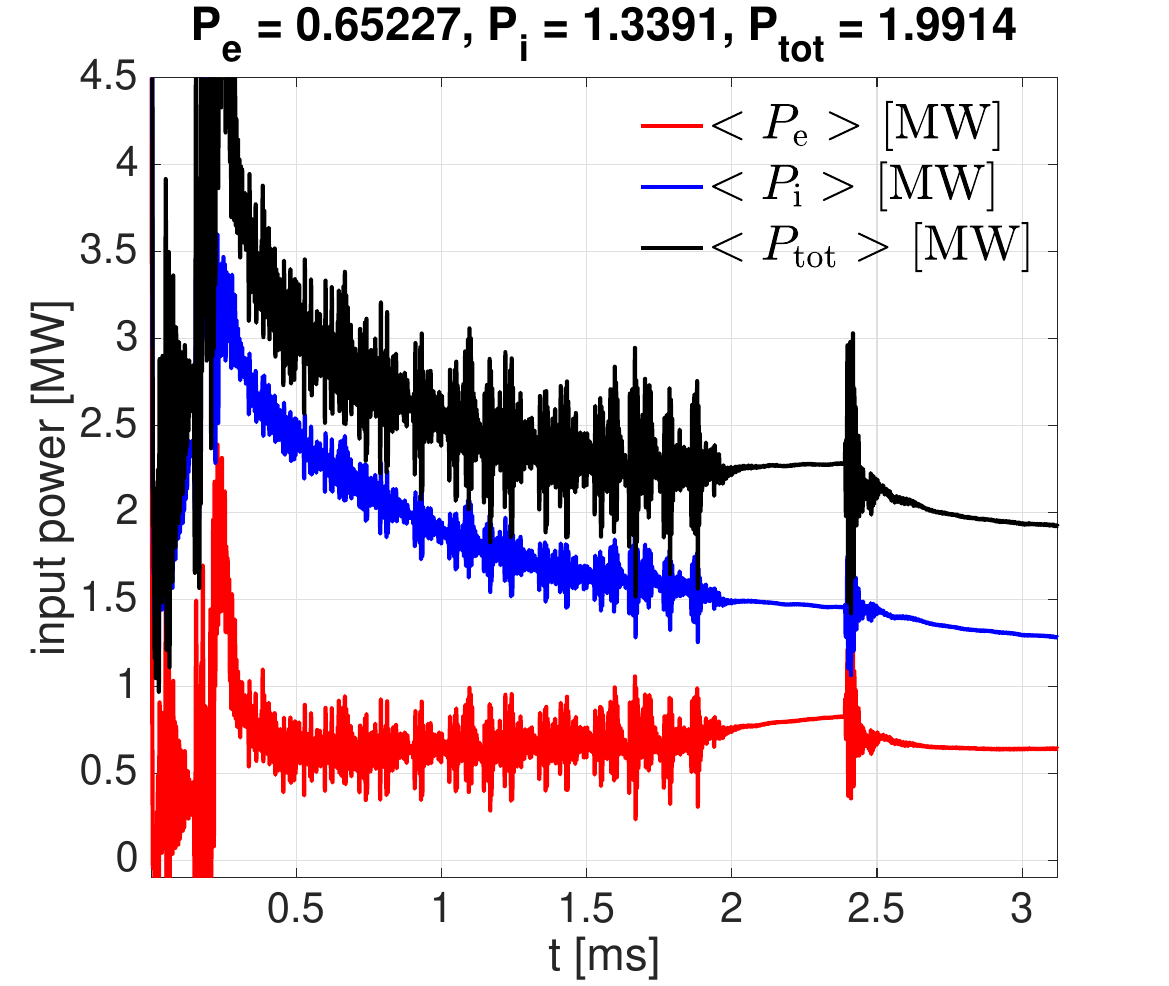}
    \caption{Time traces of input heating power for electrons and ions.}
    \label{fig:heat_in}
\end{minipage}
\begin{minipage}{0.49\textwidth}
	\centering
    \includegraphics[trim=0.8cm 0.0cm 1.0cm 0.0cm, clip, width=1.0\linewidth]{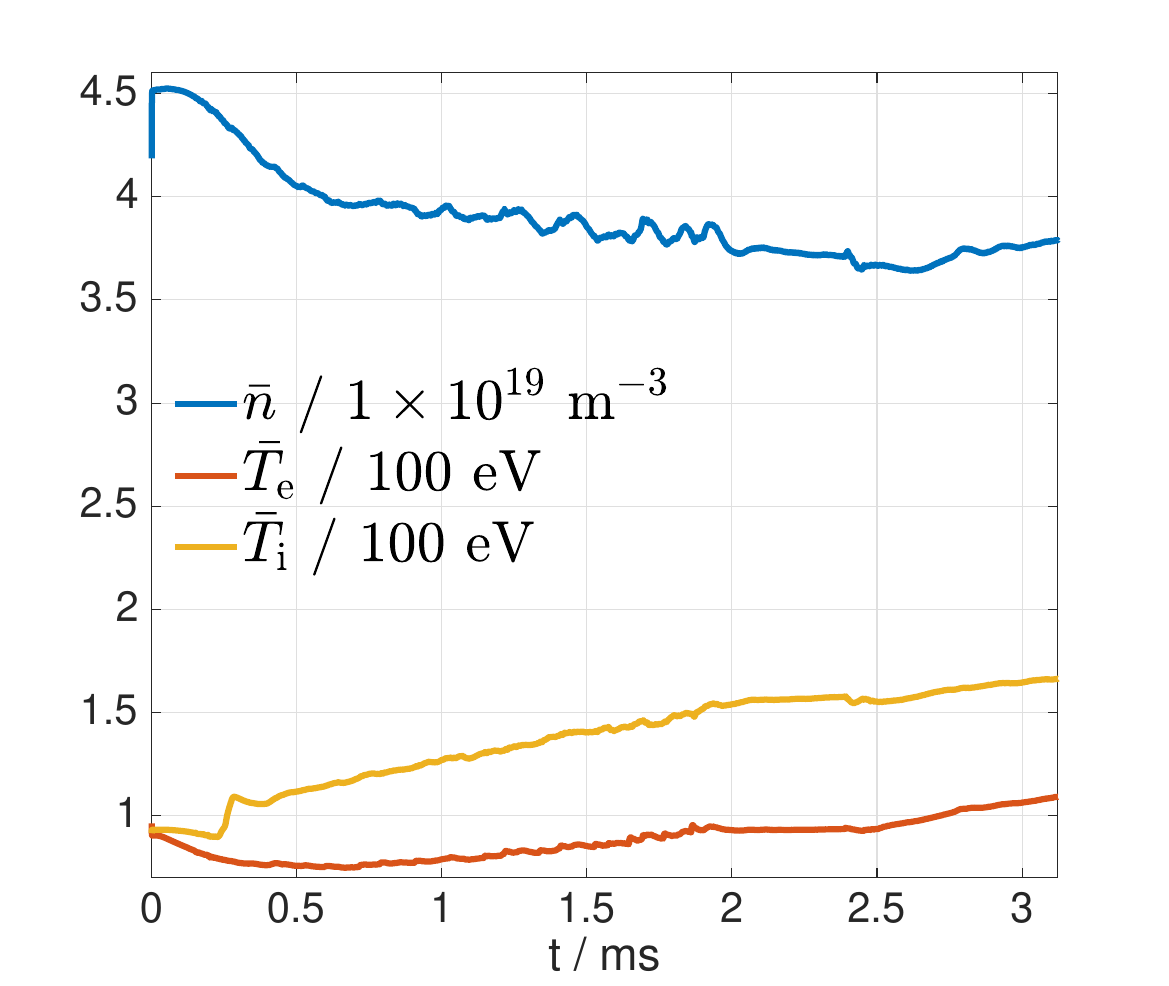}
    \caption{Time traces of zonally averaged plasma density, temperatures and electrostatic potential at the separatrix.}
    \label{fig:sep_sat}
\end{minipage}
\end{figure}

The time evolution of the adaptive heat source is displayed in figure \ref{fig:heat_in}. 
It saturates within 2 ms at around 2 MW, roughly 2/3 of what is expected from the experiment. 
However, the fluctuations in the first 2 ms of the simulation are caused by spurious electromagnetic transport, which is due to the periodic removal of the ``double Shafranov shift'', detailed in section \ref{sec:magnetic_shift}. These become more problematic with the course of the simulation, so the removed Shafranov shift was frozen after 2 ms (and updated again around 2.5 ms, leading to another short burst) -- then, the final phase of the simulation is smooth and stationary. 
Figure \ref{fig:sep_sat} shows time traces of plasma density and temperatures in a zonal average close to the separatrix ($\rho_\mathrm{pol} = 0.998$): after 2 ms of simulation time, they are quasi-stationary. The profiles are saturated on the relevant turbulence time scales, and can be evaluated down below, but they still very slowly evolve on transport time scales. 

\subsection{Resolution and cost}

For computer simulations, it is always advisable to check the numerical convergence by varying the grid resolution. Of course, this comes with a computational cost. In our lowest resolution simulation, within a poloidal plane, we employ an isotropic grid distance of $h = 1.44$ mm, corresponding to 1.9 sound Larmor radii $\rho_{s0} = \sqrt{m_\mathrm{i}T_\mathrm{e0}}/eB_0$ at reference $T_\mathrm{e0} = 100$ eV and $B_0 = 1.9$ T in deuterium. The effective resolution is better in the hot plasma core and worse in the cold divertor. Additionally, we have performed simulations with $h = 1\rho_{s0} = 0.76$ mm, also at higher toroidal resolution. 
Toroidally, we have utilised 16 and 32 poloidal planes. Such a low toroidal resolution is enabled by the FCI method in GRILLIX \cite{Stegmeir2016,Stegmeir2018,Michels2021,Stegmeir2023}, since not toroidal but magnetic-field-aligned stencils are used for the parallel discretisation, exploiting plasma anisotropy. Overall, our lowest resolution simulation has 6 million grid points, and the highest resolution simulation has 42 million. 
The time step is determined by the CFL limit on shear-Alfvén wave propagation
\begin{eqnarray}
    \Delta t \leq \frac{\mathrm{min}(\Delta s)}{\mathrm{max}(v_A)} \approx \frac{2\pi R_\mathrm{min}}{n_\mathrm{tor}B_\mathrm{max}}\sqrt{\mu_0 m_\mathrm{i} n_\mathrm{min}}.
    \label{eq:dtmax}
\end{eqnarray}
It depends on the minimal parallel grid distance in the simulation $\Delta s$ and the maximum possible Alfvén speed $v_A$. The latter is given by the maximum magnetic field and the minimal density in the simulation. The former is roughly $2\pi$ times the inboard torus radius $R_\mathrm{min}$ divided by the number of poloidal planes (points in the toroidal direction) $n_\mathrm{tor}$. This means that increasing toroidal resolution requires decreasing the time step, hence why the gain in computational performance with the low $n_\mathrm{tor}$ due to the FCI method is at least quadratic \cite{Stegmeir2023}. With $n_\mathrm{tor} = 16$, we were able to choose $\Delta t = 4.8$ ns, and half of that with $n_\mathrm{tor} = 32$. 

Thus, at least roughly a million time steps are required for a low resolution simulation, and double that for $n_\mathrm{tor} = 32$. At
low resolution, a time step takes 1 s on 8 nodes (384 Intel SkyLake processor cores) of the Marconi-A3 SKL partition at Cineca, the same for 32 planes on 16 nodes thanks to toroidal MPI decomposition, and 3 s at higher poloidal resolution. This means at best, a low resolution simulation could run in 9 days consuming 83 kCPUh, and a high resolution simulation in 54 days consuming roughly 1 MCPUh. However, we have performed many additional simulations for testing different parts of the model, mostly at lower resolution, and we have carried those out for up to 10 ms to investigate their long-term behaviour (see especially section \ref{sec:magnetic_shift}). Additionally, the Landau-fluid model \cite{Pitzal2023} as described in section \ref{sec:landau} turned out to be significantly more expensive. Solving the set of 3D implicit problems increases the cost of each time step by 30$\%$, but more critically, the solver would only converge with a factor 2-4 smaller time step. Therefore, most simulations have been carried out with the free-streaming-limited parallel heat flux closure, and only some were done with the Landau-fluid closure at lower resolution. Overall, the computational costs of this publication can be estimated as 15 MCPUh.

The physics results of the resolution scan are condensed in table \ref{tab:res_scan}: with increasing resolution, the total heat transport decreases from 3 to 2 MW. Therefore, while the simulations are not necessarily perfectly converged, it is unlikely that some amount of turbulence would be simply numerically unresolved (on `ion scales'). 
We also compare outboard-midplane profiles across the simulations with varying resolution in appendix \ref{chap:res_scan}, finding variations mostly within experimental error bars. 
Even higher resolution simulations would be desirable, but unfortunately
were not currently affordable, as explained above. 
We explicitly show the results of the resolution scan here such that the reader might assess the numerical uncertainty in the results. 
This is a typical caveat for such large scale simulations. 
Also a comparison between the Landau-fluid (LF) and free-streaming (FS) parallel heat flux closures at low resolution is included in table \ref{tab:res_scan} and appendix \ref{chap:res_scan}: we find only small deviations in most observables, justifying the cheaper FS model. The only exception is the depth of the radial electric field well, where the LF model is closer to the experiment than FS, as will be explained below. Hence, most results will be presented at the highest resolution with the FS model (the reference simulation), including the saturation figures \ref{fig:heat_in} and \ref{fig:sep_sat}.

\begin{table}[htb]
    \centering
    \begin{tabular}{|c|c|c|}
        \hline
        Resolution and type of a run & $P_\mathrm{e}$ & $P_\mathrm{i}$
        \\
        \hline 
        16 planes, $h$ = 1.9 $\rho_{s0}$, FS &
        1.6 MW & 1 MW
        \\
        \hline 
        16 planes, $h$ = 1.9 $\rho_{s0}$, LF &
        1.5 MW & 1.3 MW
        \\
        \hline
        32 planes, $h$ = 1.9 $\rho_{s0}$, FS &
        1.7 MW & 1.2 MW
        \\
        \hline 
        16 planes, $h$ = 1.0 $\rho_{s0}$, FS &
        0.8 MW & 0.9 MW
        \\
        \hline 
        32 planes, $h$ = 1.0 $\rho_{s0}$, FS &
        0.7 MW & 1.3 MW
        \\
        \hline
    \end{tabular}
    \caption{H-mode resolution scan results: total heat transport for electrons and ions. Also the free-streaming-limited (FS) and Landau-fluid (LF) parallel heat flux models are compared at low resolution. For a comparison of outboard-midplane profiles, see appendix \ref{chap:res_scan}.}
    \label{tab:res_scan}
\end{table}

We have identified one particular cause of the differences with resolution. As mentioned above, the major difficulty is to resolve turbulence locally in the SOL and divertor (simultaneously with the confined region), since temperatures become as low as 1 eV, the Larmor radius is thus up to 10 times smaller than at the separatrix, and also parallel gradients become substantial. In particular, we find that the differences in transport lead to a higher ionization rate in the higher toroidal resolution simulations (by overall $50\%$, but the majority is recycling near the divertor plates), and with higher poloidal resolution an ionization source distribution closer to the separatrix at the HFS (leading to a higher edge density source). 
In the lower resolution simulations, the global recycling rate $S_\mathrm{iz} / (S_\mathrm{iz} + S_n^\mathrm{core})$ is only $94\%$, i.e.~the core density source remains too strong. 
With increasing resolution, the separatrix plasma density increases by nearly a factor 2, and the density gradient thus decreases. This leads to an overall transport reduction and, closer to the core boundary, also to some differences in $T_\mathrm{e}$. Especially since we show in section \ref{sec:characterisation} that a large fraction of the transport is neoclassical in the confined region, the major challenge appears to be the resolution of SOL turbulence and the ionization source, which has a retroactive effect on the confined plasma. 

\subsection{Outboard mid-plane profiles}

\begin{figure}[htb]
\centering
\begin{minipage}{0.49\textwidth}
	\centering
    \includegraphics[trim=0.0cm 0.0cm 1.0cm 0.1cm, clip, width=\linewidth]{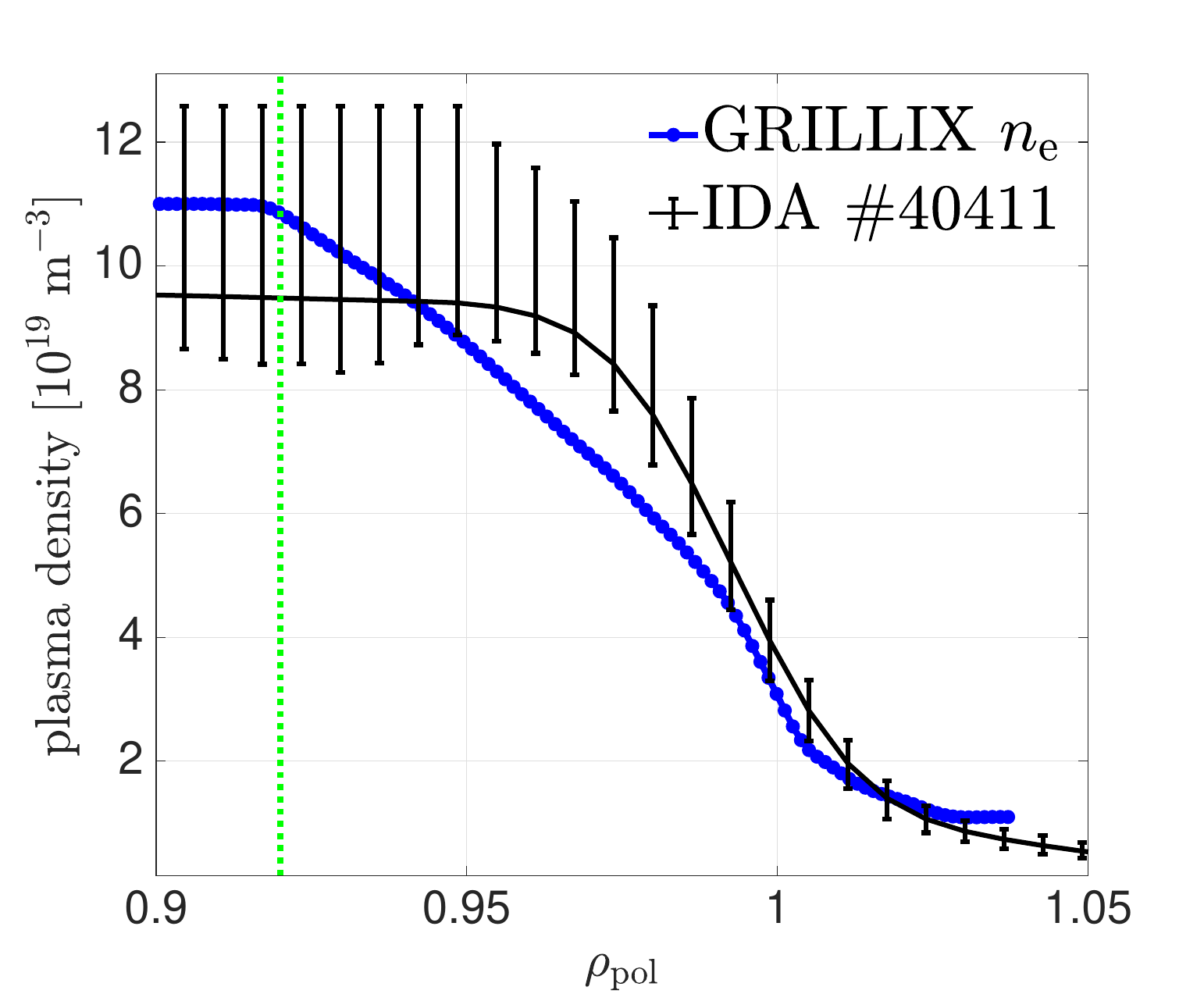}
\end{minipage}
\begin{minipage}{0.49\textwidth}
	\centering
    \includegraphics[trim=0.0cm 0.0cm 1.0cm 0.0cm, clip, width=\linewidth]{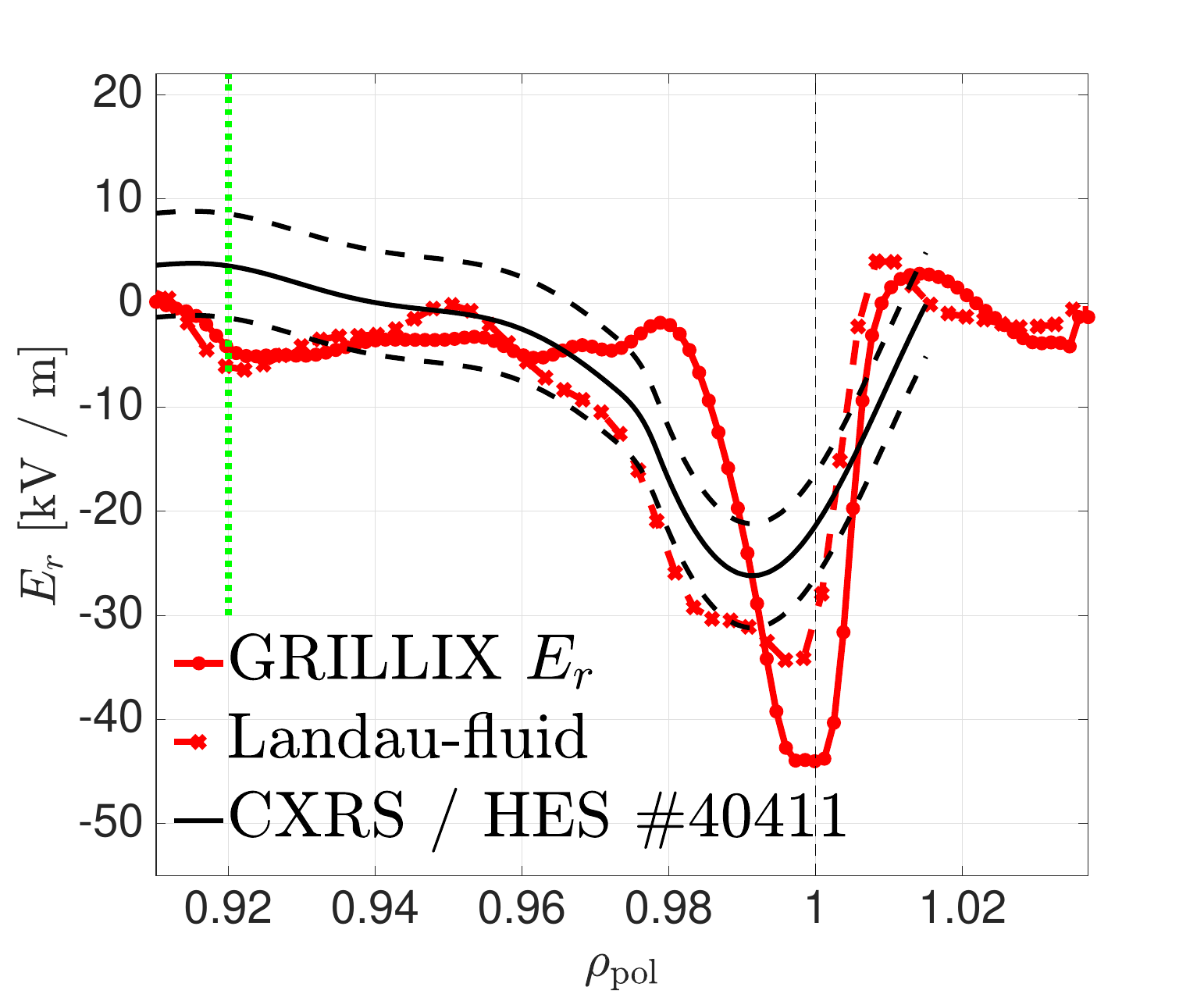}
\end{minipage}
\begin{minipage}{0.49\textwidth}
	\centering
    \includegraphics[trim=0.0cm 0.0cm 1.0cm 0.3cm, clip, width=\linewidth]{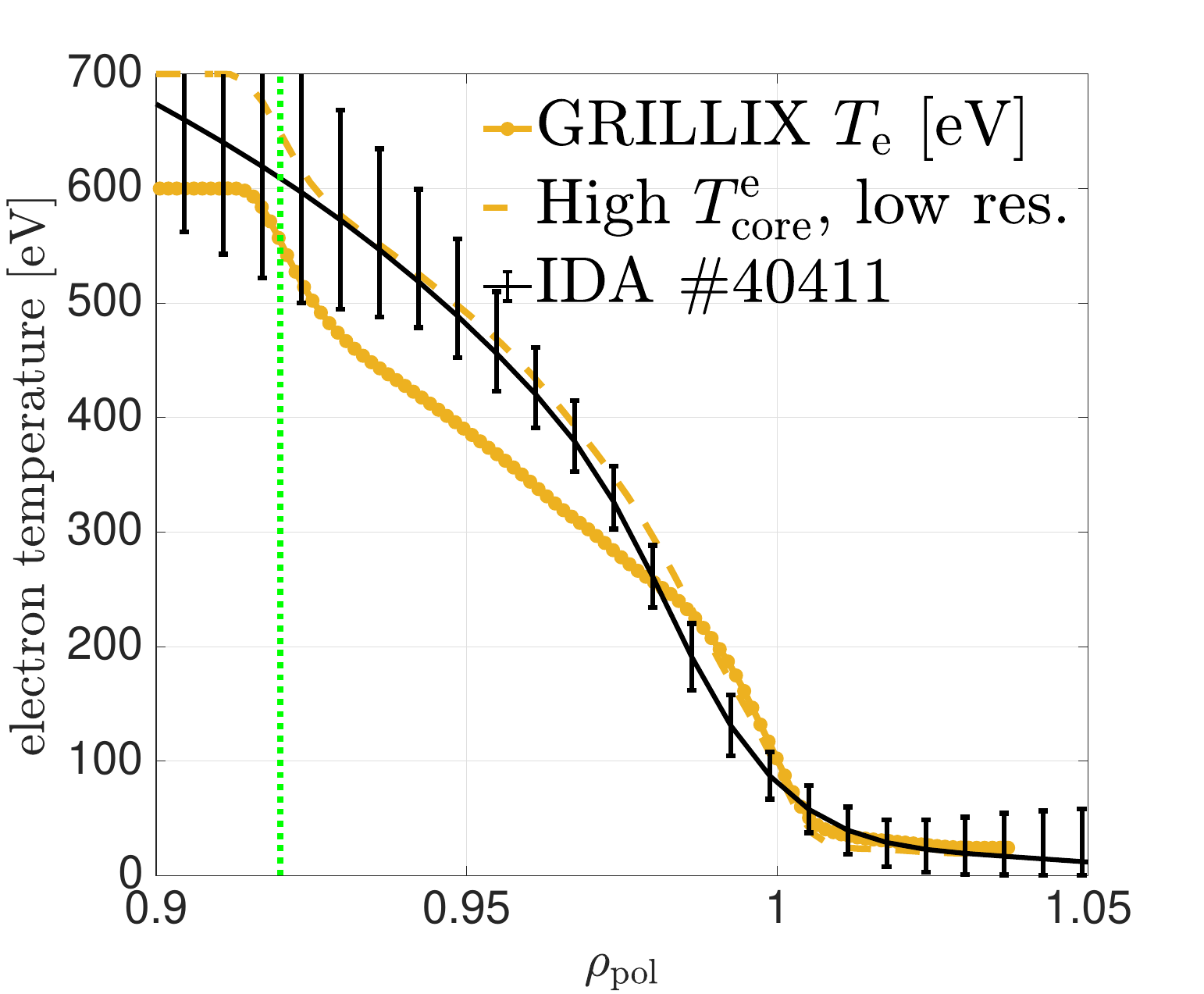}
\end{minipage}
\begin{minipage}{0.49\textwidth}
	\centering
    \includegraphics[trim=0.0cm 0.0cm 1.0cm 0.0cm, clip, width=\linewidth]{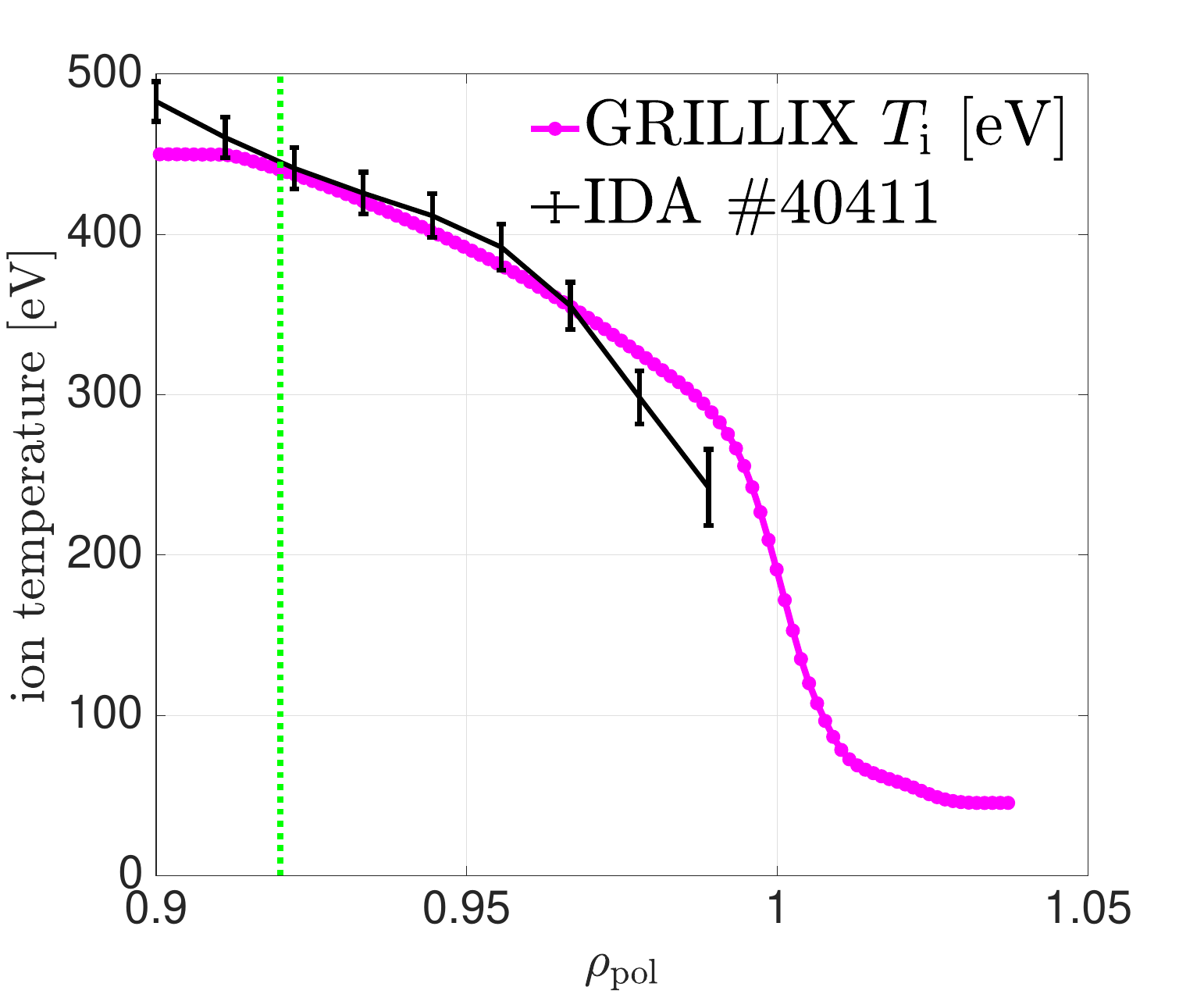}
\end{minipage}
\caption{Comparison of outboard mid-plane profiles of electron density, radial electric field, electron and ion temperatures between a GRILLIX simulation ($n_\mathrm{tor}=32$, $h=1\rho_{s0}$, FS) and ASDEX Upgrade discharge $\#40411$. The simulated profiles were averaged toroidally and over 300 µs in time, the experimental profiles were averaged from $t = 2.1-2.7$ s, where the global plasma parameters are constant. With the Landau-fluid model (see sec.~\ref{sec:landau}) at lower resolution ($n_\mathrm{tor}=16$, $h=1.9\rho_{s0}$, LF), only $E_r$ is somewhat different, as shown here. For $T_\mathrm{e}$, the reference simulation has $T^\mathrm{e}_\mathrm{core}=600$ eV, but also a simulation with $T^\mathrm{e}_\mathrm{core}=700$ eV (at lower grid resolution) is shown. A green line marks the source region. For a resolution scan, see appendix \ref{chap:res_scan}.}
\label{fig:profiles}
\end{figure}

Next, we want to compare our simulation results to experimental measurements, focusing on profiles at the outboard mid-plane (OMP). To this end, the OMP plasma profiles in the highest resolution simulation ($n_\mathrm{tor}=32$, $h=1\rho_{s0}$, FS) have been averaged toroidally and in time over 300 µs (50 snapshots). The results are compared in figure \ref{fig:profiles} with experimental measurements of electron density, electron and ion temperatures, as well as the radial electric field. The experimental profiles were averaged from $t = 2.1-2.7$ s, where the global plasma parameters were constant, with ELM burst phases filtered out. 
The electron density has been measured with the lithium and helium beams, as well as with Thomson scattering. 
The electron temperature has been measured with Thomson scattering only. 
The ion temperature has been determined by the charge exchange recombination spectroscopy (CXRS), assuming impurities (boron) have the same temperature as the main ions. 
The results have been combined with integrated data analysis (IDA), using Bayesian statistics \cite{Fischer2010}. The error bars represent the scatter of the data, both due to noise and measurement uncertainties, as well as plasma fluctuations within the time interval averaged. The black lines fitted to the data serve only for guidance, the ``real'' profiles can lie anywhere within the experimental error bars (which are also just the most probable ones). 
For the radial electric field, impurity CXRS \cite{Viezzer2013} and He\,1$^+$ spectroscopy (HES) \cite{Plank2022} were employed. Both methods determine $E_\mathrm{r}$ from the radial force balance 
\begin{equation}
    E_\mathrm{r} = \frac{\nabla p_\alpha}{n_\alpha Z_\alpha e} -  v_\mathrm{pol}^\alpha B_\mathrm{tor} + v_\mathrm{tor}^\alpha B_\mathrm{pol},
    \label{eq:imp_force_balance}
\end{equation}
which holds for any plasma species $\alpha$ separately. Here, $Z_\alpha$ is the charge number, $n_\alpha$ and $p_\alpha$ are the density and pressure, and $v_\mathrm{pol}^\alpha$ and $v_\mathrm{tor}^\alpha$ are the local poloidal and toroidal velocities of the respective impurity $\alpha$. 
In the case of CXRS, $\alpha$ corresponds to fully-ionized boron with $Z_\alpha=5$, whereas HES employs line radiation of singly ionized helium.  
The final $E_r$ profile results from a combination of these two measurements at different radial locations, CXRS at $\rho_\mathrm{pol} \lesssim 0.99$ and HES at $\rho_\mathrm{pol} \gtrsim 0.99$. 
In figure \ref{fig:profiles}, for $E_r$, the solid black line shows the best fit through the data and the dashed black lines indicate the uncertainty of $\pm 5$ kV/m. The radial measurement uncertainty is $\pm 0.005 \rho_\mathrm{pol} \approx 3$ mm, and the $E_r$ profile has already been shifted by this amount to the right.

Overall, we find a reasonable match between the simulation and the experiment, although not in every detail. 
The simulated density pedestal does not flatten as much towards the plasma core as the measured one: however, experimental data in this region are poor, indicated by the large error bars. 
Similarly, the electron temperature matches well around the separatrix, but deviates from the experiment deeper inside. 
A major difficulty is the experimental uncertainty at our core boundary, since core density and temperature are simulation input and cannot be changed a posteriori. For most of our simulations, especially due to the need of resolution scans (as described above), we have fixed $n_\mathrm{core} = 1.1 \times 10^{20}$ m$^{-3}$ and $T_\mathrm{core}^\mathrm{e} = 600$ eV. However, at lower resolution ($n_\mathrm{tor}=16$, $h=1.9\rho_{s0}$, FS), we have also carried out a simulation with $T_\mathrm{core}^\mathrm{e} = 700$ eV: it matches indeed much better the experimental $T_\mathrm{e}$ profile. Due to the high dimensionality of the input parameter space, it is impossible to find the optimum within the available computational time, so this serves just for comparison: we have carried out much more extensive scans with $T_\mathrm{core}^\mathrm{e} = 600$ eV and thus our focus remains on it. 
For the ion temperature, unfortunately, no experimental data are available near the separatrix, but in the confined region the match is good. Of course, towards the core boundary, the match is automatic due to the adaptive sourcing there. Finally, for the radial electric field, we see a different profile shape towards the plasma core: this is not surprising as we do not include a momentum source from NBI injection in the simulations. However, the $E_r$ well in the plasma edge, its width and depth, are reasonably reproduced. In particular, we find that $E_r$ matches nearly perfectly with the Landau-fluid model (see sec.~\ref{sec:landau}), similar to our previous observations in L-mode \cite{Zholobenko2021a,Pitzal2023}.

We want to comment on the role of neutral gas in these simulations, connecting to our previous L-mode investigation \cite{Zholobenko2021a}. Generally, the neutrals lead to higher density and lower temperature in the SOL and pedestal bottom. However, unlike in L-mode, the simulations did not even saturate without neutrals at all, crashing before 2 ms. 
The reason for this is in the radial electric field: we find a much higher positive radial electric field of up to 20 kV/m in the SOL without neutrals, as previously \cite{Zholobenko2021a}, due to a hotter divertor with a larger $T_\mathrm{e}$ gradient and sheath boundary conditions enforcing $E_r|_\mathrm{SOL}\approx-3\partial_r T_\mathrm{e}$ \cite{Stangeby2000,Zholobenko2021}. Together with the negative $E_r$ well in the plasma edge this leads to an overall increased $E \times B$ shear across the separatrix. In the course of the simulation this shear flow becomes unstable, ending in a large macroscopic instability.

Clearly, a more in-depth validation is desirable for the future. 
However, the comparison so far, including the total heat transport discussed above, serves to demonstrate that our simulations are reasonably realistic compared to experimentally available measurements in the investigated H-mode discharge: despite not matching every detail, it is worth to analyse 
the radial electric field formation in section \ref{sec:Er} and turbulence characteristics in section \ref{sec:characterisation}. But first, in section \ref{sec:transcollisional}, we will demonstrate that the results shown so far are already not trivial at all, requiring an involved physical model.

\section{Electromagnetic, transcollisional, global drift-fluid model}
\label{sec:transcollisional}

In this section, we discuss the extensions of the drift-fluid model \cite{braginskii65,Zeiler1997,Zholobenko2021} in GRILLIX which allowed simulations in H-mode conditions in the first place. The full set of the final equations is summarised in appendix \ref{chap:Braginskii_equations}. The most significant finding is that \textit{H-mode turbulence is deeply electromagnetic}: fluctuations of the perpendicular magnetic field are critical in controlling $E \times B$ transport, as clarified in section \ref{sec:flutter}. In section \ref{sec:magnetic_shift}, we stress that in global ``full-$f$'' simulations, the Shafranov shift must be carefully filtered out from magnetic fluctuations. Then, we elaborate on the transcollisional extensions of the fluid closure: section \ref{sec:landau} discusses the incorporation of Landau damping into the parallel heat conduction, while section \ref{sec:neocl_visc} details the neoclassical corrections for the ion viscosity.

\subsection{Electromagnetic effects}
\label{sec:flutter}

Fluctuations of the magnetic field are important for magnetised plasma turbulence \cite{Dickinson2012,Hatch2016,Eich2021,Bonanomi2024,Zhang2024} not because they are large ($B_1/B_0\sim10^{-3}$ in our present simulations), but because they redirect the fast parallel transport and parallel forces into the direction perpendicular to the equilibrium magnetic field.

Ohm's law (neglecting inertia) reads $\beta_0\partial_t A_\parallel \approx - \eta_\parallel j_\parallel - \nabla_\parallel \varphi + n^{-1} \nabla_\parallel p_\mathrm{e} + 0.71 \nabla_\parallel T_\mathrm{e}$ in normalised units (see appendix \ref{chap:Braginskii_equations}). The forces on the right-hand side push fluctuations towards adiabaticity ($\tilde{p}_\mathrm{e}\sim\tilde{\varphi}$), which would result in zero $E\times B$ transport of particles and electron heat. However, a finite non-adiabaticity can persist due to the resistivity $\eta_\parallel$ and the magnetic induction. 
Even at low $\beta$, adding magnetic induction can be beneficial for code performance: for $k_\perp \rightarrow 0$ (large scales), which is relevant for global codes simulating large machines, the wave speed diverges without induction \cite{Scott1997,Dannert2004,Dudson2021,Stegmeir2023} (without inertia, the propagation is actually diffusive, and diverges also with vanishing resistivity), requiring $\Delta t \rightarrow 0$. Induction limits the propagation of the non-adiabaticity to the Alfvén speed $v_A = B / \sqrt{\mu_0 n m_\mathrm{i}}$. The time step in an explicit scheme than follows $\Delta t < \Delta s / v_A \sim \sqrt{\beta}$, with the parallel grid distance $\Delta s$. Indeed, in our present simulations, we cannot easily decrease $\beta$ to compare to electrostatic simulations, since this increases the computational costs.

\begin{figure}[htb]
\centering
	\centering
    \includegraphics[trim=0cm 0cm 1.5cm 1.0cm, clip, width=0.49\linewidth]{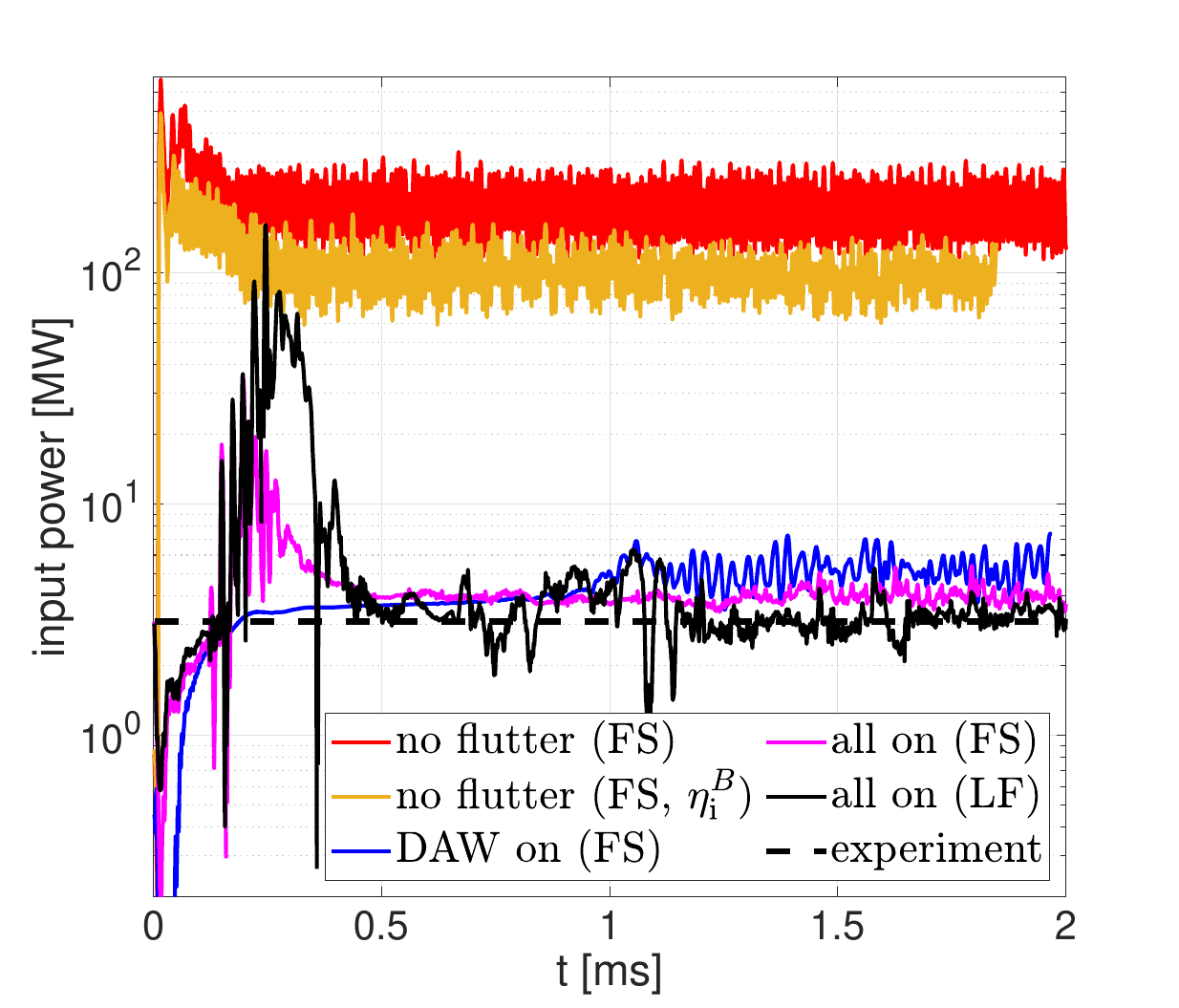}
    \caption{Total input power in simulations with and without magnetic flutter. In the experiment, the power crossing the separatrix between ELMs is estimated to be 3.1 MW. Most simulations are with the free-streaming (FS) heat conduction, as defined in section \ref{sec:landau}, since this is computationally more affordable. A simulation with the Landau fluid (LF) closure is shown in black. Most simulations are with the neoclassical ion viscosity, as discussed in section \ref{sec:neocl_visc}, except one with the Braginskii viscosity $\eta_\mathrm{i}^B$ for comparison. The black and magenta lines are from simulations with magnetic flutter included in all parallel operators, while the blue line is from a simulation with flutter only in the DAW system.}
    \label{fig:heat_sep}
\end{figure}

As $\beta$ increases, magnetic fluctuations also affect more and more the physics. It is known that magnetic induction $\partial_t A_\parallel$ destabilises drift-Alfv\'en-waves (DAWs) \cite{Scott2021,Zhang2024}. Thus, it becomes important to also include the stabilising counter-part, magnetic flutter: the perturbed magnetic field $\mathbf{\tilde{B}} = \nabla \times (A_\parallel \mathbf{b}_0)$ enters all parallel operators $\nabla_\parallel = (\mathbf{b}_0 + \mathbf{\tilde{B}}/B_0)\cdot\nabla$, redirecting parallel flows and forces. 
In L-mode AUG simulations, we have found a stabilization factor of 2 due to magnetic flutter, which is significant but can be compensated by freedom in the details of the fluid closure, as indicated in figure \ref{fig:heat_sep}.

Here, we stress that in H-mode, the stabilizing effect of magnetic flutter reaches 2 orders of magnitude: as displayed in figure \ref{fig:heat_sep}, the $E \times B$ heat transport is reduced from 200 MW to 4 MW due to flutter. 
Figure \ref{fig:snapshot} (left and center) compares density snapshots in simulations with and without flutter: we see much more violent turbulence without flutter, with vortices stretching across the whole radius at the OMP. Figure \ref{fig:heat_sep} contains data from five simulations at the lower resolution (see table \ref{tab:res_scan}): the Landau fluid one is labeled `all on (LF)', while others are with the free-streaming (FS) heat conduction limiter, as defined in section \ref{sec:landau}, to save on computational resources. The Shafranov shift is removed with the frequency of 42 kHz (see sec. \ref{sec:magnetic_shift}). A simulation without flutter and with Braginskii instead of neoclassical ion viscosity (see sec.~\ref{sec:neocl_visc}) has been performed to show that this also impacts the overall heat transport, here by roughly a factor 2. To compare with our previous, more detailed study in L-mode conditions by Zhang \textit{et al.} \cite{Zhang2024}, we have also performed a simulation where magnetic flutter was activated only in DAW equation terms (adiabatic forcing in Ohm's law and divergences of the parallel current): indeed, this is sufficient to explain the overall transport reduction. However, there remains a quantitative difference when magnetic flutter is included in the whole set of equations: this is due to heat transport by the induced magnetic fluctuations themselves, as will be discussed in section \ref{sec:characterisation}. We note that of course, when magnetic flutter is neglected and heat transport exceeds 200 MW, the OMP radial profiles as displayed in figure \ref{fig:profiles} for the reference simulation also look very different: the density and temperature profiles are flattened by the extended streamers, yielding at the separatrix for example $n = 7.3 \times 10^{19}$ m$^{-3}$, $T_\mathrm{e} = 370$ eV and $T_\mathrm{i} = 390$ eV.

\subsection{Numerical treatment of the Shafranov shift}
\label{sec:magnetic_shift}

The Shafranov shift $A_s$ results from the pressure gradient driven Pfirsch-Schlüter current (and bootstrap current in gyrokinetics) \cite{PerHelander2002}, and is thus substantial in an H-mode pedestal. It is known that $A_s$ increases pedestal MHD stability \cite{Snyder2007} by squeezing the flux surfaces at the LFS, and suppresses turbulence by effectively increasing the $E \times B$ shearing rate \cite{Lackner2000}. Therefore, a careful treatment of the Shafranov shift is particularly important in H-mode simulations. Clearly, $A_s$ is part of the MHD equilibrium, and is included in the background magnetic field $\mathbf{B}_0$ in GRILLIX. However, in global ``full-$f$'' simulations (unlike in $\delta f$) we also evolve dynamically the full plasma pressure, which leads to a diamagnetic current that is balanced by the Pfirsch-Schlüter $j_\parallel^\mathrm{PS}$, and through Ampere's law, this induces a dynamic Shafranov shift $A_s$, which enters the dynamics through magnetic flutter. Clearly, this double counting of $A_s$ needs to be removed \cite{Scott2006,Hager2020,Giacomin2022,Zhang2024}. Additionally, GRILLIX relies on local alignment of parallel operators to $\mathbf{B}_0$ for computational performance. This requires $B_1/B_0 < \Delta x / \Delta s \sim \rho_\mathrm{i} / R_0$, where $\mathbf{B}_1 = \nabla \times (A_1\mathbf{b}_0)$, $\Delta x$ is the perpendicular and $\Delta s$ the parallel grid distance, and their ratio is of the order of the ratio between the Larmor radius and the machine major radius. A large $A_s$ breaks this requirement, and thus needs to be subtracted from $A_\parallel$, $A_1 = A_\parallel - A_s$.

\begin{figure}[htb]
\centering
\begin{minipage}{0.5\textwidth}
	\centering
    \includegraphics[trim=0.5cm 1.5cm 1.6cm 0.0cm, clip, width=1.0\linewidth]{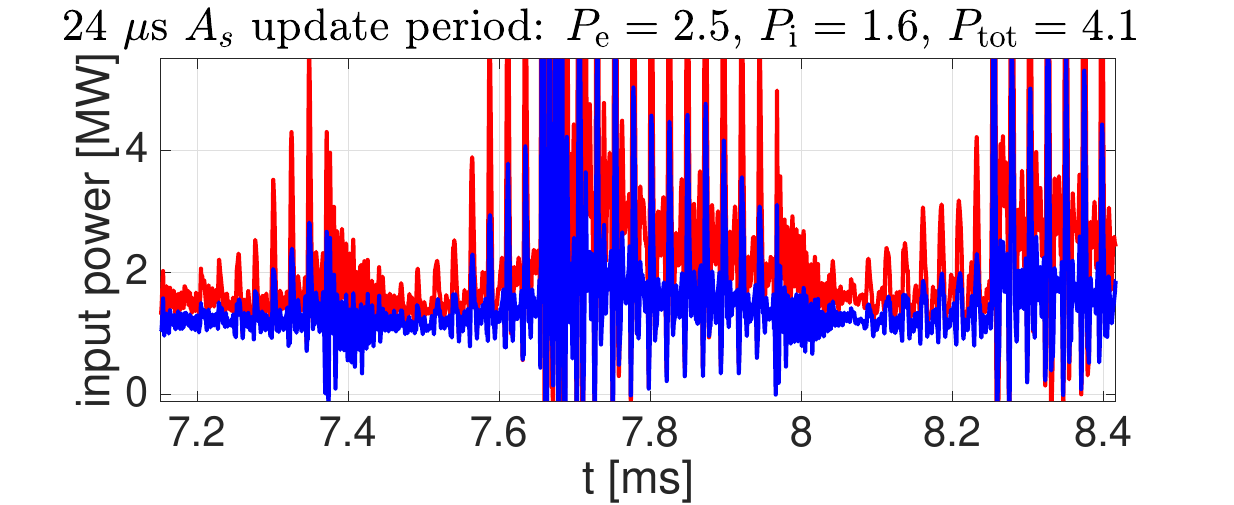}
    \label{fig:Pin_tau_update1}
\end{minipage}
\begin{minipage}{0.5\textwidth}
	\centering
    \includegraphics[trim=0.5cm 0.0cm 1.6cm 0.6cm, clip, width=1.0\linewidth]{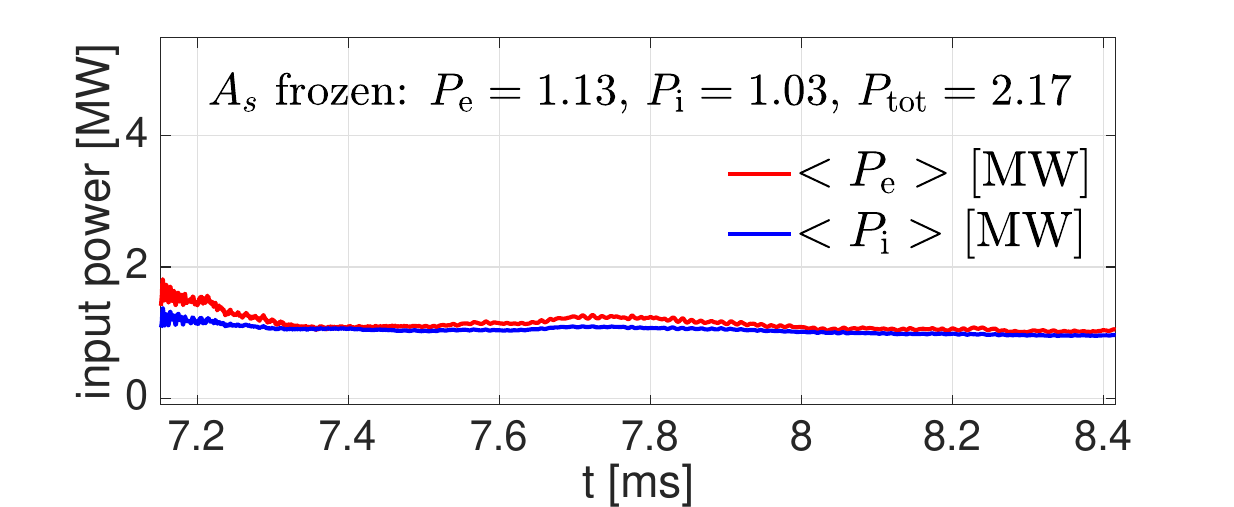}
    \label{fig:Pin_tau_update_inf}
\end{minipage}
    \caption{Input power at a later stage of the lower resolution, free-streaming simulation, with periodic updates of $A_s$ at the top and with a frozen $A_s$ at the bottom.}
    \label{fig:magnetic_shift_rm_Pin}
\end{figure}

In this work, our strategy for dealing with this is an extension of \cite{Hager2020,Giacomin2022}: we average $A_\parallel$ toroidally \textbf{and in time} to determine $A_s$. 
The removal of the toroidal average in each time step is not sufficient and on its own problematic, because it allows information to instantaneously propagate around the torus, instead of with the Alfv\'en speed 
(and is also not applicable in stellarators). 
The time-average over at least $2\pi R_0 q / v_A$ is necessary \cite{Zhang2024}, corresponding to $q n_\mathrm{tor} \Delta t_\mathrm{max}$ (see eq.~\eref{eq:dtmax}). But it is still problematic: with discrete update steps of $1t_0 = 24$ $\mu$s (5000 time steps), we found the heat transport in H-mode to synchronise with them and become artificially increased. Comparing with figure \ref{fig:heat_in}, we note that inconveniently, the small bursts that are visible there in the first 2 ms grow after some time, deteriorating the quasi-stationary simulation phase. This is shown in figure \ref{fig:magnetic_shift_rm_Pin}: in the later simulation phase, the bursts of electromagnetic transport account for half of the total heat transport in the simulation (2 MW). Therefore, once the simulations reach a steady state but before artificial electromagnetic transport is triggered, we freeze a long-time averaged $A_s$, as mentioned in section \ref{sec:setup}. Clearly, this method is not ideal, and we are working on further improving it for the future. But it works well enough for now: we see in figure \ref{fig:magnetic_shift_rm_Pin} that with this, the bursts disappear and steady heat transport is recovered. 
The reason why the removal of the Shafranov shift $A_s$ must be done more carefully in H-mode than in L-mode is likely the presence of electromagnetic transport, as discussed in section~\ref{sec:characterisation}.

\subsection{Landau damping  of the parallel heat flux}
\label{sec:landau}

\begin{figure}[htb]
\centering
	\centering
    \includegraphics[trim=1.5cm 0cm 1.0cm 0.5cm, clip, width=0.45\linewidth]{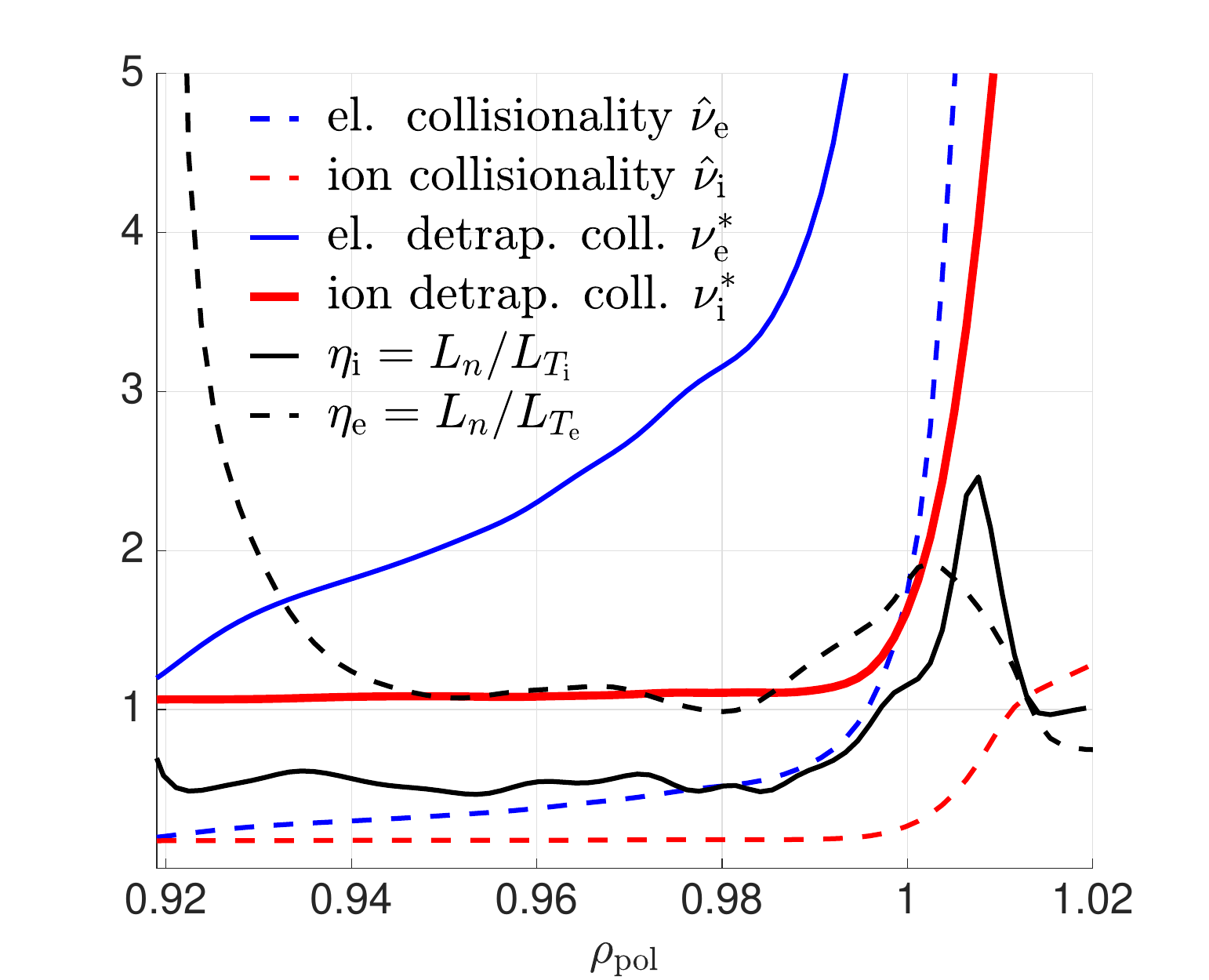}
\caption{Electron and ion collisionalities $\hat{\nu}_\mathrm{e,i}$ normalised to the parallel transit frequency, detrapping collisionalities $\nu_\mathrm{e,i}^* = \epsilon^{-3/2} \hat{\nu}_\mathrm{e,i} \approx 6 \hat{\nu}_\mathrm{e,i}$ normalised to the bounce frequency, as well as the density to ion / electron temperature gradient length ratios $\eta_\mathrm{i}$ / $\eta_\mathrm{e}$.}
\label{fig:coll_etai}
\end{figure}

The collisional Braginskii fluid closure \cite{braginskii65} is only strictly valid for $\hat{\nu}_\mathrm{e,i} \gg 1$, whereby the electron and ion collisionalities are normalised to their parallel transit frequencies \cite[sec.~8.2]{PerHelander2002}
\begin{eqnarray}
    \hat{\nu}_\mathrm{e,i} = \nu_\mathrm{e,i} / \omega_t^\mathrm{e,i},\, \mathrm{ with}\,\, \omega_t^\mathrm{e,i} = \sqrt{2T_\mathrm{e,i}/m_\mathrm{e,i}} / qR.
    \label{eq:coll_transit}
\end{eqnarray}
Taking $q \approx 4$ and $R \approx 1.65$ m, these collisionalities are computed from the density and temperature profiles in figure \ref{fig:profiles} and shown in figure \ref{fig:coll_etai} as dashed lines (the other quantities will be discussed later). Clearly, in practise, we find $\hat{\nu}_\mathrm{e,i} < 1$ in the whole confined region (and partly in the near-SOL) of our simulations. The closure is particularly problematic in equation terms that are proportional to $\hat{\nu}^{-1}$: the parallel heat flux $q_\parallel = - \kappa_\parallel \nabla_\parallel T$, with the heat conduction $\kappa_\parallel \sim nT\nu^{-1} \sim T^{5/2}$, and similarly the parallel viscosity discussed in the next section. A large heat conduction (and viscosity) suppresses turbulence \cite{Hallatschek2000,Zholobenko2021a,Pitzal2023}. We have implemented and tested two solutions, a non-local Landau-fluid (LF) closure and local free-streaming (FS) limiters. We find equally satisfactory results for both, the major effect being the reduction of the heat conductivity by more than an order of magnitude compared with Braginskii.

The problem of the fluid closure is fundamental, it arises because each moment of the kinetic equation is dependent on the next one in the hierarchy. In the collisional limit $\hat{\nu}_\mathrm{e,i}\gg1$, due to Boltzmann's H-theorem, the hierarchy can be closed self-consistently at the first few moments (typically three) by assuming only small corrections to a Maxwellian from the higher moments, and that collisional dissipation balances the driving forces ($\nabla_\parallel T$) \cite{Scott2007}. 
But at $\hat{\nu}_\mathrm{e,i} \lesssim 1$, the polynomial corrections to the Maxwellian can actually become very large, contradicting the closure \cite{Makarov2021a}. The conductive heat fluxes as well as viscosity then also become very large, diverging as $T^{5/2}$. In reality, the driving forces are actually balanced by other effects, such as Landau damping \cite{Landau_1945,hammett_perkins}, friction between trapped and passing particles \cite{Hirshman1981,PerHelander2002}, as well as non-linear mixing \cite{Scott2007}. A rigorous treatment of these processes requires (gyro-) kinetic theory. But the validity of the fluid closure can also be improved by rather simple corrections.

Landau damping \cite{Landau_1945,hammett_perkins,Mouhot2011,Schekochihin2016,Scott2021a} is particularly important for heat conduction. 
It is inherently a kinetic process, a dissipation mechanism that is independent of collisionality. But it can be mimicked by a non-local fluid closure \cite{hammett_perkins,Snyder1997} for the parallel heat flux, which in Fourier space takes the ``Hammett-Perkins'' form $q_{\parallel k}^{\mathrm{HP}} = - A \frac{i k_{\parallel}}{|k_{\parallel}|} T_{k}$, with $A = n v_{\mathrm{th}} \sqrt{8 / \pi}$ and $v_\mathrm{th}^\mathrm{e,i} = \sqrt{T_\mathrm{e,i} / m_\mathrm{e,i}}$. When inserted in the temperature equations \eref{electron_temperature_equation} and \eref{ion_temperature_equation}, this simply results in a damping rate of $v_{\mathrm{th}} \sqrt{8 / \pi} |k_\parallel|$. 
The above expression is valid in the collisionless limit. More generally, $q_{\parallel k}^{\mathrm{LF}} = - A \frac{i k_{\parallel}}{|k_{\parallel}| + \delta \nu_\mathrm{e,i} / v_{\mathrm{th}} } T_{k}$ accounts for both collisional and collisionless damping \cite{Umansky2015,Chen2019,Zhu2021}. The choice $\delta_\mathrm{e} \approx 0.5$ and $\delta_\mathrm{i} \approx 0.41$ reproduces the Braginskii closure in the collisional limit $\delta \nu_\mathrm{e,i} \gg |k_\parallel| v_\mathrm{th}$ ($\hat{\nu}_\mathrm{e,i} \gg 1$, see \eref{eq:coll_transit}). Finally, a critical step to make the closure applicable in the non-periodic domain of the tokamak SOL and thus boundary codes such as GRILLIX, is to transfer it from Fourier to real space. To this end, a fast non-Fourier method has been developed by Dimits \textit{et al.} \cite{Dimits2014}, which results in a set of elliptic equations along the magnetic field \eref{eqn:electron_landau_fluid_equation_in_grillix}-\eref{eqn:ion_landau_fluid_equation_in_grillix} (see appendix \ref{chap:Braginskii_equations} and \cite{Pitzal2023} for further details).

A key feature of the Landau-fluid (LF) closure is that it is non-local, i.e.~the heat flux does not depend on the local $\nabla_\parallel T$ in real space, but is formulated in Fourier space (and the transformation is not trivial). A detailed investigation of the behaviour of this Landau-fluid closure in GRILLIX in L-mode conditions has been recently published by Pitzal \textit{et al.} \cite{Pitzal2023}. Here, we stress that especially in H-mode conditions, but also in our previous L-mode simulations \cite{Pitzal2023}, also the commonly utilised local flux limiters yield very good results, and that they also approximate Landau damping. At low collisionality, they simply limit the local conductive heat flux to a fraction $\mathrm{f}^\mathrm{FS}$ of the free-streaming (FS) heat flux $q^\mathrm{FS}=\mathrm{f}^\mathrm{FS} nv_\mathrm{th}T$ \cite{Thyagaraja1980,Scott1997,Stangeby2000,Fundamenski2005,Xia2015}. There is a correspondence between the two models. The key approximation is $|k_\parallel| \approx 1/qR$, i.e.~the damping becomes the same for all modes and equal to the largest mode in the system. Then, taking $ik_\parallel T_k = \nabla_\parallel T$ and $\mathrm{f}^\mathrm{FS} = \sqrt{8/\pi}$, we obtain
\begin{eqnarray}
    q_\parallel = - A \frac{i k_{\parallel} T_{k} }{|k_{\parallel}| + \delta_\mathrm{e,i} \nu_\mathrm{e,i} / v_\mathrm{th}^\mathrm{e,i} } \approx -\kappa_{\parallel }^\mathrm{e,i} \left( 1 + \frac{\kappa_{\parallel }^\mathrm{e,i}}{\mathrm{f}^\mathrm{FS}_\mathrm{e,i} n\sqrt{T_\mathrm{e,i}/m_\mathrm{e,i}} R q} \right)^{-1} \nabla_\parallel T_\mathrm{e,i},
    \label{eq:LF2FS}
\end{eqnarray}
which is exactly the free-streaming limited heat flux expression \cite{Zholobenko2021a}, with the Braginskii heat conductivities $\kappa_{\parallel }^\mathrm{e,i}$. 
The FS OMP profiles in figure \ref{fig:profiles}, when compared to LF simulations in figure \ref{fig:res_scan_profiles}, are very similar, only the $E_r$ is slightly different (the difference in $E_r$ was more significant in L-mode). 
In fact, we have been using $\mathrm{f}^\mathrm{FS}_\mathrm{e,i} = 1 < \sqrt{8/\pi} \approx 1.6$, which might have lead to slightly higher saturated transport. 
Therefore, in the present setup, both approximations seem to work comparably well, which in turn suggests that heat flux non-locality is not a critical mechanism. 

\begin{figure}[htb]
\centering
	\centering
    \includegraphics[trim=0.0cm 0.0cm 0.0cm 0.0cm, clip, width=0.5\linewidth]{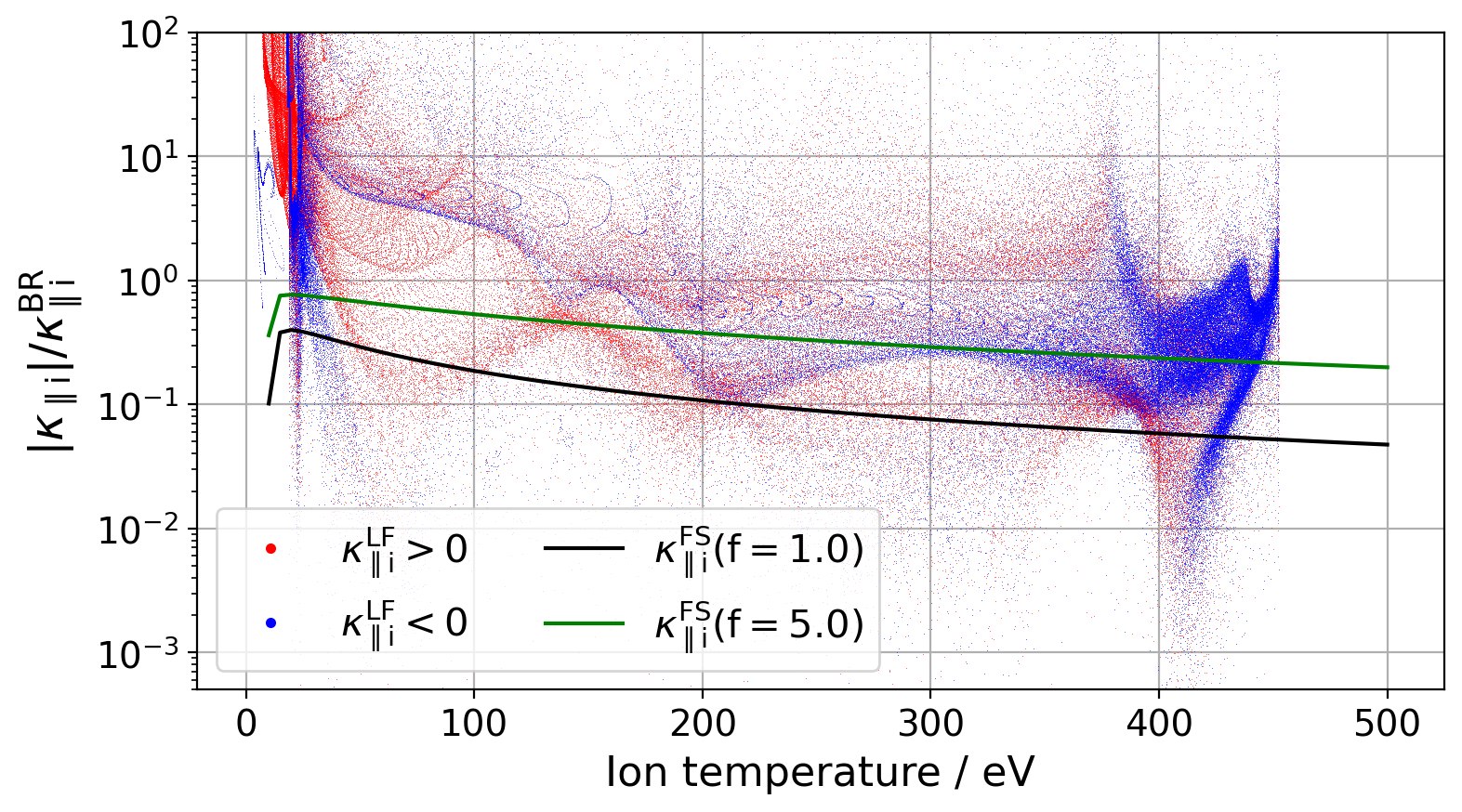}
    \includegraphics[trim=0.0cm 0.0cm 0.0cm 0.0cm, clip, width=0.5\linewidth]{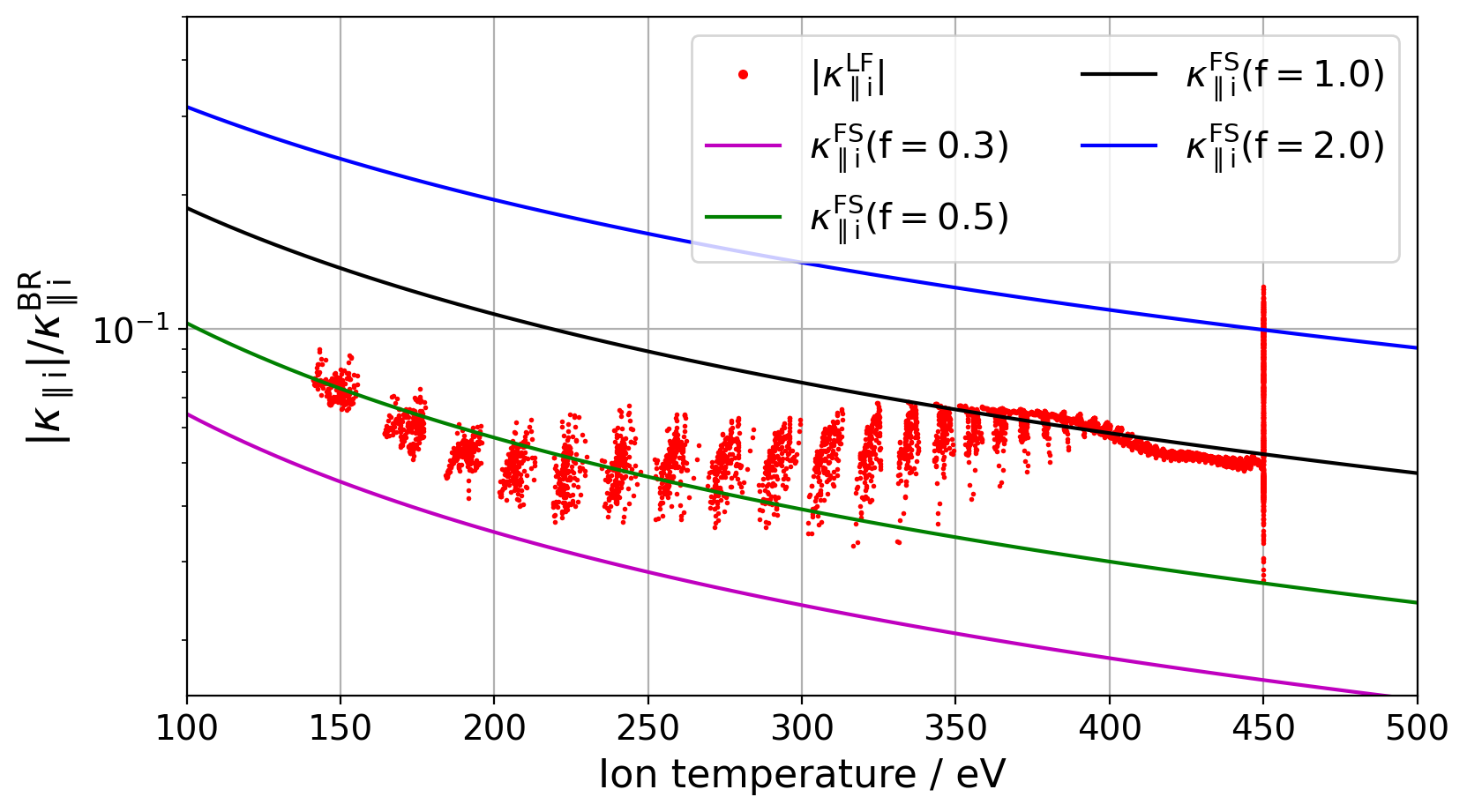}
    \caption{Effective ion heat conductivities for Landau fluid (LF) and free-streaming (FS) limited models, normalised to Braginskii. The two figures show different averages: at the top, we compute $\tilde{\kappa}_{\parallel \mathrm{i}}^{\mathrm{LF}}(R,Z) = - \left< q_{\parallel \mathrm{i}}^{\mathrm{LF}} \right>_{t,\phi} / \left< \nabla_{\parallel} T_\mathrm{i} \right>_{t,\phi}$ and plot it as a function of the local average temperature $\left< T_\mathrm{i} \right>_{t,\phi}(R,Z)$. At the bottom, we compute $\tilde{\kappa}_{\parallel \mathrm{i}}^{\mathrm{LF}}(\rho_\mathrm{pol},t) = - \left< | q_{\parallel \mathrm{i}}^{\mathrm{LF}} | \right>_{\theta,\phi} / \left< | \nabla_{\parallel} T_\mathrm{i} | \right>_{\theta,\phi}$ and plot it versus $\left< T_\mathrm{i} \right>_{\theta,\phi} (\rho_\mathrm{pol},t)$. The absolute value was taken because the flux-surface average annihilates $\nabla_\parallel T$.}
    \label{fig:kappa_LF}
\end{figure}

To support this argument, instead of only comparing the resulting plasma profiles, we can make a more direct comparison of the effective heat conductivities $\tilde{\kappa}_\parallel =  - q_\parallel / \nabla_\parallel T$ between the two models. For FS, it is simply a function of local parameters ($n(T)$ is taken from fig.~\ref{fig:profiles}), which we plot vs temperature for different $f^\mathrm{FS}$ as lines in figure \ref{fig:kappa_LF}. For the LF model, we take the results from our simulations, which in principle can be arbitrary. Indeed, in the top figure, in an average of $\tilde{\kappa}_{\parallel\mathrm{i}}^\mathrm{LF}$ toroidally and in time, there seems to be no correlation between the LF heat conductivity and the FS one for any temperature -- a signature of the LF non-locality (it is the same for $\tilde{\kappa}_{\parallel\mathrm{e}}^\mathrm{LF}$). However, in the bottom figure, we show the $\tilde{\kappa}_{\parallel\mathrm{i}}^\mathrm{LF}$ in a flux-surface average -- finding a very good correlation between FS (with $\mathrm{f}^\mathrm{FS}_\mathrm{e,i} \approx 0.5 - 1$) and LF. 
The flux-surface average is used only in the confined region, so results are shown only for $T_\mathrm{i} \geq 150$ eV, and at 450 eV the correlation is destroyed by the buffer zone. Thus, it is clear that in a flux-surface average the two closures yield similar heat conductivities, which seems to lead to overall similar results, rendering the non-locality less important. The major effect is rather the reduction of the heat conduction by more than an order of magnitude in comparison to Braginskii. The difference along the flux surfaces can still affect global phenomena though, such as zonal flows and GAMs, and might explain why particularly the $E_r$ differs with the two closures. This might motivate further studies, but so far, the effect seems to be stronger in L-mode \cite{Pitzal2023} than in our H-mode case.

\subsection{Neoclassical ion viscosity}
\label{sec:neocl_visc}

In a 3D drift-fluid model, contrary to typical gyro-fluid treatments \cite{Scott2021a}, a closure is required not only in the highest moment. Since collisions act to make the distribution function isotropic, at high collisionality, an isotropic pressure (temperature) can be assumed to lowest order. The deviation from isotropy is the viscous stress. 
As detailed in appendix \ref{chap:Braginskii_equations}, ion viscosity enters in the vorticity, parallel momentum and ion heat equations. 
However, by far the dominant contribution is in the parallel momentum balance \cite[sec.~12.1]{PerHelander2002}: by coupling parallel motion and perpendicular drifts, it acts as a damping of the poloidal flow. Additionally, by dissipating sound waves, viscosity also couples to geodesic acoustic modes and the turbulence \cite{Scott2005}, and through this way as well as through the poloidal rotation to the zonal flows \cite[sec.~13]{PerHelander2002}. The impact on the radial electric field will be detailed below, in section \ref{sec:Er}. Here, we discuss the most dramatic problems of the Braginskii viscosity at low collisionality, and its modifications by neoclassical theory.

The parallel viscosity can be written \cite[sec.~12.1]{PerHelander2002} as
\begin{equation}
    \mathbf{B}\cdot\nabla\cdot\Pi = (p_\perp - p_\parallel)\nabla_\parallel B + \frac{2}{3} B \nabla_\parallel (p_\perp - p_\parallel) = \frac{2}{3}B^{5/2}\nabla_\parallel \frac{G}{B^{3/2}}.
    \label{eq:viscos_anisotrop}
\end{equation}
In drift-fluid models which only evolve the total pressure $p = (2p_\perp+p_\parallel)/3$, the anisotropy is assumed to be small and approximated in a polynomial expansion as a balance between collisional relaxation and the driving forces \cite{Scott2007}, namely the flow gradients. Importantly, unlike in the original Braginskii closure, also heat flows enter \cite{Hirshman1981,PerHelander2002,Rozhansky2009}. Then, after drift-reduction \cite{zeiler:habil99,Makarov2021}, we can write the viscous function $G$ for ions as
\begin{eqnarray}
    {G} = &- \eta_\mathrm{i} \left[ \frac{2}{{B}^{3/2}}{\nabla}\cdot\left({u}_\parallel {B}^{3/2}\mathbf{b}\right) - \frac{{C}({\varphi})}{2} - \frac{{C}({p}_\mathrm{i})}{2en} \right] \nonumber\\
    &- \eta^\mathrm{heat}_\mathrm{i} \left[ \frac{2}{n {T}_\mathrm{i} {B}^{3/2}}{\nabla}\cdot\left(q_{\parallel \mathrm{i}}{B}^{3/2}\mathbf{b}\right)  - \frac{5{C}({T}_\mathrm{i})}{4e} \right].
\end{eqnarray}
Note that for the parallel heat flux, we use whatever the heat flux model is, either LF or FS. The Braginskii ion parallel flow viscosity coefficient is $\eta_\mathrm{i} = 0.96 n T_\mathrm{i} \nu_\mathrm{i}^{-1} \sim T^{5/2}_\mathrm{i}$, and the heat viscosity in the collisional limit is $\eta^\mathrm{heat}_\mathrm{i} \approx 0.71 \eta_\mathrm{i}$. Electron viscosity is usually neglected \cite{PerHelander2002}: since $\nu_\mathrm{e} / \nu_\mathrm{i} =\sqrt{2m_\mathrm{i}/m_\mathrm{e}} \approx 86$, it is formally small. 

\begin{figure}[htb]
\centering
	\centering
    \includegraphics[trim=0.0cm 0.0cm 0.0cm 0.0cm, clip, width=0.49\linewidth]{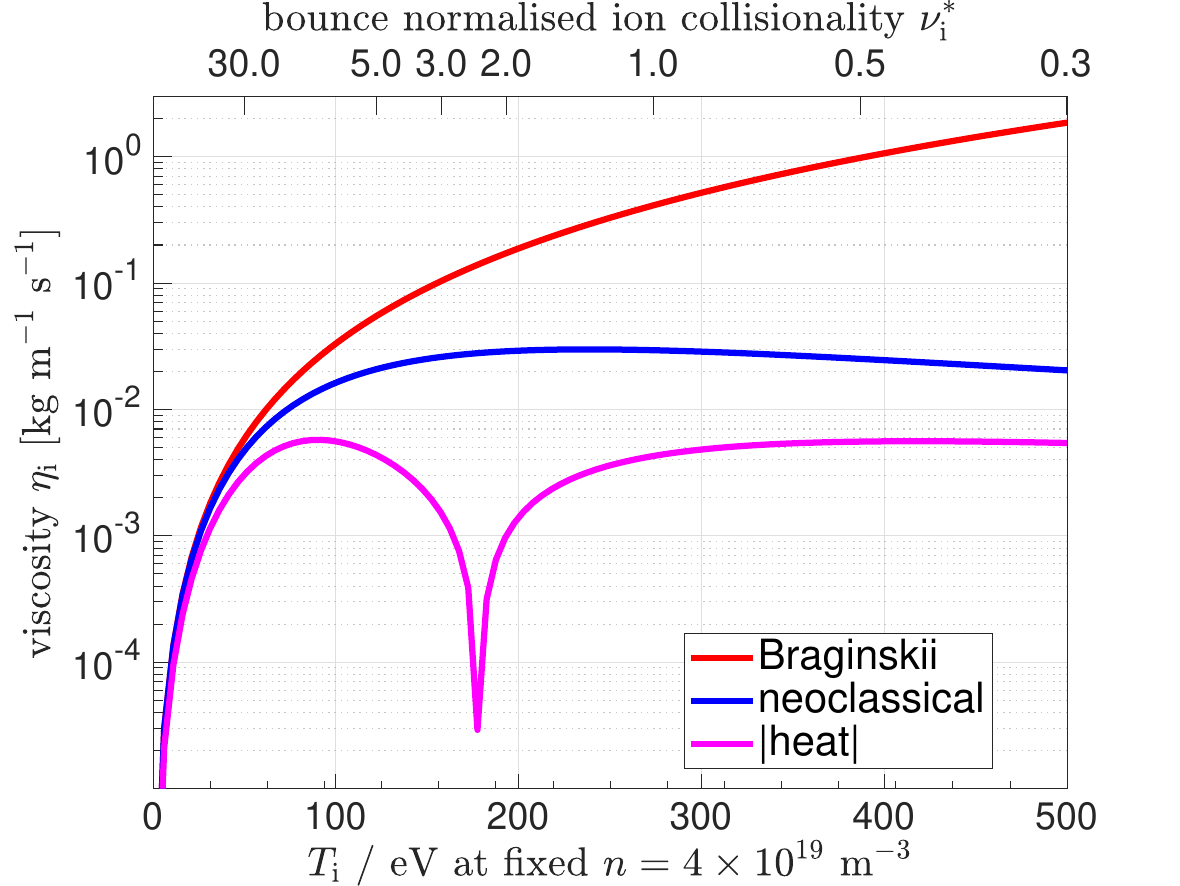}
    \caption{Ion flow viscosity $\eta_\mathrm{i}$ according to the Braginskii closure, with neoclassical corrections according to eq.~\eref{eq:neocl_visc}, and the absolute value of the heat viscosity $|\eta_\mathrm{i}^\mathrm{heat}|$. The latter is positive at $\nu_\mathrm{i}^* \gtrsim 2.4$, and negative at $\nu_\mathrm{i}^* \lesssim 2.4$.}
    \label{fig:viscosity}
\end{figure}

As for the heat conduction closure, the above approximation for the ion viscosity is only strictly valid for $\hat{\nu}_\mathrm{e,i} \gg 1$, which typically, as we have seen in figure \ref{fig:coll_etai}, is not the case. Then, the viscosity closure yields too large damping of the plasma flows and fluctuations. A rigorous treatment would require to extend the fluid hierarchy to separate parallel and perpendicular temperatures and heat fluxes \cite{Scott2021a}, i.e.~to compute the anisotropy in eq.~\ref{eq:viscos_anisotrop} explicitly. However, the procedure is complicated by the necessity to consider such intrinsically kinetic effects as friction between trapped and passing ions \cite{Hirshman1981,PerHelander2002}. Instead, motivated by the solution implemented in SOLPS by Rozhansky \textit{et al.} \cite{Rozhansky2009}, and by turbulence simulations with EMEDGE3D by De Dominici \textit{et al.} \cite{Dominici2019}, we simply modify the viscosity in accordance with neoclassical theory \cite{Hirshman1981,PerHelander2002,Rozhansky2009}. 
The resulting ion viscosity coefficient, computed according to eq.~\eref{eq:neocl_visc} in appendix \ref{chap:Braginskii_equations}, is depicted in figure \ref{fig:viscosity}.

While the Braginskii flow viscosity diverges as $T_\mathrm{i}^{5/2}$, the neoclassical flow viscosity scales like $\eta \sim \nu_*^{-1}$ only in the collisional limit. The dependence is on the bounce averaged ion collisionality, or ion detrapping rate, $\nu_\mathrm{i}^* = \epsilon^{-3/2} \hat{\nu}_\mathrm{i} \approx 6 \hat{\nu}_\mathrm{i}$, with the inverse aspect ratio $\epsilon = a/R$ ($\approx 0.3$ for AUG). It has no collisionality dependence ($\eta \sim \nu_*^{0}$) in the plateau regime $\epsilon = 0$, and it becomes $\eta \sim \nu_*$ in the banana regime. This collisionality is also shown in figure \ref{fig:coll_etai}: in our simulations we have $\nu^*_\mathrm{i} \approx 1$ in the whole confined region, meaning we are in the plateau-banana regime. Even though this collisionality is not too small yet, owing to the high density achievable in H-mode, the neoclassical flow viscosity is already an order of magnitude lower than the Braginskii viscosity. This correction becomes much more important at even lower collisionality, which occurs not only deeper in the plasma core, but also at lower density high confinement regimes such as the I-mode \cite{Whyte2010}, and during the L-H transition, due to the L-mode density limit \cite{Eich2021}.

Below, in section \ref{sec:Er}, we discuss the importance of these corrections for the radial electric field formation. However, before that, we want to stress an even more severe necessity to include the correct ion viscosity in global ``full-$f$'' turbulence simulations: the poloidal background asymmetry. We have observed it already in our previous L-mode simulations \cite{Zholobenko2021a}. Since at $\nu^*_\mathrm{i} \approx 1$, in our current H-mode simulations, we are in a marginally collisionless regime, we demonstrate this with a simulation at a lower collisionality of $\nu^*_\mathrm{i} \approx 0.05$, which corresponds e.g. to $T_\mathrm{i} = 1$ keV and $n = 2 \times 10^{19}$ m$^{-3}$. The 2D poloidal density profile in such a simulation is shown in figure \ref{fig:snapshot} (right): it uses the Braginskii viscosity, and was run in the L-mode AUG equilibrium from $\#36190$ (for historical reasons). 
Running such simulations until saturation is not possible since the formation of the immense in-out density asymmetry ultimately leads to a severe flow instability.

\begin{figure}[htb]
	\centering
    \includegraphics[trim=0cm 0cm 1.0cm 0.3cm, clip, width=0.5\linewidth]{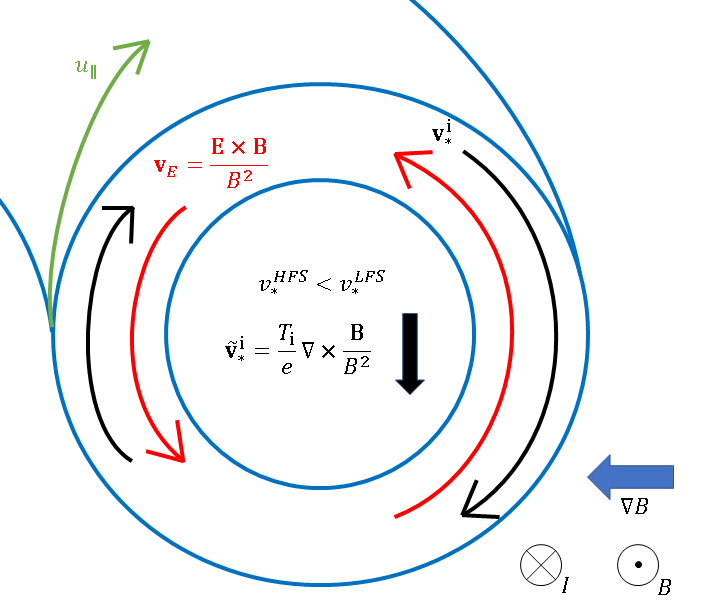}
    \caption{Cartoon of $E\times B$ and ion diamagnetic ($\mathbf{v}_*$) velocities on a closed flux surface. The toroidal magnetic field is out of the plane, the toroidal plasma current is into the plane, and $B \approx B_0 R_0 / R$. The helical parallel flow is indicated in green. Note that only the divergent part of the diamagnetic and mean $E\times B$ velocity matters \cite{Zholobenko2021}.}
    \label{fig:rot_cartoon}
\end{figure}

The reason why this problem occurs is illustrated in figure \ref{fig:rot_cartoon}. Intrinsically, in a toroidal device, the drifts are larger on the outside than on the inside of the torus due to their $1/B$ dependence. To lowest order, the $E\times B$ drift can be assumed to be balanced by the diamagnetic drift, leading to $E_r \approx \partial_r p_\mathrm{i} / en$. However, deviations from this balance occur both in the stationary state, due to neoclassical rotation and zonal flows  (see sec.~\ref{sec:Er} below), and transiently. Then, the particle balance $\left<\nabla\cdot(n\mathbf{v}_\mathrm{e,i})\right>_t=0$ is not automatically maintained: this can lead to a significant density asymmetry opposite to the velocity asymmetry. According to neoclassical theory, plasma density is supposed to be nearly constant on a flux surface. This is not the case at low collisionality when using the Braginskii ion viscosity. The reason is the excessive damping of the parallel velocity and poloidal rotation: to maintain the particle balance and the density flux-surface symmetric, the parallel plasma velocity has to become larger on the inboard than on the outboard side of the torus \cite[sec.~8.5]{PerHelander2002}. Indeed, with the corrections above, much more flux-surface symmetric densities are obtained (as in fig.~\ref{fig:snapshot}). Additionally, the flow damping affects the overall turbulent transport \cite{Scott2005}, as shown in figure \ref{fig:heat_sep}, and the radial electric field as discussed below.

\section{Composition of the radial electric field}
\label{sec:Er}

The radial electric field is a critical mechanism for macroscopic plasma stability as well as the regulation of turbulent flows \cite{Biglari1990,Groebner1990,Burrell1997,Stroth2011,Cavedon2016,Diamond2005,DifPradalier2009,Zholobenko2021,Stangeby2000,Brida2020}. As indicated in figure \ref{fig:rot_cartoon}, it leads to a poloidal rotation of the plasma. Largely, it arises in the confined region as a stabilisation of the background plasma rotation by counter-acting the diamagnetic flow, $E_r \approx \partial_r p_\mathrm{i} / en$ \cite{Zholobenko2021}. But due to diamagnetic cancellation, only $E \times B$ advection and no diamagnetic advection acts on fluctuations with poloidal gradients, so these are advected with the $E \times B$ flow. A radial $E_r$ variation thus leads to a radially sheared poloidal flow, which can squeeze, strain-out or even break vortices, regulating their radial propagation \cite{Biglari1990,Zholobenko2021,Manz2009b}. 

Because of its importance in turbulence regulation, the formation of the radial electric field $E_r$ is of major interest. Different mechanisms can compete: the balance between diamagnetic and $E \times B$ flow compression, poloidal and toroidal rotation regulated by neoclassical viscosity, parallel currents and sheath physics in the SOL, and finally zonal flows driven by the turbulence itself through the Reynolds stress. We have discussed all these mechanisms and their interaction in our previous work \cite{Zholobenko2021}, which was focused on L-mode ASDEX Upgrade. Here, we apply the analysis to H-mode conditions, to determine the dominant contributions there.

The composition of the mean radial electric field (time averaged, denoted by the overbar) in the confined region \cite{Zholobenko2021,Zholobenko2023} can be summarized as
\begin{eqnarray}
    \bar{E}_r \approx \frac{\partial_r \bar{p}_\mathrm{i}}{e\bar{n}} + (k_T - 1) \frac{1}{e} \partial_r \bar{T}_\mathrm{i} + \left< u_\parallel B_\theta \right>_{t,\theta,\phi} + \frac{m_\mathrm{i}}{e} \left< \mathbf{u} \cdot \nabla\mathbf{u} \right>_t \cdot \mathbf{e}_r.
\end{eqnarray}
This equation resembles the often cited radial force balance from the radial ion momentum equation, similar to \eref{eq:imp_force_balance}, where the second and third term on the right-hand side represent poloidal and toroidal rotation ($\theta$ is the poloidal and $\phi$ is the toroidal angle). Importantly, we stress the fourth term which is often neglected: the ion inertia. Evaluated with the mean plasma velocity, this term would vanish, because there is no mean radial component. However, since it is a quadratic quantity, the fluctuations do not vanish in a time-average: they form the Reynolds stress, driving a polarisation current, which sustains the zonal flow. Evaluated with the local poloidal and toroidal rotation, this equation would hold in general. Here, we use it in a somewhat ad-hoc fashion to connect to neoclassical theory \cite{Hinton1976,Hirshman1981,PerHelander2002}: we replace the poloidal rotation with the ion temperature gradient term, and toroidal rotation with the parallel velocity term. The coefficient $k_T$ is given in eq.~\eref{eq:k_T}. This would strictly hold only in a flux-surface average (and without ion inertia). However, the electric potential and the ion density and temperature are to lowest order flux-surface quantities. Only the parallel velocity term we write and evaluate explicitly as a flux-surface average. Others are evaluated at the outboard mid-plane, averaging them toroidally and in time, to be able to connect them to experimental measurements in figure \ref{fig:profiles}.

\begin{figure}[htb]
\centering
\begin{minipage}{0.49\textwidth}
	\centering
    \includegraphics[trim=1.0cm 0.0cm 1.0cm 0.0cm, clip, width=1.0\linewidth]{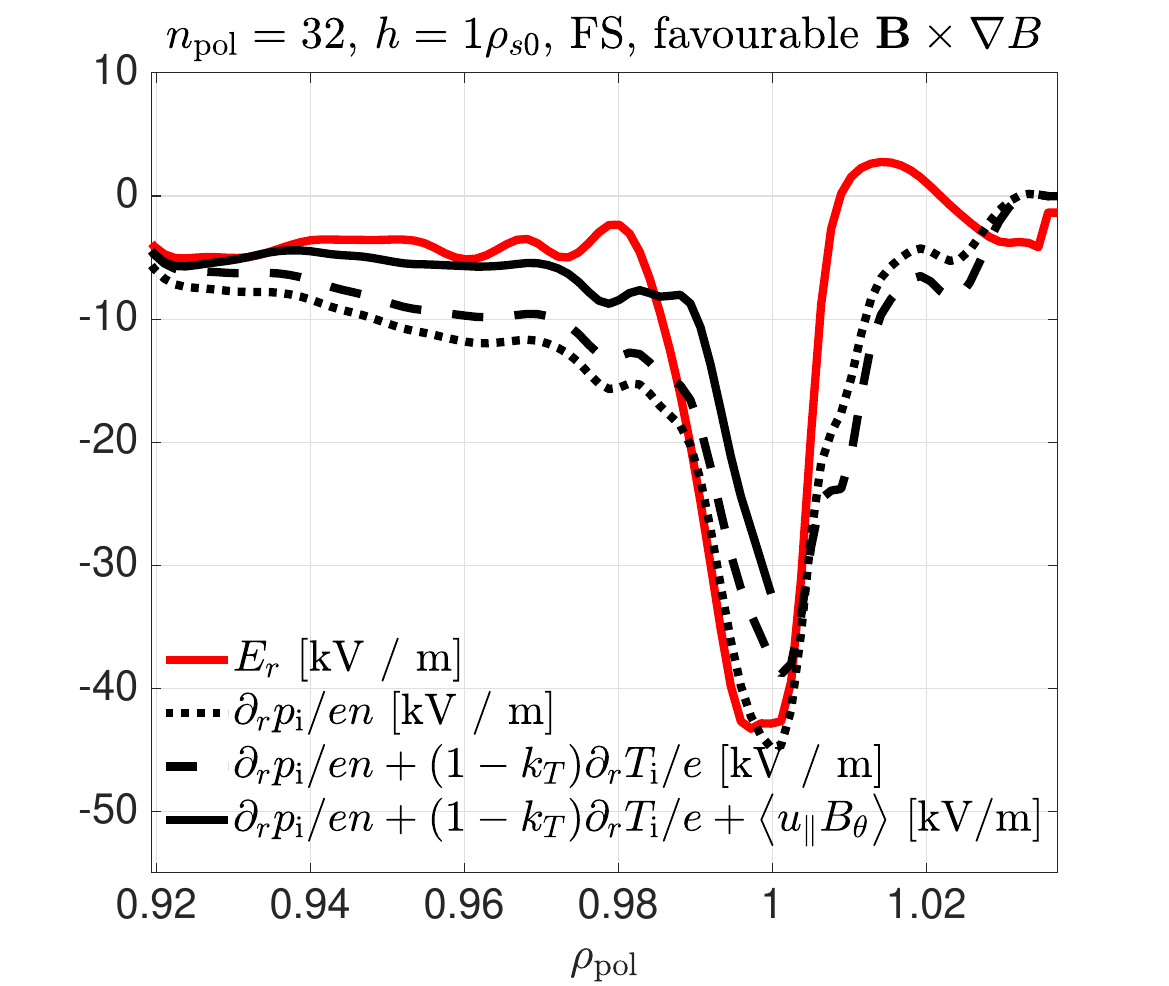}
\end{minipage}
\begin{minipage}{0.49\textwidth}
	\centering
    \includegraphics[trim=1.0cm 0.0cm 1.0cm 0.0cm, clip, width=1.0\linewidth]{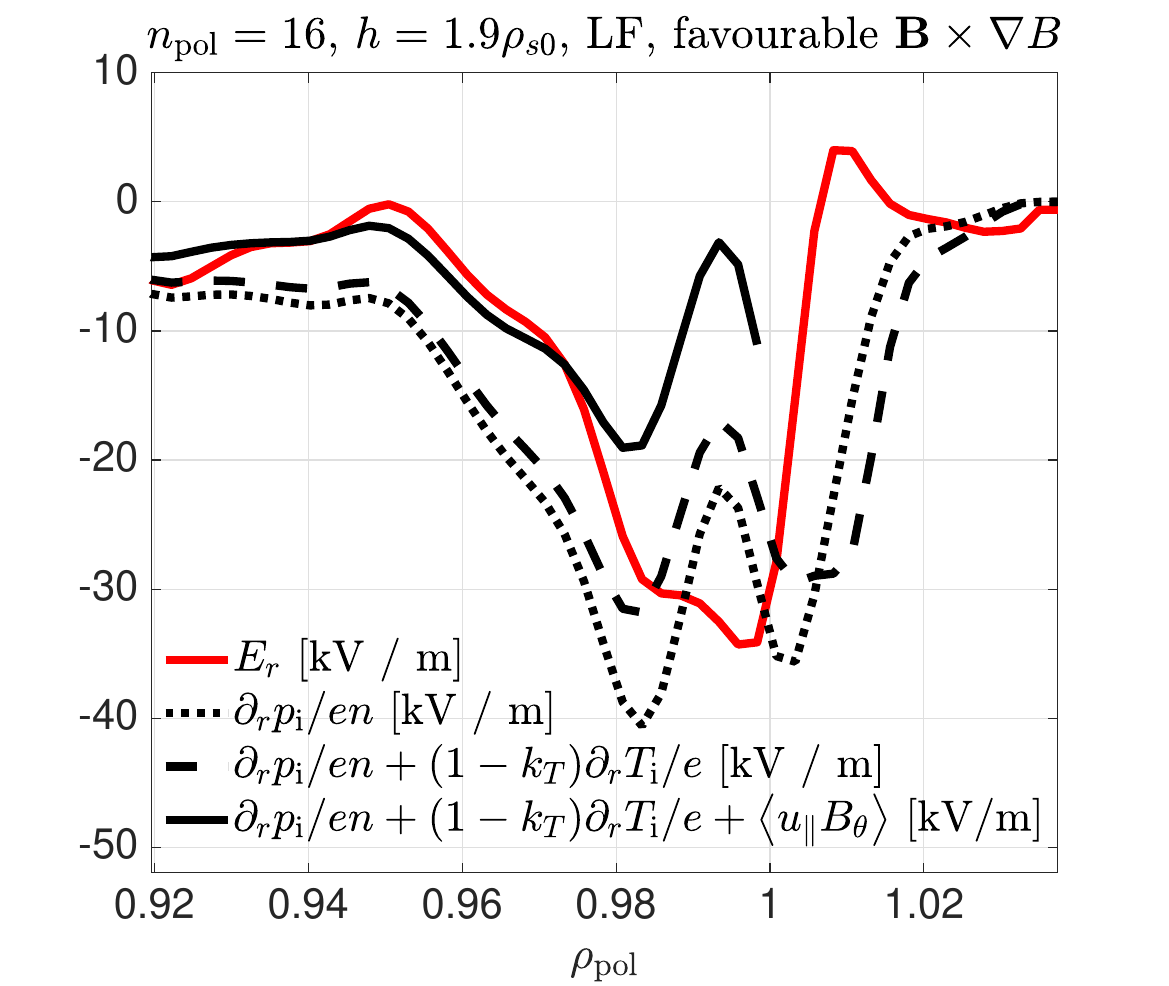}
\end{minipage}
\caption{The radial electric field at the OMP and its composition: ion pressure gradient due to the diamagnetic flow balance, neoclassical poloidal rotation and toroidal rotation. Toroidal rotation $\left< u_\parallel B_\theta \right>$ is flux-surface and time averaged, everything else is averaged only toroidally and in time over 300 µs. The residual between the electric field and the sum of these terms indicates the presence of a zonal flow. Left: high resolution case with free-streaming-limited parallel heat flux. Right: low resolution case with Landau-fluid closure.
}
\label{fig:Er_comp}
\end{figure}

The contributions are plotted in figure \ref{fig:Er_comp} for two cases: the high resolution free-streaming case, and the lower resolution Landau-fluid case (both in favourable $\mathbf{B}\times\nabla B$ configuration). Let us begin the analysis with the high resolution (reference) simulation: a key finding is that particularly around the minimum of the $E_r$ well, it is mostly governed by the ion pressure gradient term, in agreement with previous findings from ASDEX Upgrade \cite{Viezzer2013,Cavedon2016}. The mean ion pressure gradient thus plays a key role in sustaining the H-mode radial electric field, and is a good proxy for it. However, this does not prove the absence of other contributions. Both poloidal and toroidal rotation lead to an up-shift of $E_r$. Deeper in the confined plasma, $E_r$ only matches well when both rotational contributions are added. We have confirmed this by carrying out simulations without neoclassical heat viscosity, $\eta_\mathrm{i}^\mathrm{heat} = k_T - 1 = 0$, finding indeed an even lower (larger in magnitude) $E_r$ in the confined region: this confirms the need for including it, as argued in sec.~\ref{sec:neocl_visc}. The reason for $E_r \approx \partial_r p_\mathrm{i} / en$ in the well is the zonal flow, the difference between the red and the black solid curves: it is largest around the separatrix, and happens to balance the poloidal and toroidal rotation. This is similar to our L-mode simulations \cite{Zholobenko2021}, the zonal flow (polarisation current / vorticity) has even larger absolute amplitude here, but it is lower relative to the mean-field diamagnetic flow. 
What is different from L-mode is that there are no geodesic acoustic mode \cite{Conway2021} oscillations here (\textbf{no GAMs}), as we will discuss in section \ref{sec:characterisation}: this might happen precisely when the zonal flow becomes balanced by plasma rotation instead of the ion pressure gradient, because unlike the plasma pressure, rotation can be statically poloidally asymmetric, avoiding the need for periodic GAM relaxations. 
Lastly, we note that the rather small $E_r$ peak in the SOL is well described by $-\Lambda \partial_r T_\mathrm{e}$ \cite{Brida2022,Plank2023}. But in the near-SOL, $E_r < 0$ indicates a polarisation halo (an extension of the zonal flow into the SOL) \cite{Loizu2017}.

The radial electric field composition in the lower resolution case with the Landau fluid closure is interesting for two reasons. Firstly, as discussed in section \ref{sec:validation}, it matches better the experiment: the $E_r$ well is reduced, but widened into the confined region. 
Secondly, we see more clearly the zonal flow: particularly in the ion pressure gradient, the radial oscillations are larger in amplitude, but also in the $E_r$ there is an additional radial wiggle inside the larger well. This is similar to experimental observations on the JET tokamak \cite{Hillesheim2016}. However, we have to stress that in parts, this is due to numerical resolution: the zonal flow amplitude is larger in the lower resolution case already with the FS closure. Thus, at this point, we can only highlight the qualitative reproduction of experimentally observed features. Even higher resolution simulations with advanced fluid closures are required for exact predictions, and they have to be precisely validated against more experiments. But also, kinetic effects in general, like trapped particles, ion orbit losses, arbitrary wavelength finite Larmor radius effects (gyro-averaging) as well as better conducting sheath boundary conditions could influence the results.

\section{Characterisation of H-mode transport}
\label{sec:characterisation}

\begin{figure}[htb]
\centering
	\centering
    \includegraphics[trim=0.0cm 0.0cm 1.0cm 0.5cm, clip, width=0.49\linewidth]{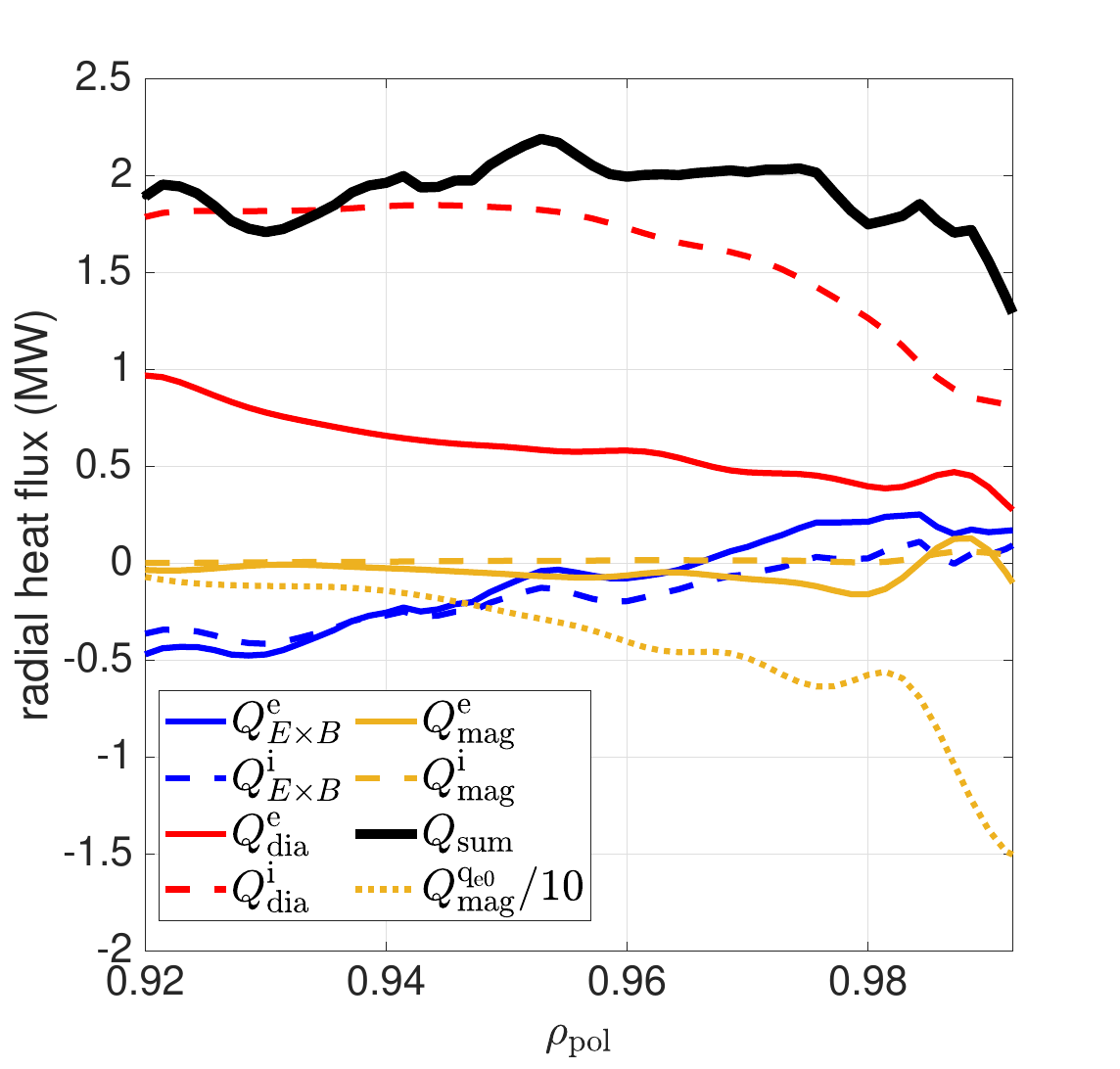}
    \caption{Flux-surface integrated and time-averaged radial heat flows, defined in eq.~\eref{eq:Q_ExB}-\eref{eq:Q_magi}. 
    }
    \label{fig:Qzonal}
\end{figure}

In this section, we characterise the heat transport in our H-mode simulations in more detail. To this end, multiple observables can be investigated: the different transport channels, fluctuation amplitudes and phase relations between quantities. We will focus the analysis on our highest resolution simulation ($n_\mathrm{tor}=32$, $h=1\rho_{s0}$, FS). Let us begin with the three different radial heat flux channels: $E\times B$, diamagnetic and magnetic,
\begin{eqnarray}
    \label{eq:Q_ExB}
    Q_{E \times B}^\mathrm{e,i} = \frac{3}{2}nT_\mathrm{e,i} \mathbf{v}_{E \times B}\cdot\mathbf{e}_r, \\
    \label{eq:Q_dia}
    Q_\mathrm{dia}^\mathrm{e,i} = \frac{5}{2}nT_\mathrm{e,i} \mathbf{v}_\mathrm{dia}^\mathrm{e,i}\cdot\mathbf{e}_r = \mp \frac{5}{2}n T_\mathrm{e,i}^2 \left( \nabla \times \frac{\mathbf{B}}{eB^2} \right)\cdot\mathbf{e}_r = \mp \frac{5 n T_\mathrm{e,i}^2}{eB_0R_0} \mathbf{e}_Z \cdot \mathbf{e}_r , \\
    \label{eq:Q_mage}
    Q_\mathrm{mag}^\mathrm{e} = \left( \frac{5}{2}nT_\mathrm{e} v_\parallel 
    + q_{\parallel \mathrm{e}}  - 0.71 j_\parallel T_\mathrm{e} \right) \mathbf{b}_1\cdot\mathbf{e}_r, \\
    \label{eq:Q_magi}
    Q_\mathrm{mag}^\mathrm{i} = \left( \frac{5}{2}nT_\mathrm{i} u_\parallel 
    + q_{\parallel \mathrm{i}} \right) \mathbf{b}_1\cdot\mathbf{e}_r.
\end{eqnarray}
$\mathbf{e}_r$ is the radial unit vector orthogonal to the flux surfaces. $\mathbf{b}_1$ is the perturbed radial magnetic field unit vector. $v_\parallel$ and $u_\parallel$ are the electron and ion parallel velocities, and $q_{\parallel \mathrm{e,i}}$ are the parallel heat fluxes. The $E\times B$ heat flux has a factor 3/2 instead of 5/2 due to the Poynting cancellation \cite{Scott2021}. For the diamagnetic velocity, we consider only the divergent part \cite{Stangeby2000,Zholobenko2021}, which includes the curvature and the $\nabla B$ drifts: $B^2\nabla\times\frac{\mathbf{B}}{B^2} = B \nabla \times \mathbf{b} + \mathbf{b} \times \nabla B$. Under the large aspect ratio, toroidal magnetic field approximation $\mathbf{B} = B_0 \frac{R_0}{R} \mathbf{e}_\phi$, this becomes a purely vertical drift. These three channels, integrated over closed flux-surfaces and averaged in time, are plotted in figure \ref{fig:Qzonal}. Note that the total radial power flux $Q_\mathrm{sum}$ matches the input power in figure \ref{fig:heat_in} and table \ref{tab:res_scan}.

The number one conclusion here is that $E\times B$ turbulence contributes rather little to the total heat flux. This is the opposite of our previous L-mode simulations, where it was absolutely dominant \cite{Zhang2024}. At the pedestal top, close to the core boundary, $Q_{E \times B}$ is even inwards: this simply balances a part of the diamagnetic outward flux. 
By far, the dominant transport channel at the pedestal top is diamagnetic, in particular for the ions: this can be regarded as neoclassical transport, because it does not depend on electric or magnetic fluctuations. 
A detailed analysis of neoclassical transport would however require to separate the background and fluctuating contributions also for the $E \times B$ drift, which is beyond our present scope. 
In future, we could investigate how well our drift-fluid model captures neoclassical transport in the collisionless limit \cite{Rozhansky2009}, and how fluctuations and turbulence can  induce flux surface asymmetries and enhance neoclassical transport: in-out due to ballooned fluctuations, and top-bottom due to the zonal flow \cite{Manz2016,Zholobenko2021}. 

The next important observation is that transport by magnetic fluctuations becomes noticeable, particularly at the pedestal foot (outer plasma edge). This is again in contrast to our previous L-mode simulations \cite{Zhang2024}, where magnetic transport did not exceed $1\%$. Now, at $\rho_\mathrm{pol} = 0.99$, it is around $10\%$. 
The mechanism by which magnetic fluctuations cause transport is somewhat different from that of electrostatic fluctuations. The $E\times B$ drift is a fluid velocity by itself, so it directly transports the plasma pressure (energy). The magnetic fluctuations $\mathbf{b}_1 = \nabla A_1\times\mathbf{b}_0/B$ have a similar form, but they rather redirect the parallel heat flows, as can be seen in equations \eref{eq:Q_mage} and \eref{eq:Q_magi}. Therefore, magnetic heat transport is typically larger for electrons, since $v_\parallel > u_\parallel$ and $q_{\parallel\mathrm{e}} > q_{\parallel\mathrm{i}}$. Further, heat conduction $q_{\parallel\mathrm{e,i}}$ along the magnetic field is typically larger than heat convection. This is problematic for a drift-fluid model which uses a closure on the heat conduction. Note that both for the Landau-fluid \cite{Pitzal2023} and the free-streaming limited heat flux models, $q_{\parallel\mathrm{e,i}}$ itself has two components, driven by the equilibrium and the perturbed magnetic fields. In the FS case, it is
\begin{equation}
    q_{\parallel \mathrm{e}} = -\tilde{\kappa}^\mathrm{e}_{\parallel0} T_\mathrm{e}^{5/2} \left( \mathbf{b}_0 + \mathbf{b}_1 \right) \cdot \nabla T_\mathrm{e} = q_{\mathrm{e}0} + q_{\mathrm{e}1}.
\end{equation}
We have plotted separately in figure \ref{fig:Qzonal} the electron radial magnetic heat flux due to the heat conduction along the equilibrium field, $Q_\mathrm{mag}^\mathrm{qe0} = q_{\parallel \mathrm{e} 0} \mathbf{b}_1\cdot\mathbf{e}_r$. Close to the separatrix, it is around 10 times larger than the total heat flux, but it is balanced by $Q_\mathrm{mag}^\mathrm{qe1} = q_{\parallel \mathrm{e} 1} \mathbf{b}_1\cdot\mathbf{e}_r$ (which is a cubic non-linearity \cite{Villa2022}!). Clearly, electromagnetic fluctuations play a critical role in H-mode turbulence: not just by stabilising $E \times B$ transport as discussed in section \ref{sec:flutter}, but also by causing radial transport themselves. However, since it involves the parallel heat conduction, an exact determination of radial magnetic transport depends on details of the fluid closure.

\begin{figure}[htb]
\centering
	\centering
    \includegraphics[trim=0.0cm 0.0cm 1.0cm 0.5cm, clip, width=0.49\linewidth]{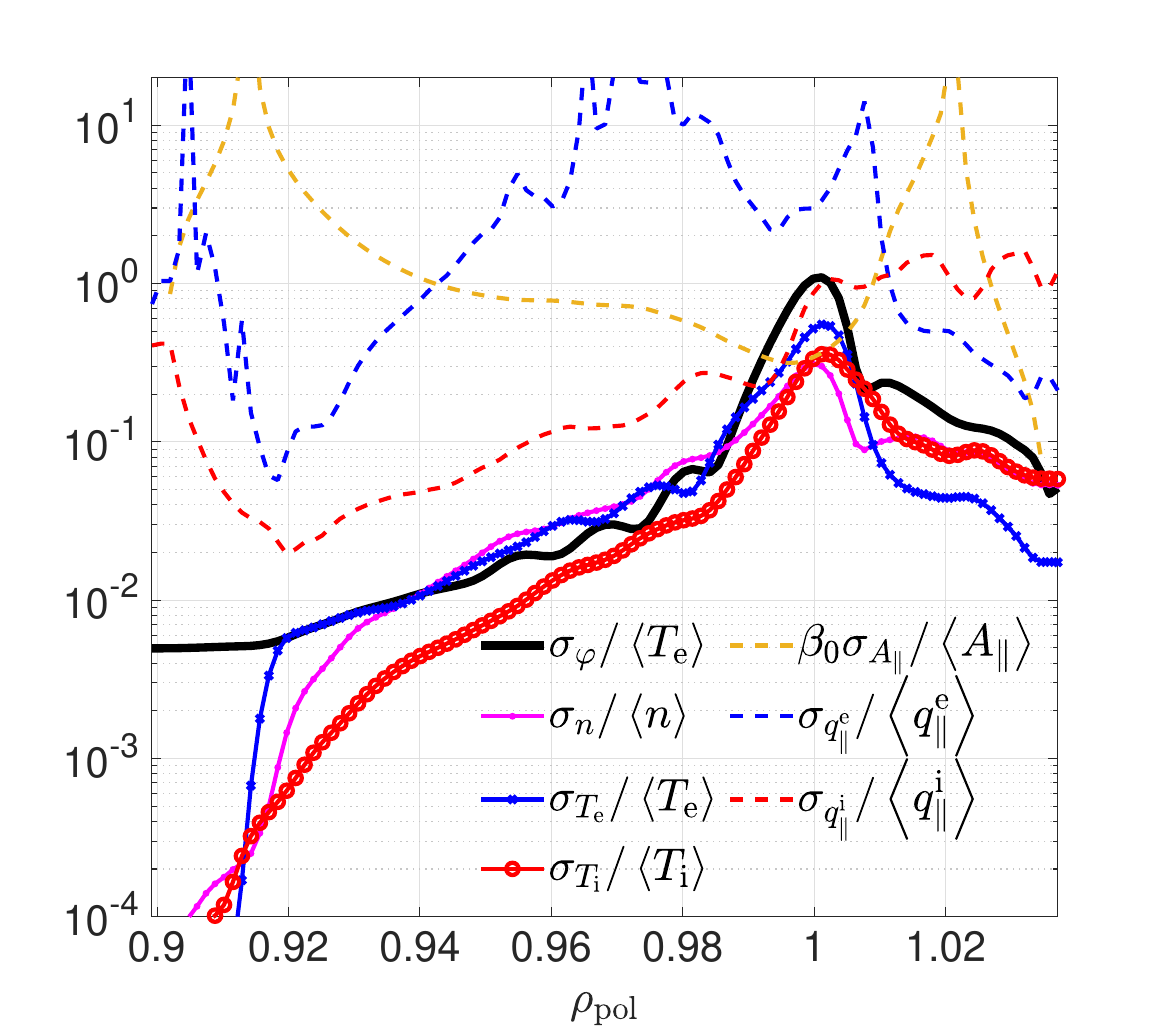}
    \caption{OMP fluctuation amplitudes.}
    \label{fig:fluct}
\end{figure}

Now, let us examine additionally the fluctuation amplitudes at the OMP in figure \ref{fig:fluct}. They are computed from standard deviations $\sigma_f^2 = \left< f^2 \right>_{t,\phi} - \left< f \right>^2_{t,\phi}$, averaged toroidally and in time. All fluctuations are normalised to their mean values, except for the electrostatic potential $\varphi$, which is usually normalised to the mean electron temperature. We also show parallel heat flux and electromagnetic potential fluctuation amplitudes for the discussion of electromagnetic transport. $A_\parallel$ is additionally multiplied by the reference electron beta $\beta_0=\mu_0 n_0T_\mathrm{e0}/B_0^2=5.6\times10^{-5}$, just to make it fit on the same scale as the other fluctuations.

Let us begin by discussing the fluctuations relevant for $E \times B$ transport (solid lines): firstly, they are indeed all small (few $\%$) deeper in the confined region. This is consistent with dominant neoclassical transport, but could also be partly caused by our artificial core boundary. On the other hand, the fluctuation amplitudes peak at the separatrix, reaching $30-100\%$. This shows that turbulent fluctuations are nonetheless quite significant in the plasma edge, despite not causing that much transport. 
The second key observation is actually something that is \textbf{not} there: geodesic acoustic modes (GAMs). 
If we subtract the toroidal average of the electrostatic potential in the computation of its fluctuation amplitudes, the result remains the same! Such a procedure can be used to isolate small-scale turbulence from GAMs, i.e.~toroidally (zonally) symmetric fluctuations of the zonal flow. In our past L-mode simulations \cite{Zholobenko2023}, indeed, a difference has been observed -- but not in H-mode. This serves to highlight that despite the presence of a zonal flow, GAMs are absent. As noted in section \ref{sec:Er}, this might be because the zonal flow is balanced by asymmetric plasma rotation, avoiding the need for relaxations of the pressure asymmetry due to zonal flow compression \cite{Conway2021}. 
Next, we note that ion temperature fluctuations are the smallest: this means that the ion temperature gradient (ITG) mode is insignificant here. This is consistent with $\eta_\mathrm{i} = \partial_r \ln T_\mathrm{i} / \partial_r \ln n < 1$ \cite{Guzdar1983,Mosetto2015}, as shown in figure \ref{fig:coll_etai}. The fact that electrostatic potential, density and electron temperature fluctuations are comparable is a strong indication for drift-wave turbulence. At $\rho_\mathrm{pol}>0.99$, $\varphi$ fluctuations become particularly large though, indicating a possibly stronger impact of ballooning modes.

\begin{figure}[htb]
\begin{minipage}{0.315\textwidth}
	\centering
    \includegraphics[trim=0.0cm 0.0cm 1.0cm 0.0cm, clip, width=1.0\linewidth]{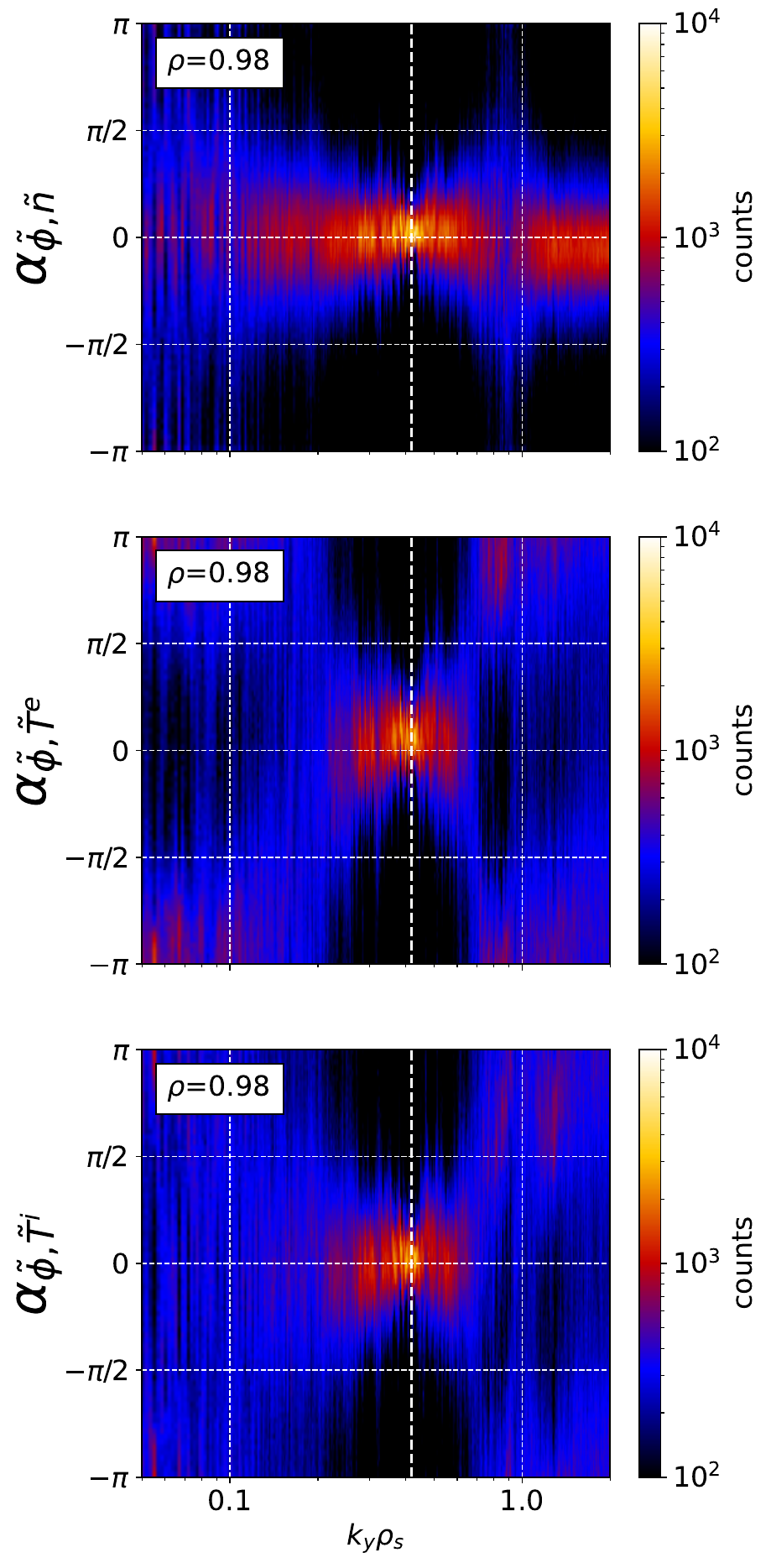}
\end{minipage}
\begin{minipage}{0.315\textwidth}
	\centering
    \includegraphics[trim=0.0cm 0.0cm 1.0cm 0.0cm, clip, width=1.0\linewidth]{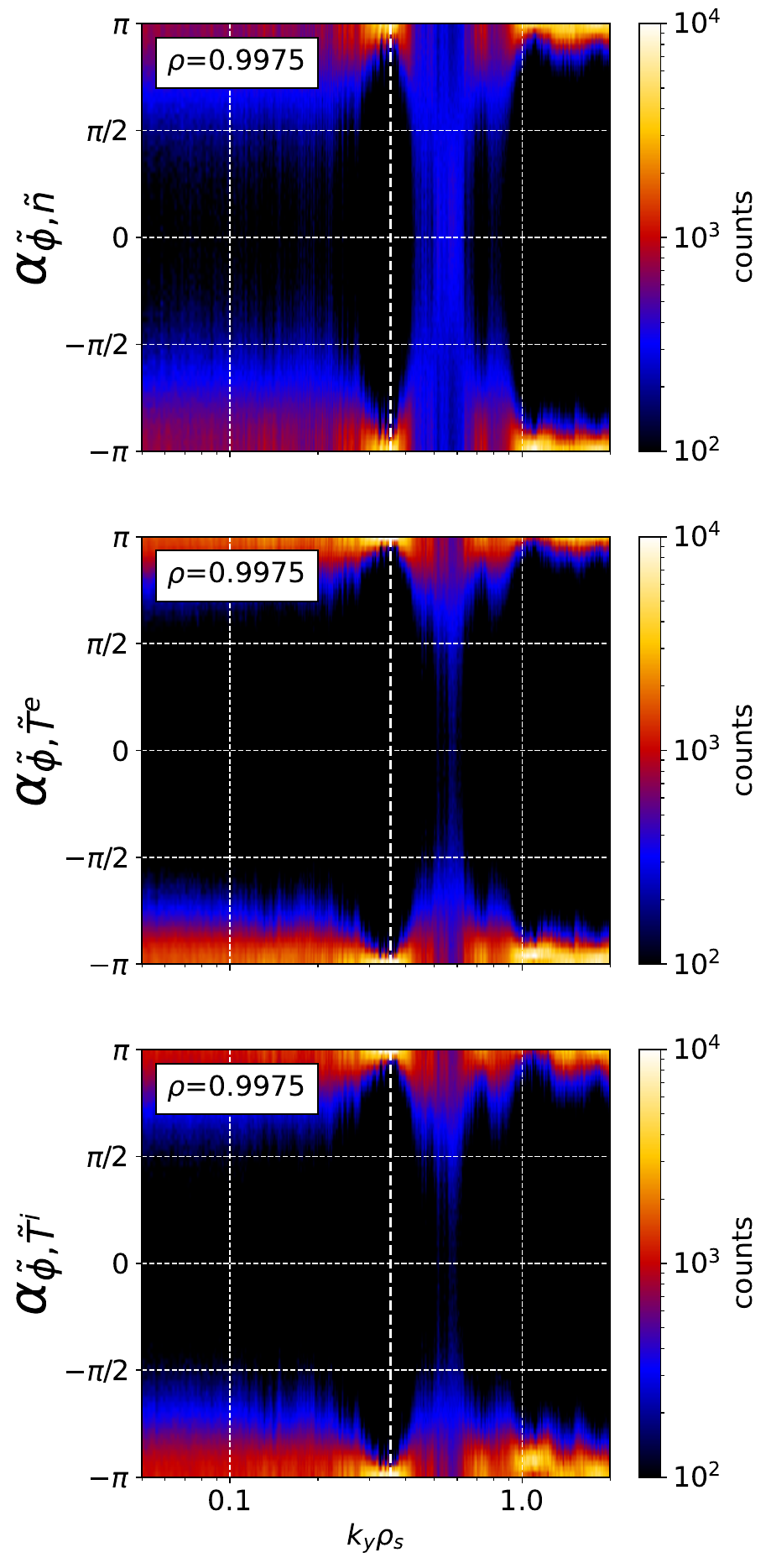}
\end{minipage}
\begin{minipage}{0.36\textwidth}
	\centering
    \includegraphics[trim=0.0cm 0.0cm 0.0cm 0.0cm, clip, width=1.0\linewidth]{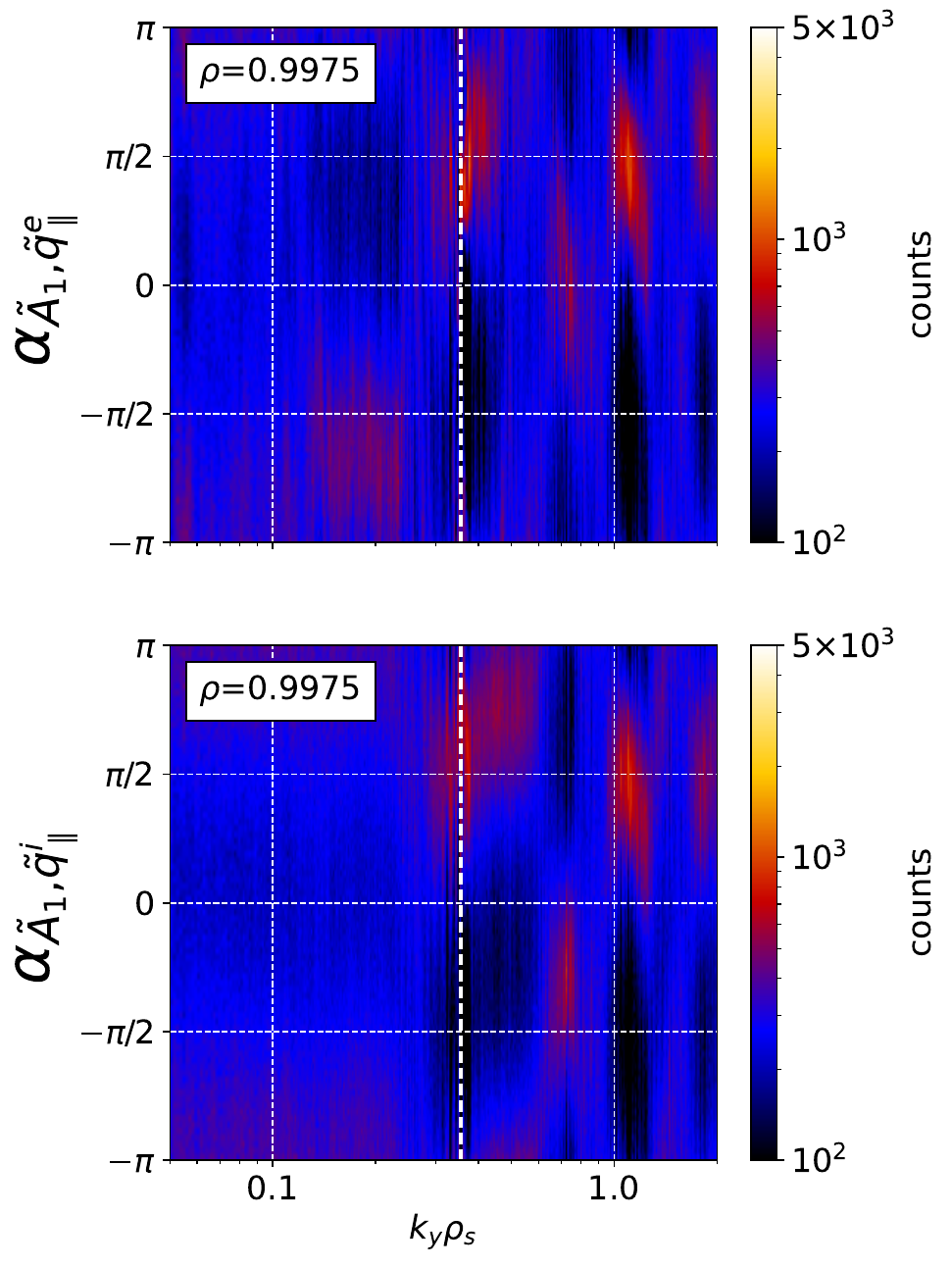}
\end{minipage}
\caption{Poloidal phase shift distributions at $\rho_\mathrm{pol} = 0.98$ (left) and $\rho_\mathrm{pol} = 0.9975$ (center) between the electrostatic potential, density and temperatures, sampled toroidally and in time. On the right are phase shifts between the electromagnetic potential and electron and ion parallel heat fluxes at $\rho_\mathrm{pol} = 0.9975$.}
\label{fig:phases}
\end{figure}

Finally, the fluctuations relevant for magnetic transport (dashed lines) are shown just to highlight that they are significant. However, we must admit that this situation is not significantly different from our simulations of L-mode turbulence. For identifying the actual difference, we must rather examine the correlations between fluctuating quantities \cite{Zholobenko2023, Ulbl2023}. For the mean radial $E \times B$ heat flux, we can write
\begin{eqnarray}
    \left< Q_{E\times B}^\mathrm{e,i} \right> = \frac{3}{2} \left< \frac{n T_\mathrm{e,i}}{B} \partial_y \varphi \right> \approx \frac{3}{2} \frac{\left<T_\mathrm{e,i}\right>}{B} \left< n \partial_y \varphi \right> + \frac{3}{2} \frac{\left<n\right>}{B} \left< T_\mathrm{e,i} \partial_y \varphi \right>.
\end{eqnarray}
The radial $E \times B$ velocity is given by the poloidal electric field $-\partial_y \varphi$. In the second step, we split the heat flux into a convective and a conductive component: this is valid when $\left< \partial_y \varphi \right> = 0$ and triple correlations are negligible. Then, integrating poloidally, we can Fourier transform the quantities, obtaining
\begin{eqnarray}
    \oint Q_{E\times B}^\mathrm{e,i} \mathrm{d}y \approx 
    \frac{3}{4\pi} \frac{\left< T_\mathrm{e,i} \right>_y}{\left< B \right>_y} \int k |\tilde{n}| |\tilde{\varphi}| \sin (\alpha_{\tilde{\varphi},\tilde{n}}) \mathrm{d}k + \frac{3}{4\pi} \frac{\left< n \right>_y}{\left< B \right>_y} \int k |\tilde{T}_\mathrm{e,i}| |\tilde{\varphi}| \sin (\alpha_{\tilde{\varphi},\tilde{T}_\mathrm{e,i}}) \mathrm{d}k
    \label{eq:Q_fourier}.
\end{eqnarray}
The radial turbulent transport is thus given by the product of the individual Fourier fluctuation amplitudes of the different quantities times the phase difference between them, $\alpha_{\tilde{\varphi},\tilde{n}}(k) = \mathrm{Im} \log ( \tilde{\varphi} \tilde{n}^* )$. Even if fluctuations are arbitrarily large, if they are in phase with each other, they cause on average no transport. Therefore, it is important to also analyse the phase shifts, which we do in figure \ref{fig:phases}.

At inner flux surfaces $\rho_\mathrm{pol} \leq 0.98$, we find typical drift-wave mode structure: all quantities have a phase difference close to zero \cite{Scott2021,Zhang2024}. The transport peaks where the phase distribution is most coherent, around $k_y \rho_s \approx 0.5$. A phase slightly above zero means transport radially outward, but it is indeed small. The reason for the small $E \times B$ radial transport is thus that drift-waves are very stable, due to the efficient adiabatic coupling between pressure and potential fluctuations, despite fluctuation amplitudes of up to $10\%$. So far, this is similar to our L-mode simulations \cite{Zholobenko2023,Zhang2024}. But closer to the separatrix, things change.

At $\rho_\mathrm{pol} \geq 0.99$, we find that $\alpha_{\tilde{\varphi},\tilde{n}}$ and $\alpha_{\tilde{\varphi},\tilde{T}_\mathrm{e,i}}$ become centered around $\pm \pi$. Quantitatively, this leads to little transport, similarly to $\alpha = 0$. However, it serves as an indication that there is a qualitative change in the character of the turbulence \cite{Doerk2016,Grenfell2024}. The reason becomes apparent when we also characterise correlations for the electromagnetic transport
\begin{eqnarray}
    Q_\mathrm{mag}^\mathrm{e,i} \sim q_{\parallel \mathrm{e,i}} \mathbf{b}_1 \cdot \mathbf{e}_r = - \frac{q_{\parallel \mathrm{e,i}}}{B} \partial_y A_1 \sim - k |\tilde{q}_{\parallel \mathrm{e,i}}| |\tilde{A}_1| \sin (\alpha_{\tilde{A}_1,\tilde{q}_{\parallel \mathrm{e,i}}}).
\end{eqnarray}
Even though the fluctuation amplitudes of the electromagnetic potential and parallel heat fluxes are not insignificant on inner flux surfaces, their phases are rather uncorrelated, leading to little electromagnetic transport. This changes towards the separatrix: at $\rho_\mathrm{pol} \geq 0.99$, $\alpha_{\tilde{A}_1,\tilde{q}_{\parallel \mathrm{e,i}}}$ becomes more coherent at $k_y \rho_s \approx 0.35$, with a value close to $\pi/2$ (the peaks in the spectrum at $k_y \rho_s > 0.5$ lead to no transport due to the low fluctuation amplitudes there). This is why electromagnetic transport becomes substantial. The phase close to $\pi/2$ indicates a ballooning mode, an Alfvén wave destabilized by curvature \cite{Scott1997}. These are neither resistive electrostatic nor ideal ballooning. Often they are referred to as kinetic ballooning modes (KBM), a kinetic extension of ideal MHD ballooning \cite{Tang1980,Aleynikova2018}, whereby the main effect is a diamagnetic modification \cite{Zocco2018} which is captured by our two-fluid model. The presence of KBMs in H-mode pedestals has been predicted previously by local models \cite{Dickinson2012}, it is even an integral part of the EPED pedestal stability model \cite{Snyder2011}. We observe it most clearly in the very vicinity of the separatrix. However, we must note that this observation is not unconditional: so clearly, we only find this mode in our highest resolution free-streaming simulation. As indicated above, electromagnetic fluctuations involve the conductive parallel heat flux $q_{\parallel \mathrm{e,i}}$, on which we apply a closure in our drift-fluid model. Indeed, the mode is less clearly pronounced (less coherent) in lower resolution (also LF) simulations. Thus, even higher resolution and ideally gyrokinetic simulations are required for a more definitive analysis.

\begin{figure}[htb]
\centering
	\centering
    \includegraphics[trim=0.0cm 0.0cm 0.0cm 0.0cm, clip, width=0.49\linewidth]{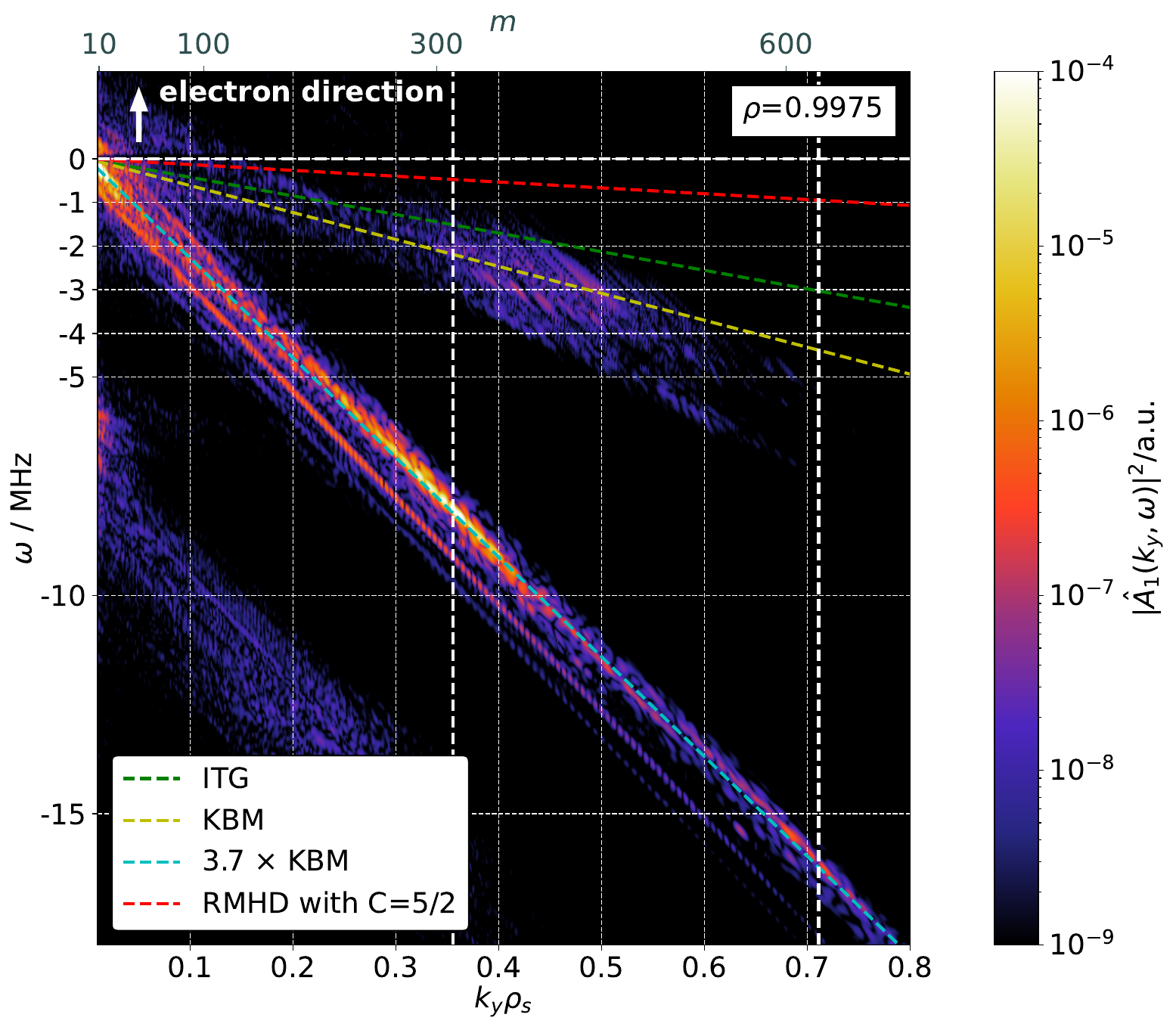}
    \caption{Temporal and poloidal Fourier power spectrum of magnetic potential $A_1$ fluctuations at $\rho_\mathrm{pol}=0.9975$. The top axis shows the corresponding poloidal mode number. The cyan line is the same as yellow but scaled by a constant factor of 3.7.}
    \label{fig:apar_dispersion}
\end{figure}

We can do one more thing to corroborate the observation of KBMs in our simulations: transform the $k_y$ spectrum also in time. The procedure, including the removal of the Doppler shift due to the mean $E \times B$ rotation, is described in more detail by Ulbl \textit{et al.} in \cite{Ulbl2023}. 
The resulting power spectrum of the vector potential fluctuations $|\hat{A}_1(k_y,\omega)|^2$ at $\rho_\mathrm{pol}=0.9975$ is displayed in figure \ref{fig:apar_dispersion}.

The dominant mode is found at $k_y \rho_s \approx 0.35$ ($m=314$ and $n=71$ with $q\approx4.4$), as indicated by the first, vertical, dashed, white line. The second line indicates the presence of a lower amplitude, higher harmonic. We can compare the spectrum with analytical dispersion relations. The yellow dashed line describes the (collisionless) gyrokinetic KBM in the flux-tube limit \cite{Zocco2018}. The mode observed here has approximately four times higher frequency than that, but as Zocco \textit{et al.} \cite{Zocco2018} point out, geometry can substantially modify the flux-tube dispersion relation, which is not exactly applicable to a mode localised right at the separatrix. Also collisionality (resistivity), $E \times B$  shear and other model differences can lead to discrepancies. The main finding is that the dominant mode propagates in the ion diamagnetic direction with a frequency of 5-10 MHz. This excludes micro-tearing (MTM) modes, which would propagate in the electron direction. Further evidence is provided by the observations that 1) the electrostatic heat flux is $Q_\mathrm{ES} \sim Q_\mathrm{EM}$ and 2) the mode is MHD-like since the normalized parallel electric field $\hat{E}_{||} = \int |-\partial_z \phi_1 + \partial_t A_{1}| \mathrm{d}z / (\int |\partial_z \phi_1| \mathrm{d}z + \int |\partial_t A_{1}| \mathrm{d}z) \lesssim 0.1$ is small\cite{Hatch2016}. Here $z$ denotes the coordinate along the magnetic field. To check for the possibility of resistive ballooning modes, we compared the dispersion relation of resistive fluid modes given in the appendix B of Kotschenreuther \textit{et al.} \cite{Kotschenreuther2019}. Figure \ref{fig:apar_dispersion} contains one of these dispersion relations in red. We find that for different resistive modes, the frequencies are in a similar order of magnitude, much lower than the analytical KBM and the dominant mode found in the spectrum. Together, these different observations provide strong evidence that the observed mode is a KBM. 

A fundamental property of the KBM is a threshold in the beta $\beta_\mathrm{e} = 2\mu_0 n T_\mathrm{e} / B^2$ \cite{Pueschel2008}: above a critical value the mode is found to be linearly unstable \cite{Terry2015}. While the radial localization of the mode at $\rho_\mathrm{pol}\approx0.997$ means that plasma pressure and beta are generally small, this does not universally rule out the possible growth of a KBM. Using a local estimation in sheared geometry, the MHD ideal ballooning limit is $\beta_\mathrm{crit}^\mathrm{MHD} = 0.6\hat{s}/\big[q^2(2w_n+w_{T,\mathrm{i}}+w_{T,\mathrm{e}})\big]$, where $\hat{s}=r (\partial{q}/{\partial r}) / q$ is the normalized shear and $w_g = -R_0 (\partial{g}/{\partial r}) / g$ with $g \in \{n, T_\mathrm{i}, T_\mathrm{e}\}$ are normalized gradients \cite{Pueschel2008}. We find a local minimum of this quantity at around $\rho_\mathrm{pol}\approx0.995$. With nominal values of $q\approx4.1$, $\hat{s}\approx15.2$, $w_n\approx130$, $w_{T,\mathrm{i}}\approx192$, $w_{T,\mathrm{e}}\approx104$, this results in $\beta_\mathrm{crit}^\mathrm{MHD}\approx0.1\%$. At this location, $\beta \approx 0.07\%$, which is approximately 25\% below the ideal limit. Previous studies have found the KBM threshold typically below the ideal limit in similar orders of magnitude as reported here \cite{Pueschel2008,Snyder2001}. 
Further inside, at $\rho_\mathrm{pol}<0.993$, $E \times B$ shear seems to prevent the KBM from growing \cite{Chen2023}, explaining its localisation close to the separatrix where $\partial_r E_r \approx  0$.

Figure \ref{fig:apar_dispersion} additionally reveals the presence of a low amplitude, low frequency mode close to the ITG dispersion relation (green line) \cite{DannertPhd}, seen just like this also in the $|\varphi(k_y,\omega)|^2$ spectrum. Indeed, the phase shift diagnostics (figure \ref{fig:phases} center column) show a gap between $0.4 \lesssim k_y\rho_s \lesssim 0.6$ where the phase shifts are qualitatively different but overshadowed by the dominating KBM in the plot. In this gap, density phase shifts are close to 0, while temperature phase shifts tend to be closer to $\pi/2$, typical for ITG. The corresponding fluctuation amplitudes are rather small, and the transport caused by this mode is insignificant. Thus, consistently with $\eta_\mathrm{i}$ reaching 1 just at the separatrix (figure \ref{fig:coll_etai}), there is indeed a subdominant ITG mode.

In summary, we find an interesting composition of transport channels in our H-mode simulations. Overall, turbulence is mostly of drift-Alfvén type, which is why $E \times B$ transport is actually rather low. In the upper part of the pedestal, transport is mostly neoclassical through the diamagnetic (curvature and $\nabla B$) drift. At the pedestal foot, on the other hand, electromagnetic transport becomes critical in the form of KBMs. Thus, magnetic fluctuations play a critical role both in stabilising $E \times B$ transport, as well as driving magnetic transport themselves. The latter could become even more important in other H-mode scenarios, such as the small-ELM quasi-continuous exhaust (QCE) regime \cite{Faitsch2021} and the no-ELM enhanced $D_\alpha$ (EDA) regime \cite{Stroth2022}.

\section{Conclusions}
\label{sec:conclusions}

First global turbulence simulations have been carried out across the edge and SOL of the ASDEX Upgrade tokamak in (ITER baseline like) attached H-mode conditions, in an inter-ELM phase. Away from the inner (core) boundary of the simulations, the background profiles are evolved self-consistently together with and according to radial turbulent and neoclassical transport, allowing predictions for transport across the separatrix. The results have been compared to experimental measurements of outboard mid-plane profiles: a satisfactory agreement is obtained for the plasma density, electron and ion temperature, as well as the radial electric field. 

The total heat transport is somewhat underpredicted: experimentally, 3.1 MW of turbulent and neoclassical transport have been expected (balancing heating power and losses due to radiation and ELMs), while 2\,MW were found in the highest resolution simulation. For this validation, to obtain faster saturation times, the simulations were adaptively flux driven to maintain prescribed density and temperature at the core boundary: a better match with the experiment might be obtainable just by modifying the boundary conditions within experimental error bars. 
We also have not exploited the freedom to adjust transport levels with free-streaming fractions (see eq.~\eref{eq:LF2FS}) as in \cite{Fundamenski2005,Xia2015,Zholobenko2021a,Zhang2024}. 
This could be further explored in future. For now, we consider such an agreement with the experiment satisfactory, considering that a drift-fluid model has been used. To build confidence in the results, we have carried out resolution scans within computational feasibility. We have analysed in more detail the composition of the radial electric field. And we determined the character of the radial transport: our turbulence is of drift-Alfvén type in the upper part of the pedestal, but most transport is neoclassical. At the pedestal foot, on the other hand, turbulence is strongly electromagnetic, with clear kinetic ballooning mode (KBM) characteristics.

For the radial electric field, the picture is qualitatively similar to our previous L-mode simulations \cite{Zholobenko2021}: in the (mid-) SOL, it is predominantly determined by sheath boundary conditions. The neutral gas cooling of the divertor is thus important for keeping shear flows and simulations overall stable. In the confined region, the radial electric field mostly balances the diamagnetic compression and is proportional to the ion pressure gradient. At higher resolution and at $\rho_\mathrm{pol} \geq 0.99$, $E_r$ is even nearly exactly equal to $\partial_r p_\mathrm{i}/en$, consistently with experimental findings on ASDEX Upgrade \cite{Viezzer2013,Cavedon2016}. However, we argue that this is not due to the absence of any flows, but rather because the zonal flow balances neoclassical poloidal and toroidal rotation. In L-mode simulations, the zonal flow instead led to a perturbation of the pressure gradient \cite{Zholobenko2021}. This difference could be the reason why GAMs \cite{Zholobenko2019,Conway2021} were observed in L-mode, but not in H-mode. On the other hand, at lower resolution and with the Landau-fluid closure, more pronounced zonal flow oscillations inside the $E_r$ well are found, similar to observations at JET \cite{Hillesheim2016}. The exact amplitude of the zonal flow depends on numerical resolution and physical model details. As a caveat, we stress that our current simulations employ the long-wavelength polarisation limit and insulating sheath boundary conditions. Nonetheless, it can be stated that some amount of zonal flows seems to survive even in H-mode conditions, albeit it is smaller than the ion pressure gradient part. 

We stress the improvements of our drift-fluid model that have been necessary to obtain current results. These are first and foremost electromagnetic extensions of the code \cite{Zhang2024}. The inclusion of electromagnetic induction in Ohm's law \cite{Stegmeir2019} was necessary because the propagation of the parallel current on larger scales is determined by Alfvén waves \cite{Scott2021}. This is not only important in terms of physics, but it also improves the code performance \cite{Scott1997,Dannert2004,Dudson2021,Stegmeir2023} 
-- otherwise, realistic tokamak simulations were not feasible. In this work, we stress that besides magnetic induction, particularly in H-mode, it is also critical to include magnetic flutter: it has a stabilising effect on drift-waves by making them more adiabatic \cite{Scott2021}. In L-mode conditions \cite{Zhang2024}, we were finding flutter stabilisation factors of around 2, within the uncertainty introduced by the fluid closure. However, in the higher beta H-mode conditions, the stabilisation factor is close to two orders of magnitude: without magnetic flutter, we observed utterly unrealistic turbulent transport ($>100$ MW). Finally, we stress a particular challenge when magnetic flutter is included in global (``full-$f$'') turbulence models: the treatment of the background Shafranov shift \cite{Lackner2000,Snyder2007,Scott2006,Hager2020,Giacomin2022,Zhang2024}. Due to the need for field-alignment, dynamic magnetic field fluctuations must be small. However, since we evolve the full plasma pressure, our parallel current includes the background Pfirsch-Schlüter current, which leads to the Shafranov shift. Such a large-scale, quasi-stationary magnetic field shift has to be included in the fixed background magnetic equilibrium, and subtracted from magnetic flutter. We have presented our current solution to this, but we also stress the need for further improvements.

The second set of critical model improvements involves the correction of the collisional Braginskii closure for regimes of low collisionality. In particular, terms which diverge as $nT\nu^{-1}\sim T^{5/2}$ must be limited. For the parallel heat conduction, an often employed method is limiting the heat flux to a free-streaming (FS) fraction \cite{Thyagaraja1980,Scott1997,Stangeby2000,Fundamenski2005,Xia2015}. This method is simple and effective, but it introduces the free-streaming fraction $\mathrm{f^{FS}_{e,i}}$ as a free parameter. Recently, a Landau-fluid (LF) model \cite{hammett_perkins,Umansky2015,Chen2019,Zhu2021} has been implemented in GRILLIX \cite{Pitzal2023}. Here, we show that despite the significantly different (non-local) form of the heat flux, for the turbulent transport, results are very similar between the LF and FS models with $\mathrm{f^{FS}_{e,i}}=1$. A possible reason is that H-mode collisionality is not too low ($\nu\sim1$) due to the high density. The largest impact of the LF model is on the radial electric field, similar as in our previous L-mode simulations \cite{Pitzal2023}, which is likely due to the non-locality (see sec.~\ref{sec:landau}). 
Besides parallel heat conduction, in drift-fluid models, also (ion) viscosity requires limitation. It represents thermal anisotropy which is not explicitly included in our model. The ion viscosity is important for the regulation of poloidal and toroidal rotation. Most importantly, with the Braginskii closure, the damping of flows is too strong at low collisionality, resulting in large poloidal asymmetries. A remedy is found by adjusting the viscosity coefficient according to neoclassical theory \cite{Hirshman1981,PerHelander2002,Rozhansky2009}: the neoclassical flow viscosity scales like $\eta \sim \nu_*^{-1}$ only in the collisional limit. It has no collisionality dependence ($\eta \sim \nu_*^{0}$) in the plateau regime $\epsilon = 0$, and it becomes $\eta \sim \nu_*$ in the banana regime. Additionally to flow viscosity, we also include heat viscosity, which induces additional poloidal rotation and yields a more realistic radial electric field.

The characterisation of radial transport helps to understand to what extent drift-fluid models can capture H-mode turbulence, putting our work into perspective with established (mostly local) gyrokinetic studies of the plasma edge \cite{Dickinson2012,Hatch2016,Ku2018,Bonanomi2019,Kotschenreuther2019,Scott2021a,Leppin2023,Bonanomi2024}. One key finding is that in the upper half of the pedestal, in these simulations, $E\times B$ transport is small. Diamagnetic (neoclassical) transport dominates, possibly enhanced by turbulence. 
Turbulent fluctuations of a few percent persist (they increase to up to $100\%$ towards the separatrix), but they are mostly in phase with each other and cause little transport, corresponding to a very stable drift-wave regime. The ITG mode is insignificant due to $\eta_\mathrm{i}<1$. Naturally, the pedestal top has the lowest collisionality ($\nu_\mathrm{e}^*\approx1$), so ion-scale gyrokinetics is expected to play the largest role there, in particular due to collisional trapped electron modes (TEMs). Also electron scale (ETG) turbulence may be important \cite{Hatch2016,Kotschenreuther2019,Scott2021a,Leppin2023}. Note that multi-scale simulations would increase the computational cost of global simulations by at least another factor $m_\mathrm{i}/m_\mathrm{e} \approx 4000$, so it is of great interest to construct and use reduced ETG models instead. 
We stress that H-mode turbulence is strongly electromagnetic. On one hand, magnetic fluctuations (flutter) have a strongly stabilising effect on drift-wave turbulence. But on the other hand, they cause some amount of transport themselves. In particular, just inside the separatrix, we find clearly the signature of the kinetic ballooning mode (KBM). This is another reason to continue these studies with gyrokinetic models: magnetic heat transport is largely caused by parallel heat conduction along the perturbed radial magnetic field, which is the highest fluid moment that we currently apply a closure on. 
The freedom to adjust transport levels with free-streaming fractions \eref{eq:LF2FS} as in \cite{Fundamenski2005,Xia2015,Zholobenko2021a,Zhang2024} was not exploited because $\mathrm{f_{e,i}^{FS}}=1$ matches rather well the Landau-fluid closure results. However, our current model only captures linear, parallel Landau damping \cite{hammett_perkins}, and results may vary if toroidal effects \cite{Beer1996} and higher dynamic fluid moments \cite{Schekochihin2016} were included. Also heat anisotropy is expected to be important. Note that there can be other reasons for insufficient radial transport in our simulations than gyrokinetics. It is possible that extending the simulation domain towards the core (e.g. to $\rho_\mathrm{pol}=0.8$) could increase the transport. Including current gradient driven peeling modes \cite{Snyder2002,Li2022} could be important. ELMs might not only cause magnetic transport themselves, but also trigger an increased turbulence activity \cite{Kendl2010}, which would require to study the interaction between turbulence and larger scale MHD events. It is also possible that including impurities would modify the transport level \cite{Tokar2000}, despite the low $Z_\mathrm{eff}<1.5$.

Despite the significant motivation to study edge-SOL turbulence gyrokinetically, we conclude that H-mode conditions can be simulated reasonably well also with a transcollisional drift-fluid model. This is important due to the high computational cost of gyrokinetics at high collisionality \cite{Maeyama2019}, which is found in detached divertor conditions that are mandatory for a fusion reactor: while collisionality is roughly between 1 and 10 at the OMP separatrix (see fig.~\ref{fig:coll_etai}), it can be $>10^4$ in the divertor. Thus, we suggest that fluid models can be employed for such conditions. Of particular interest are core-edge-divertor integrated regimes which combine good confinement, absence of ELM transients and manageable heat exhaust \cite{Viezzer2018}. Among these is the quasi-continuous exhaust (QCE) regime \cite{Faitsch2021}, which has a particularly large SOL width and is obtained at high outer edge collisionality, and the X-point radiator (XPR) regime \cite{Bernert2020}, which is obtained with feedback-control of full detachment, reliably avoids ELMs and may allow for a compact radiative divertor \cite{Lunt2023}. These reactor attractive regimes are of prime interest for our future studies. With further code performance optimizations \cite{Stegmeir2023}, such simulations could help to extrapolate these regimes to ITER and DEMO fusion reactors.

\section*{Acknowledgments}

For additional experimental data analysis, we thank Rainer Fischer, Davide Silvagni, Michael Faitsch, Sebastian H\"{o}rmann, Thomas P\"{u}tterich and Elisabeth Wolfrum. 
Further, we thank Sergei Makarov, Clarisse Bourdelle, Emiliano Fable and Per Helander for discussions on neoclassical theory. We thank Andres Cathey for discussions about MHD and ELMs. We thank Ondrej Grover and Garrard Conway for discussions about the radial electric field, the Reynolds stress and GAMs. We thank Leonhard Leppin and Tobias G\"{o}rler for discussions about pedestal turbulence. We thank David Tskhakaya for discussing sheath boundary conditions. Finally, we thank Baptiste Frei for discussing fluid closures.

This work has been carried out within the framework of the EUROfusion Consortium, funded by the European Union via the Euratom Research and Training Programme (Grant Agreement No 101052200 -- EUROfusion). Views and opinions expressed are however those of the author(s) only and do not necessarily reflect those of the European Union or the European Commission. Neither the European Union nor the European Commission can be held responsible for them.
This work has been granted access to the HPC resources of the EUROfusion High Performance Computer (Marconi-Fusion) under the project TSVV3. 
Lidija Radovanovic is a fellow of the Friedrich Schiedel Foundation for Energy Technology.

\begin{appendices}

\section{Global drift-reduced Braginskii equations with transcollisional extensions}
\label{chap:Braginskii_equations}

The physical model in GRILLIX is based on global drift-reduced Braginskii equations \cite{braginskii65,Zeiler1997,zeiler:habil99,Zholobenko2021}.  The model is fully electromagnetic \cite{Stegmeir2019,Zhang2024}. Fluid closure terms which are anti-proportional to the collisionality, viscosity and heat conduction, and thus diverge at vanishing collisionality, have been extended: the ion viscosity includes neoclassical corrections \cite{Hirshman1981,PerHelander2002,Rozhansky2009}, and the heat conduction is approximated with a Landau-fluid closure \cite{hammett_perkins,Umansky2015,Chen2019,Zhu2021,Pitzal2023}. Since the plasma background is evolved together with turbulent fluctuations, it must be maintained by realistic sources. To this end, the plasma is coupled to a diffusive neutral gas model \cite{Zholobenko2021a}. It has now been extended to a three-moment model \cite{Rensink1998,Uytven2020}, evolving the neutral gas density, parallel momentum and pressure. However, this extension is not critical to the present manuscript, and thus will be detailed elsewhere. The full set of plasma equations in normalised form is summarised below.

In the following, time scales are normalised to~$R_0/c_{s0}$, with $R_0$ the major radius. $c_{s0}=\sqrt{T_\mathrm{e0}/m_\mathrm{i}}$ is the sound speed at reference electron temperature $T_\mathrm{e0}$ and ion mass $m_\mathrm{i}$. Perpendicular scales are normalised to the sound Larmor radius $\rho_{s0}=\sqrt{T_\mathrm{e0}m_\mathrm{i}}/(eB_0)$ (in SI units) and parallel scales to $R_0$. The dynamical fields evolved in GRILLIX are the plasma density $n$ normalised to a reference density $n_0$, the electrostatic potential $\varphi$ normalised to $T_\mathrm{e0}/e$, the parallel ion velocity $u_\parallel$ normalised to $c_{s0}$, the electron and ion temperatures $T_\mathrm{e}$ and $T_\mathrm{i}$ normalised to reference values $T_\mathrm{e0}$ respectively $T_\mathrm{i0}$, the parallel current $j_\parallel$ normalised to $en_0c_{s0}$, and the parallel component of the perturbed electromagnetic potential $A_\parallel$ normalised to $\beta_0B_0\rho_{s0}$, with $\beta_0=\mu_0 n_0T_\mathrm{e0}/B_0^2$ and the vacuum permeability $\mu_0$.

With this normalisation, the system is determined by 9 dimensionless parameters. The four collisionless parameters are the drift scale $\delta_0 = R_0/\rho_{s0}$, dynamical plasma beta $\beta_0=\mu_0 n_0T_\mathrm{e0}/B_0^2$, and electron to ion mass and temperature ratios $\mu = m_\mathrm{e}/m_\mathrm{i}$ respectively $\zeta = T_\mathrm{i0}/T_\mathrm{e0}$. For the collisional parameters, we require $\tau_\mathrm{e0}$ and $\tau_\mathrm{i0}$: the electron respectively ion collision times \cite{braginskii65} evaluated at reference temperature and density and normalised to $R_0 / c_{s0}$. Then, the remaining five dimensionless collisional parameters of the system are the electron collisionality $\nu_\mathrm{e0} = 1 / \tau_\mathrm{e0}$, normalised parallel resistivity $\eta_{\parallel0} = 0.51 \mu \nu_\mathrm{e0}$, normalised parallel electron and ion heat conductivities $\kappa_{\parallel 0}^\mathrm{e} = 3.15\tau_\mathrm{e0} / \mu$ respectively $\kappa_{\parallel 0}^\mathrm{i} = 3.9\tau_\mathrm{i0}\zeta$, and normalized ion viscosity $\eta_\mathrm{i0} = 0.96\tau_\mathrm{i0}$. 

The six dynamical equations for the plasma read
\begin{equation}
    \frac{\mathrm{d}}{\mathrm{d}t} n = n C(\varphi) - C(p_\mathrm{e}) + \nabla \cdot \left[ \left( j_{\parallel} - n u_{\parallel} \right) \mathbf{b} \right] +  D_{n} + S_{n},
\label{continuity_equation}
\end{equation}
\begin{eqnarray}
    \label{vorticity_equation}
    \nabla \cdot &\left[ \frac{n}{B^2} \left( \frac{\mathrm{d}}{\mathrm{d}t} + u_{\parallel} \nabla_{\parallel} \right) \left( \nabla_{\perp} \varphi  + \zeta \frac{\nabla_{\perp} p_\mathrm{i}}{n}\right)  \right] \\ \nonumber
    & = -C(p_\mathrm{e} + \zeta p_\mathrm{i}) + \nabla \cdot \left( j_{\parallel} \mathbf{b} \right) - \frac{\zeta}{6} C\left( G \right) +  D_{\Omega} + S_\Omega,
\end{eqnarray}
\begin{equation}
    \left( \frac{\mathrm{d}}{\mathrm{d}t} + u_{\parallel} \nabla_{\parallel} \right) u_{\parallel} =  - \frac{\nabla_{\parallel} \left( p_\mathrm{e} + \zeta p_\mathrm{i} \right)}{n} + \zeta T_\mathrm{i} C \left( u_{\parallel} \right) - \frac{2}{3} \zeta \frac{B^{3/2}}{n} \nabla_{\parallel} \frac{G}{B^{3/2}} +  D_{u_{\parallel}} + S_{u_\parallel},
\label{momentum_equation}
\end{equation}
\begin{equation}
    \frac{\partial}{\partial t} \Psi_m + \mu \left(\mathbf{v}_E\cdot\nabla + v_{\parallel} \nabla_{\parallel} \right) \frac{j_{\parallel}}{n} + \left( \frac{\eta_{\parallel 0}}{T_e^{3/2}} \right) j_{\parallel} = - \nabla_{\parallel} \varphi + \frac{\nabla_{\parallel} p_\mathrm{e}}{n} + 0.71 \nabla_{\parallel} T_\mathrm{e} +  D_{\Psi_{m}},
\label{ohms_law}
\end{equation}
\begin{eqnarray}
    \label{electron_temperature_equation}
    \frac{3}{2} &\left( \frac{\mathrm{d}}{\mathrm{d}t} + v_{\parallel} \nabla_{\parallel} \right) T_\mathrm{e} = T_\mathrm{e} C(\varphi) - \frac{T_\mathrm{e}}{n} C(p_\mathrm{e}) - \frac{5}{2}T_\mathrm{e} C(T_\mathrm{e}) - T_\mathrm{e} \nabla \cdot \left( v_{\parallel} \mathbf{b} \right) + 0.71 \frac{T_\mathrm{e}}{n} \nabla \cdot (\mathbf{b} j_{\parallel}) \\ \nonumber
    &- \frac{1}{n} \nabla \cdot \left( q_{\parallel \mathrm{e}} \mathbf{b} \right) - 3 \nu_{e0} \mu \left( \frac{n}{T_\mathrm{e}^{3/2}} \right) \left( T_\mathrm{e} - \zeta T_i \right) + \left( \frac{\eta_{\parallel 0}}{T_\mathrm{e}^{3/2}} \right) \frac{j_{\parallel}^2}{n} + \frac{3}{2} \left( D_{T_\mathrm{e}} + S_{T_\mathrm{e}} \right),
\end{eqnarray} 
\begin{eqnarray}
    \label{ion_temperature_equation}
    \frac{3}{2} &\left( \frac{\mathrm{d}}{\mathrm{d}t} + u_{\parallel} \nabla_{\parallel} \right) T_\mathrm{i} = T_\mathrm{i} C(\varphi) - \frac{T_\mathrm{i}}{n} C(p_\mathrm{e}) + \frac{5}{2} \zeta T_\mathrm{i} C(T_\mathrm{i}) - T_\mathrm{i} \nabla \cdot (u_{\parallel} \mathbf{b}) + \frac{T_\mathrm{i}}{n} \nabla \cdot \left( j_{\parallel} \mathbf{b} \right) \\ \nonumber
    &- \frac{1}{n} \nabla \cdot \left( q_{\parallel \mathrm{i}} \mathbf{b} \right) + 3 \nu_{e0} \mu \left( \frac{n}{T_\mathrm{e}^{3/2}} \right) \left( \frac{1}{\zeta} T_\mathrm{e} - T_\mathrm{i} \right) + \frac{1}{3 \hat{\eta}_{\mathrm{flow}}} \frac{1}{n T_\mathrm{i}^{5/2}} G^2_\mathrm{flow} + \frac{3}{2} \left( D_{T_\mathrm{i}} + S_{T_\mathrm{i}} \right).
\end{eqnarray}
They are supplemented by Ampere's law for the parallel current, $\nabla_{\perp}^2 A_{\parallel} = -j_{\parallel}$, and the following definitions: the advective derivative is $\frac{\mathrm{d}}{\mathrm{d}t} = \frac{\partial}{\partial t} + \mathbf{v}_E\cdot\nabla$, with $\mathbf{v}_E = \frac{\delta_0}{B}\left(\mathbf{b}_0\times\nabla\varphi\right)$, and the curvature operator is $C(f)=-\delta_0\left(\nabla\times\frac{\mathbf{B_0}}{B^2}\right)\cdot\nabla f$. The parallel electron velocity is $v_\parallel = u_\parallel-j_\parallel/n$, the electron and ion pressures are  $p_\mathrm{e} = nT_\mathrm{e}$ respectively $p_\mathrm{i}=nT_\mathrm{i}$, the generalized vorticity is $\Omega = \nabla\cdot\ \left[ \frac{n}{B^2}\left(\nabla_\perp\varphi+\zeta \frac{\nabla_\perp p_\mathrm{i}}{n}\right)\right]$ and the generalised electromagnetic potential is $\Psi_m = \beta_0 A_\| + \mu \frac{j_\|}{n}$. The magnetic field unit vector contains both the equilibrium component and the turbulence-induced perturbation from $A_1$, $\mathbf{b} = \mathbf{b}_0 + \mathbf{b}_1$, with $\mathbf{b}_1 =\frac{\beta_0}{B}\nabla\times(A_1 \mathbf{b}_0)\approx \frac{\beta_0}{B} \nabla A_1 \times \mathbf{b}_0$. We made sure that the latter approximation in the implementation, which is typically used in literature \cite{Kendl2010,Xia2015,Zhu2018,Giacomin2022,Michels2022}, has indeed no influence on our results, although it formally breaks $\nabla\cdot\mathbf{B}=0$. The parallel gradient is then defined as $\nabla_\parallel = \mathbf{b}\cdot\nabla$. For the divergence, we approximate $\nabla \cdot (f\mathbf{b}_1) \approx \mathbf{b}_1 \cdot \nabla f$, but we keep $\nabla \cdot (f\mathbf{b}_0) = \mathbf{B}_0 \cdot \nabla (f/B)$. 
For numerical performance, $A_1$ should be a purely fluctuating quantity, while all stationary background components should be contained in the equilibrium magnetic field $\mathbf{b}_0$. The discretization of the equilibrium operators $\mathbf{b}_0 \cdot \nabla f$ and $\nabla \cdot (f \mathbf{b}_0)$ is done in a locally field-aligned manner, by field line tracing within the FCI framework  \cite{Stegmeir2016,Stegmeir2017,Stegmeir2018,Michels2021,Stegmeir2023}. However, our global model evolves not just the fluctuations of the plasma pressure, but also the background. Therefore, $A_\parallel$ contains not only fluctuations, but also the Pfirsch-Schlüter current. Since the latter is already contained in $\mathbf{b}_0$, it is subtracted via $A_1=A_\parallel-\langle A_\parallel \rangle_{\varphi,t}$, with the brackets indicating a toroidal and time average. More details on $\mathbf{b}_1$ can be found in \cite{Zhang2024} and section \ref{sec:magnetic_shift}.

Next, we define the (normalised) ion viscous stress function $G$ as
\begin{eqnarray}
{G} = &- \hat{\eta}_{\mathrm{flow}} {T}_\mathrm{i}^{5/2} \left[ \frac{2}{{B}^{3/2}}{\nabla}\cdot\left({u}_\parallel {B}^{3/2}\mathbf{b}\right) - \frac{1}{2}{C}({\varphi}) - \frac{\zeta}{2n}{C}({p}_\mathrm{i}) \right] \nonumber\\
&- \hat{\eta}_{\mathrm{heat}} {T}_\mathrm{i}^{5/2} \left[ \frac{2}{n {T}_\mathrm{i} {B}^{3/2}}{\nabla}\cdot\left(q_{\parallel \mathrm{i}}{B}^{3/2}\mathbf{b}\right)  - \frac{5\zeta}{4}{C}({T}_\mathrm{i}) \right].
\end{eqnarray}
The form of the flow viscosity is the same as the pure drift-reduced Braginskii expression \cite{Zholobenko2021}. Neoclassical extensions \cite{Hirshman1981,PerHelander2002,Rozhansky2009} enter here in two ways: firstly, the second row introduces the heat viscosity, i.e.~a viscous force due to the parallel and diamagnetic heat fluxes. Secondly, while the Braginskii viscosity coefficient $\eta_\mathrm{i0} = 0.96\tau_\mathrm{i0}$ is simply proportional to the collision time and thus diverges at vanishing collisionality, the neoclassical corrections modify it to
\begin{equation}
\hat{\eta}_{\mathrm{flow}} = \eta_{i0} \frac{1}{1 + \nu_*^{-1}} \frac{1}{1 + \epsilon^{-3/2} \nu_*^{-1}} \qquad \mathrm{and} \qquad 
\hat{\eta}_{\mathrm{heat}} = \frac{0.75}{0.96} \frac{8}{15} (k^T - 1) \hat{\eta}_{\mathrm{flow}}.
\label{eq:neocl_visc}
\end{equation}
Here, $\epsilon = a/R$ is the inverse aspect ratio. The ion collisionality normalised to the bounce frequency is defined as
\begin{eqnarray}
\nu_* = \frac{0.96 q_{95}}{\sqrt{2\zeta}\eta_\mathrm{i0}\epsilon^{3/2}} \frac{n}{T_\mathrm{i}^2}.
\end{eqnarray}
The safety factor $q_{95} \approx 4$ and $\epsilon \approx 0.3$, typical for AUG, are taken as constant in this work. Finally, we have 
\begin{eqnarray}
k^T = \frac{ -0.17 + 1.05\sqrt{\nu_*\sqrt{2}} + 2.7\left(\nu_*\sqrt{2}\right)^2\epsilon^3 }{ 1 + 0.7\sqrt{\nu_*\sqrt{2}} + \left(\nu_*\sqrt{2}\right)^2\epsilon^3 }.
\label{eq:k_T}
\end{eqnarray}
The precise form of this ion viscosity is taken from the SOLPS code \cite{Rozhansky2009}, whereby we found useful the coordinate-free multi-species representation in \cite{Makarov2021}. We remark that for technical reasons, only the flow viscosity part is currently included in the ion temperature equation, with which we checked that it has no effect anyways. The main effect of the modified viscosity coefficient is that it scales like $\eta \sim \nu_*^{-1}$ in the collisional limit, like $\eta \sim \nu_*^{0}$ in the plateau regime $\epsilon = 0$, and like  $\eta \sim \nu_*$ in the banana regime. 

It remains to define the conductive heat flux $q_{\parallel \mathrm{e,i}}$. With the Landau-fluid model, using the fast non-Fourier method developed by Dimits \textit{et al.} \cite{Dimits2014}, it is $q_{\parallel \mathrm{e,i}} = \sum_l q^{\mathrm{LF}}_{\parallel \mathrm{e,i}, l}$, determined by a set of elliptic equations along the magnetic field,
\begin{equation}
    \frac{\kappa_{\parallel 0}^\mathrm{e} {\mu}}{3.16 \delta_\mathrm{e} \alpha_l \sqrt{8 / \pi}} \left( \left( \frac{3.16 \delta_\mathrm{e}}{\kappa_{\parallel 0}^\mathrm{e} \sqrt{\mu}} \frac{{n}}{{T}_\mathrm{e}^2} \beta_l \right)^2 \mathbf{b}_0 - \mathbf{b}_0\cdot{\nabla} {\nabla}\right) \cdot {q}_{\parallel \mathrm{e},l}^\mathrm{LF} \mathbf{b}_0 = - \frac{{n}^2}{\sqrt{{T}_\mathrm{e}}} {\nabla}_{\parallel} \log {T}_\mathrm{e},
\label{eqn:electron_landau_fluid_equation_in_grillix}
\end{equation}
\begin{equation}
    \frac{\kappa_{\parallel 0}^\mathrm{i} \zeta^{-1}}{3.9 \delta_\mathrm{i} \alpha_l \sqrt{8 / \pi}} \left( \left( \frac{3.9 \delta_\mathrm{i}}{\kappa_{\parallel 0}^\mathrm{i} \zeta^{-1/2}} \frac{{n}}{{T}_\mathrm{i}^2} \beta_l \right)^2 \mathbf{b}_0 - \mathbf{b}_0\cdot{\nabla} {\nabla} \right) \cdot {q}_{\parallel \mathrm{i},l}^\mathrm{LF} \mathbf{b}_0 = - \frac{{n}^2}{\sqrt{{T}_\mathrm{i}}} {\nabla}_{\parallel} \log {T}_\mathrm{i}.
\label{eqn:ion_landau_fluid_equation_in_grillix}
\end{equation}
In this work, we use $l\in(1,7)$, with the coefficients $\alpha_l$ and $\beta_l$ defined in \cite{Chen2019}, $\delta_\mathrm{e} = 0.5$ and $\delta_\mathrm{i} = 0.41$. Details about the model can be found in \cite{hammett_perkins,Umansky2015,Chen2019,Zhu2021}, and specifically about the implementation in GRILLIX in \cite{Pitzal2023}. This system of equations poses a stiff 3D elliptic problem, which we solve with the same iterative Krylov methods as previously the Braginskii (or free-streaming-limited) heat conduction \cite{Zholobenko2019}. Importantly, magnetic flutter is included on the right-hand side of the equation as well as in the parallel divergence of the heat flux in the temperature equations \eref{electron_temperature_equation} and \eref{ion_temperature_equation}, but not on the left-hand side, as justified in \cite{Pitzal2023}.

Alternatively, as justified in section \ref{sec:landau} and mostly used throughout this work, the free-streaming-limited heat flux \cite{Thyagaraja1980,Scott1997,Stangeby2000,Fundamenski2005,Xia2015} has the form of Fick's law $q_{\parallel \mathrm{e,i}} = -\tilde{\kappa}^\mathrm{e,i}_{\parallel0} T_\mathrm{e,i}^{5/2} \nabla_\parallel T_\mathrm{e,i}$ with the limited heat conductivities
\begin{eqnarray}
    \tilde{\kappa}^\mathrm{e}_{\parallel0} = \kappa^\mathrm{e}_{\parallel0} \left( 1 + \frac{\kappa^\mathrm{e}_{\parallel0}\sqrt{\mu}}{\mathrm{f}^\mathrm{FS}_\mathrm{e}} \frac{T_\mathrm{e}^2}{nq_{95}} \right)^{-1}  \qquad \mathrm{and} \qquad 
    \tilde{\kappa}^\mathrm{i}_{\parallel0} = \kappa^\mathrm{i}_{\parallel0} \left( 1 + \frac{\kappa^\mathrm{i}_{\parallel0}}{\mathrm{f}^\mathrm{FS}_\mathrm{i}\sqrt{\zeta}} \frac{T_\mathrm{i}^2}{nq_{95}}  \right)^{-1}.
\end{eqnarray}

The $S$ terms are source functions. They include both the particle and energy sources at the core boundary of the simulations, as well as neutral particle interaction terms. For numerical reasons, respectively in order to cut the turbulent spectrum, a dissipation is added to all equations of the form
\begin{equation}
	\mathcal{D}_f=\nu_{f\perp}\nabla_\perp^{2N}f + \nu_{f\parallel}\nabla\cdot \left(\mathbf{b}_0\nabla_{\parallel0} f\right) + \nabla\cdot\left(\nu_{f,\mathrm{buffer}}\nabla_\perp f\right),
	\label{dissperp}
\end{equation}
with constants $\nu_{f\perp}$, $\nu_{f\parallel}$ for every field. $\nu_{f,\mathrm{buffer}}$ is zero in most of the domain, but is high in the last few grid points towards the radial boundaries. For hyperviscosity, $N = 3$ is chosen.

Finally, boundary conditions are required to close the system. At the inner and outer limiting flux surfaces, the following homogeneous boundary conditions are applied
\begin{equation}
\partial_\rho n=0, \qquad  \partial_\rho T_\mathrm{e,i} = 0, \qquad   \partial_\rho u_\parallel = 0, \qquad  \Omega=0, \qquad  A_\parallel = 0, \qquad  \partial_\rho j_\parallel=0,
\end{equation} 
to prevent particles and energy fluxes through the boundary. For the potential, the sheath boundary condition $\left.\phi\right|_{\rho_\mathrm{max}}=\Lambda T_\mathrm{e}$ is applied at the outer wall boundary. At the inner (core) boundary $\rho_\mathrm{min}$, the zonal homogeneous Neumann boundary condition $\partial_\rho \left< \varphi \right>_\theta = 0$ and $\varphi = \left< \varphi \right>_\theta$ is applied, which allows the potential to float but prevents net $E \times B$ flux through that boundary.

At the divertor, insulating sheath boundary conditions are applied,
\begin{eqnarray}
    u_\parallel \gtrless \sqrt{T_\mathrm{e} + \zeta T_\mathrm{i}}, \qquad j_\parallel = 0, \qquad \varphi =\Lambda T_\mathrm{e}, \qquad \nabla_\parallel n = 0, \qquad \nabla_\parallel \Omega = 0, \\
    \nabla_\parallel T_\mathrm{e} = 0, \qquad \nabla_\parallel T_\mathrm{i} = 0, \qquad q^{\mathrm{LF}}_{\parallel \mathrm{e,i}} = \pm \gamma_\mathrm{e,i} {n} {T}_\mathrm{e,i} \sqrt{{T}_e + \zeta {T}_i},
    \label{eq:Bohm_bc}
\end{eqnarray}
where $\Lambda = 2.69$, $\gamma_\mathrm{e} = 1$ and $\gamma_\mathrm{i} = 0.1$, implemented with the volume penalisation method \cite{Stegmeir2019,Pitzal2023}. In the free-streaming model, the heat flux boundary condition has to be set on the temperature gradient, $\nabla_\parallel T_\mathrm{e,i} = \pm \gamma_\mathrm{e,i} {n} {T}_\mathrm{e,i} \sqrt{{T}_e + \zeta {T}_i} / (-\tilde{\kappa}^\mathrm{e,i}_{\parallel0} T_\mathrm{e,i}^{5/2})$. 
We have tested that neither setting $\nabla_\parallel j_\parallel = 0$ as in \cite{Oliveira2022} nor drift-corrections to $u_\parallel$ made a significant difference in the present work, likely due to penalisation. Implementing better (conducting sheath) boundary conditions, especially for $\varphi$, remains a challenge for FCI codes \cite{Stegmeir2023,Wiesenberger2024}.

\section{Resolution scan: OMP profiles}
\label{chap:res_scan}

\begin{figure}[htb]
\centering
\begin{minipage}{0.49\textwidth}
	\centering
    \includegraphics[trim=0.0cm 0.0cm 1.0cm 0.0cm, clip, width=1.0\linewidth]{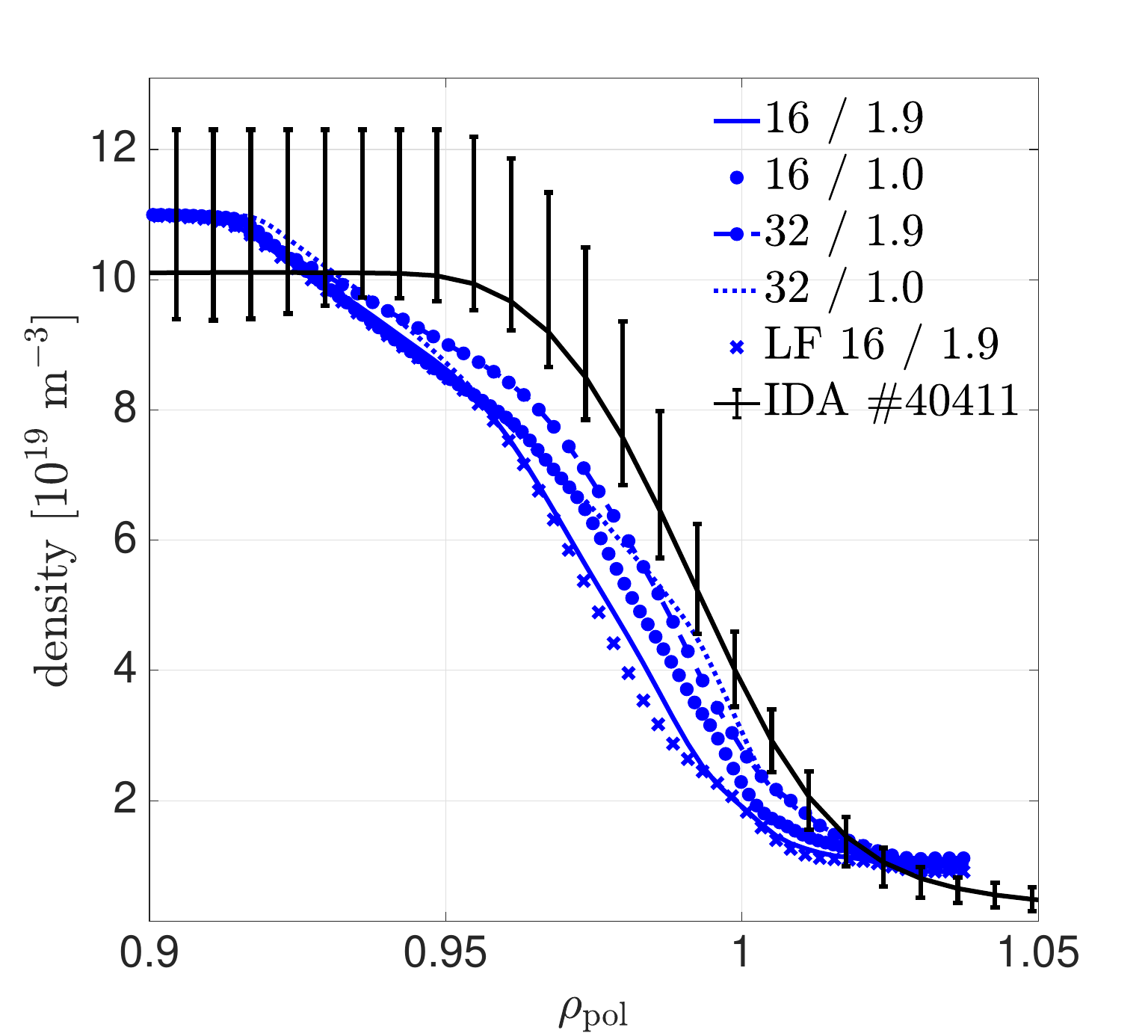}
\end{minipage}
\begin{minipage}{0.49\textwidth}
	\centering
    \includegraphics[trim=0.0cm 0.0cm 1.0cm 0.0cm, clip, width=1.0\linewidth]{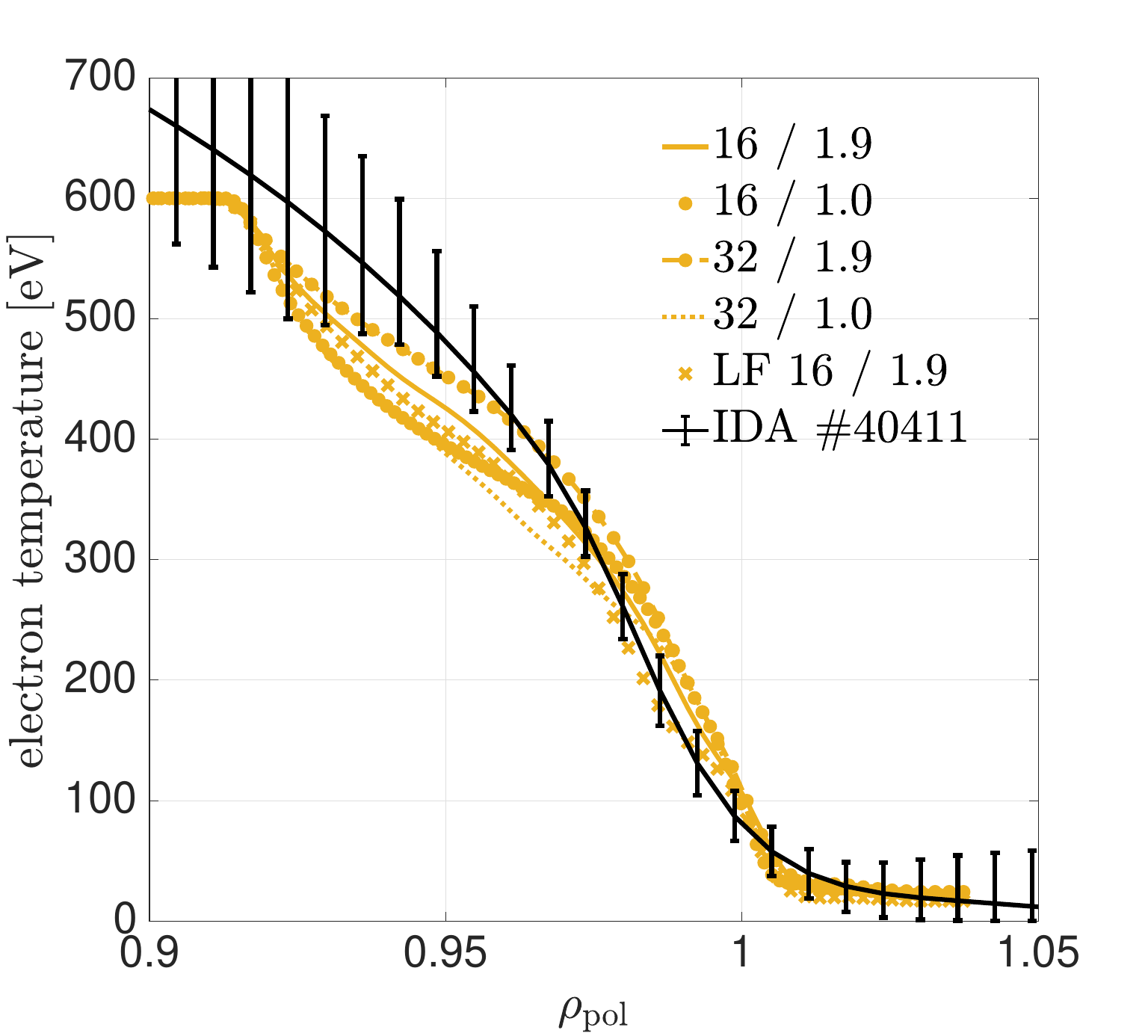}
\end{minipage}
\begin{minipage}{0.49\textwidth}
	\centering
    \includegraphics[trim=0.0cm 0.0cm 1.0cm 0.0cm, clip, width=1.0\linewidth]{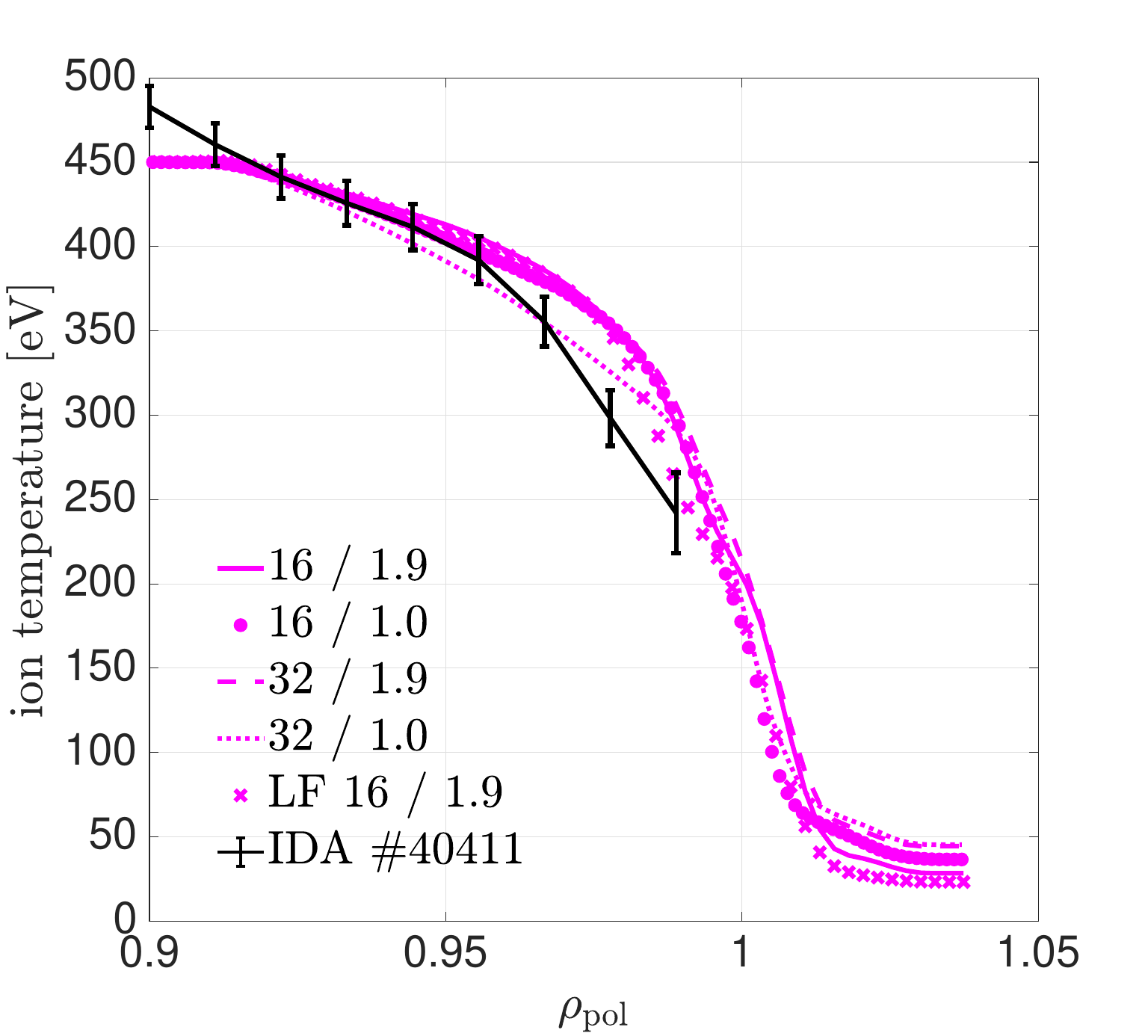}
\end{minipage}
\begin{minipage}{0.49\textwidth}
	\centering
    \includegraphics[trim=0.0cm 0.0cm 1.0cm 0.0cm, clip, width=1.0\linewidth]{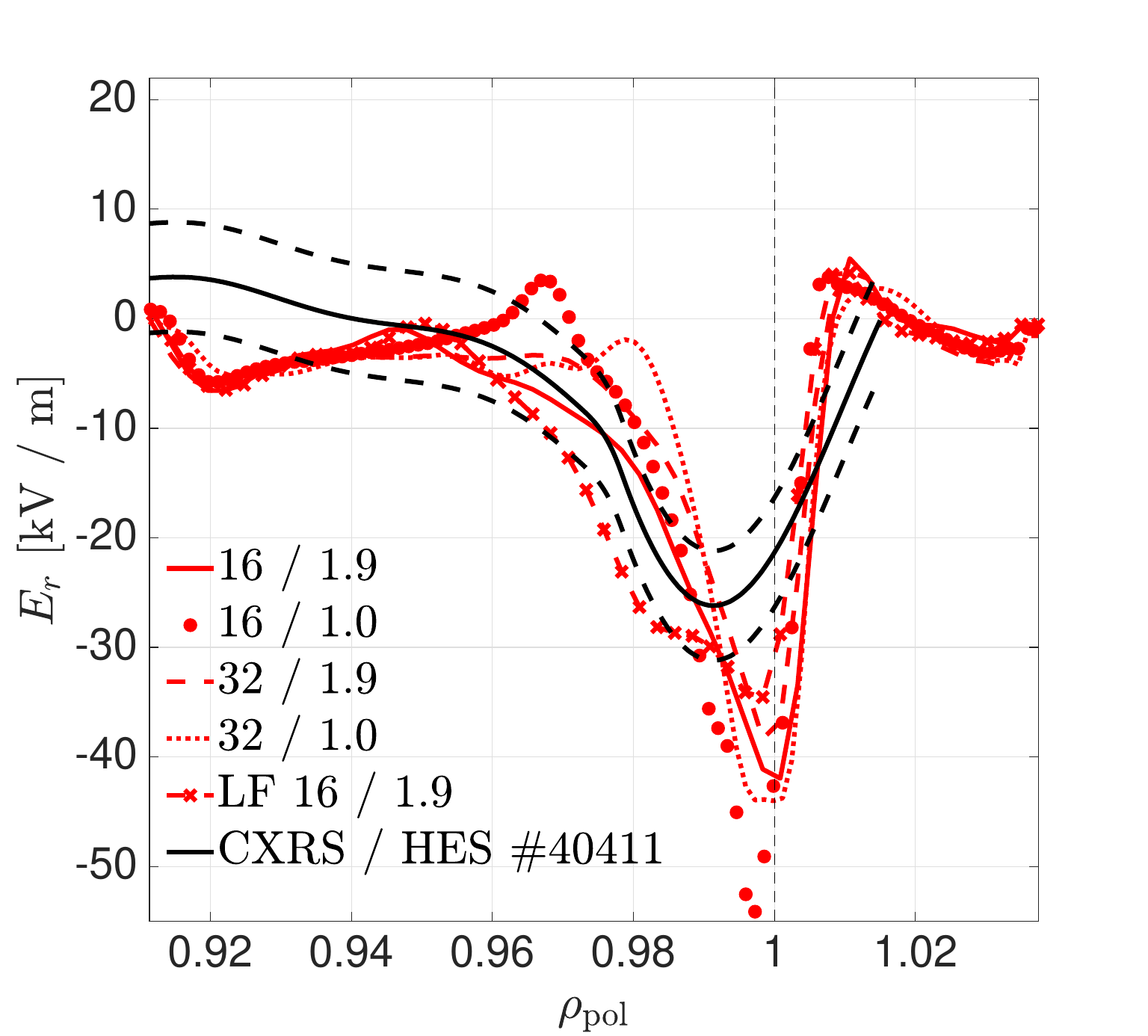}
\end{minipage}
\caption{Comparison of OMP profiles for simulations with varying resolution. In the legend, the first column corresponds to the toroidal resolution $n_\mathrm{tor}$, and the second column to the poloidal resolution $h / \rho_{s0}$. For the lower resolution case 16 / 1.9, also a comparison with a Landau-fluid simulation is given (all others are with the free-streaming-limited parallel heat conduction). }
\label{fig:res_scan_profiles}
\end{figure}

Numerical convergence results are presented in figure \ref{fig:res_scan_profiles} for outboard-midplane profiles of plasma density, electron and ion temperature, and the radial electric field. The differences are mostly within experimental error bars. For $E_r$, the largest deviation is for 16 / 1.0 ($n_\mathrm{tor}$ / $h / \rho_{s0}$), which disappears again for 32 / 1.0, indicating that both poloidal and toroidal resolution need to be increased. The most significant deviations are for the plasma density, where a higher resolution yields results closer to the experiment.

Most simulations have been performed with the cheaper free-streaming-limited parallel heat flux model (FS). For the lower resolution 16 / 1.9, results have also been obtained with the Landau-fluid model (LF, see sec.~\ref{sec:landau} and appendix \ref{chap:Braginskii_equations}). LF and FS are not significantly different, except that for $E_r$ the LF result is slightly closer to the experiment. This is similar to our previous L-mode observations \cite{Pitzal2023}. A possible explanation is given in sec.~\ref{sec:landau}.

\end{appendices}

\section*{References}
\bibliographystyle{rackstyle}
\bibliography{H2023}

\end{document}